\pdfoutput=1

\documentclass[11pt]{article}
\usepackage[utf8]{inputenc}
\usepackage{natbib}

\usepackage{ifpdf}
\usepackage{array}
\usepackage{amsfonts}
\usepackage{bm}
\usepackage[mathscr]{euscript}
\usepackage{enumerate}
\usepackage{amsmath}
\usepackage{graphics}
\usepackage{graphicx}
\usepackage{subfig}
\usepackage{algorithm}
\usepackage{algpseudocode}
\usepackage{authblk}
\usepackage{booktabs}
\usepackage[labelsep=period]{caption}
\usepackage[super]{nth}
\usepackage{multirow}
\usepackage{hyperref}
\usepackage{amsthm}
\usepackage{linegoal}
\usepackage{titlesec}
\usepackage{chngcntr}
\usepackage{apptools}
\usepackage[margin=1in]{geometry}
\usepackage{pdflscape}
\usepackage{afterpage}

\graphicspath{{./figures/}}

\makeatletter
\newcommand{\algmargin}{\the\ALG@thistlm}
\makeatother
\newlength{\whilewidth}
\algdef{SE}[parWHILE]{parWhile}{EndparWhile}[1]
  {\parbox[t]{\dimexpr\linewidth-\algmargin}{%
     \hangindent\whilewidth\strut\algorithmicwhile\ #1\ \algorithmicdo\strut}}{\algorithmicend\ \algorithmicwhile}%
\algnewcommand{\parState}[1]{\State%
  \parbox[t]{\dimexpr\linewidth-\algmargin}{\strut #1\strut}}

\algrenewcommand\algorithmicforall{\textbf{for each}}

\newcolumntype{L}[1]{>{\raggedright\arraybackslash}p{#1}}
\newcolumntype{C}[1]{>{\centering\arraybackslash}p{#1}}
\newcolumntype{R}[1]{>{\raggedleft\arraybackslash}p{#1}}

\makeatletter
\def\BState{\State\hskip-\ALG@thistlm}
\makeatother
\newcommand{\argmin}{\mathop{\rm arg~min}\limits}

\title{\bf Insights from Machine Learning for Evaluating Production Function Estimators on Manufacturing Survey Data}

\author[1]{José Luis Preciado Arreola}
\author[1]{Daisuke Yagi}
\author[1,2]{Andrew L. Johnson}
\affil[1]{Department of Industrial and Systems Engineering\\Texas A\&M University}
\affil[2]{Department of Information and Physical Sciences\\Osaka University}

\date{\today}

\begin{document}
\def\spacingset#1{\renewcommand{\baselinestretch}%
	{#1}\small\normalsize} \spacingset{1}

\maketitle

\begin{abstract}
Organizations like U.S. Census Bureau rely on non-exhaustive surveys to estimate industry-level production functions in years in which a full Census is not conducted. When analyzing data from non-census years, we propose selecting an estimator based on a weighting of its in-sample and predictive performance. We compare Cobb-Douglas functional assumptions to existing nonparametric shape constrained estimators and a newly proposed estimator. For simulated data, we find that our proposed estimator has the lowest weighted errors. For actual data, specifically the 2010 Chilean Annual National Industrial Survey, a Cobb-Douglas specification describes at least 90\% as much variance as the best alternative estimators in practically all cases considered providing two insights: the benefits of using application data for selecting an estimator, and the benefits of structure in noisy data. 
\end{abstract}



\spacingset{1.5}
\section{Introduction}
\label{sec:intro}
 
The production function literature has focused on Monte Carlo simulations to demonstrate the quality of proposed estimators and has used application data to illustrate the output of the estimator. In contrast, the machine learning literature typically uses a set of analysis methods (described below) to allow the evaluation of estimation methods using application data. A key benefit of the machine learning methods is the best estimator for a particularly application can be identified. We implement the machine learning methods for the assessment of production functions using non-exhaustive survey data\footnote{We will primarily focus on establishment censuses and surveys performed by census bureaus where an establishment is defined as a single physical location where business is conducted or where services or industrial operations are performed. An example is the U.S. annual survey of manufacturers. This survey is conducted annually, except for years ending in 2 and 7 in which a Census is performed, \citet{foster2008reallocation}. This survey includes approximately 50,000 establishments selected from the census universe of 346,000, or approximately a 15\% sampling, \citet{fort2013firms}.}. We extend the machine learning methods to allow investigation of estimation methods to characterize production functions for both the set of surveyed establishments and the set of un-surveyed establishments for a particular industry. 

Production function estimators such as Stochastic Frontier Analysis (SFA, \citet{aigner1977formulation}), as well as more recent developments such as Stochastic DEA (\citet{simar2011stochastic}) and Convex Nonparametric Least Squares (CNLS, \citet{kuosmanen2008representation}), have used Monte Carlo simulation results taking random samples drawn from a known DGP to justify their use or benefit\footnote{A body of economic literature of a less computational and more aggregate in nature is the growth accounting literature, \citet{solow1957technical} and \citet{barro2004economic}. These methods rely on price information, a cost minimization assumption and parametric forms to deterministically compute the coefficients of a first order approximation of a production function using observed input cost shares (see for example \citet{syverson2011determines}). This literature foregoes any model adequacy check of the production model.}. The developers of the methods evaluate estimator performance on the same input vectors of as the sample used for estimation. The literature provides independent comparisons between some of these methods considering a more ample set of scenarios in a Monte Carlo simulation framework, \citet{Andor2014}, but still using the same dataset to fit the production function/frontier and test its goodness-of-fit, see \citet{kuosmanen2013best} as an example. Numerous applied studies to fit production functions/frontiers have been conducted without comparing the performance of multiple estimators on the actual application datasets. 

To address the issues above, we consider a model selection strategy that improves on the current paradigm for selecting a production function models and can be implemented for application datasets. Our primary interest is estimating industry population-level production functions from potentially non-exhaustive manufacturing survey data. Accordingly, we focus on estimating the production function for the observed production units and the unobserved production units that we know exist, but are not observed in the survey. The framework we apply to both simulated and real data encompasses three elements: estimation of the optimism-corrected in-sample error (defined in Section (\ref{sec3:simulations})) for the observed establishment set, use of a learning set-testing set context to estimate the predictive error on the unobserved establishment set \citet[p.~222]{hastie2009elements}, and a finite-sample weighting, which acknowledges the potential existence of only a finite set of establishments, thus weighting the in-sample and predictive errors proportionally to the survey size. 
For the simulated datasets, we take advantage of the practically infinite data-generating capability in the Monte Carlo context. We estimate the expected optimism-corrected in-sample error and the predictive error by computing mean squared errors for our fitted estimators on previously unobserved testing sets.  These important error measures provide estimates of the functional estimator's expected predictive power for a full census of firms, of which some are unobserved. We also compute the performance against the known true DGP in both the in-sample and learning-to-testing set contexts.

For manufacturing survey datasets, we use different estimators to calculate the expected in-sample and predictive errors because we do not have the ability to generate new, unobserved datasets from the underlying DGP. For the in-sample error, given the different natures of the estimators considered, we cannot simply compute Mallow's - $C_p$ (\citeyear{mallows1973some}), Akaike Information Criterion (AIC, \citeyear{akaike1974new}), or similar optimism penalization, which are specific to linear models. Instead, we employ the parametric bootstrap approach by \citet{efron2004estimation} to estimate in-sample optimism for potentially non-linear functional estimators. For our predictive error estimation, we estimate production functions using random subsamples of the survey data and assess the predictive error of the fit, thus following a cross-validation strategy (see \citet{stone1974cross}; \citet{allen1974relationship} and \citet{geisser1975predictive} for seminal work on cross-validation). 

As is standard in the analysis of Census data, we are focused on estimating a conditional-mean production function. Therefore, in our application, we focus on characteristics of the production function such as most productive scale size, marginal productions, and elasticity of substitutions\footnote{Because of our focus on a conditional-mean estimator, we do not model inefficiency and therefore do not report inefficiency levels.}. Thus, existing literature reporting Monte Carlo simulation results for monotone and concave estimators of production frontiers, such as \citet{kuosmanen2013best} or \citet{Andor2014}, are not directly comparable. However, these two papers characterize nicely two alternative methods for comparing estimators. \citet{kuosmanen2013best} perform Monte Carlo simulations using a calibrated data generation process where parameters of the simulation are selected to closely match the empirical data for the application of interest. Alternatively, \citet{Andor2014} develop Monte Carlo simulations using commonly used parameters values in previous literature. Because both abstract away from the actual application of interest, the external validity of the simulations indicating the preferred estimator may not be apparent.

The main contribution of this paper is the proposed optimism-corrected model selection method, which allows evaluation of estimator performance in both simulated and actual manufacturing survey data. The proposed method will benefit organizations unable to collect census data on an annual basis. Further, because applications may have characteristics that favor the use of that particular estimator, we propose that relative performance of an estimator on the real application dataset should be the main criterion to follow when choosing a production function estimation method for an application. Furthermore, as an additional contribution we propose a functional estimator, Convex Adaptively Partitioned Nonparametric Least Squares (CAPNLS), which integrates the idea of adaptive partitioning from Convex Adaptive Partitioning (CAP), \citet{hannah2013multivariate}, with the global optimization strategy of the CNLS estimator. CAPNLS imposes monotonicity and concavity, two widely implemented assumptions in production, \citet{shephard1970theory,fare1994production,kumbhakar2003stochastic}.  

The remainder of this paper is organized as follows. Section 2 discusses Convex Adaptively Partitioned Nonparametric Least Squares (CAPNLS), a method that integrates CAP and CNLS using an adaptive partitioning strategy, the \citet{afriat1967construction, afriat1972efficiency} inequalities, and global optimization which greatly mitigates the overfitting issues of CNLS.  Section 3 describes a Monte Carlo simulation analysis to demonstrate the performance of the proposed estimator for both in-sample and learning set-to-testing set scenarios. Section 4 describes fitting production data for the five industries with the largest sample sizes in the 2010 Chilean Annual National Industrial Survey, compares the proposed method to the performance of other estimation methods, and discusses the results. Section 5 discusses the implications of our research, summarizes the contributions to the production/cost function estimation literature, and suggests future research.

\section{Production Model and Convex Adaptively Partitioned Nonparametric Least Squares}
\label{sec2:CAPNLS}

The CNLS estimator is an example of a sieve estimator which is extremely flexible and is optimized to fit the observed data set, \citet{white1991some} and \citet{chen2007large}. Alternatively, the adaptive least squares-based CAP, developed in the machine-learning literature, has demonstrated good predictive performance. CAP integrates model estimation and selection strategies, thus resulting in parsimonious functional estimates as opposed to optimizing fit on the observed dataset. Specifically, \citet{hannah2013multivariate} recognize that the CNLS estimator overfits the observed dataset at the boundaries of the data, thus affecting the quality of prediction for the true underlying function. Other researchers, such as \citet{huang2014finite} and \citet{balazs2015near} build examples in which CNLS estimation results in infinite Mean Squared Error due to overfitting of the sample. 

\subsection{Production Function Model}
We define the regression model for our semiparametric estimation procedure as
\begin{equation}
    \label{eq:prod1}
    Y = f(\bm{X}) + \epsilon,
\end{equation}
where $Y$ represents observed output, $f(\bm{X})$ denotes the production function, $\bm{X} = (X_1, \cdots, X_d)'$ is the input vector, $d$ is the dimensionality of the input vector, and $\epsilon$ is a symmetric random term, which we call noise, assuming the expected value is zero, $E(\epsilon) = 0$. For our estimator, we use the establishment-specific Equation \ref{eq:prod2} to derive our objective function:
\begin{equation}
    \label{eq:prod2}
    Y_i = f(X_{1i}, \cdots, X_{1i}) + \epsilon_i, i = 1, \cdots, n.
\end{equation}
For notational simplicity, we let $f_i = f(X_{1i}, \cdots, X_{1i})$ and $\bm{X}_i = X_{1i}, \cdots, X_{di}$, where $i$ is an establishment index and there are $n$ observations. We describe the decreasing marginal productivity (concavity) property in terms of $\nabla f(\bm{X})$, i.e., the gradient of $f$ with respect to $\bm{X}$, as
\begin{equation}
    \label{eq:prod3}
    f(\bm{X}_i) \leq f(\bm{X}_j) + \nabla f(\bm{X})^T (\bm{X}_i - \bm{X}_j) \; \forall i, j.
\end{equation}
Given that the constraints in Equation \ref{eq:prod3} hold, the additional constraint $\nabla f(\bm{X}_i) > 0 \; \forall i$ imposes monotonicity.

Several of the estimators discussed in this paper will take monotonicity and concavity as maintained axioms and impose them during the estimation procedure. However, there are alternative axiomatic properties, \citet{frisch1964theory}. For estimators that relax the concavity property, see for example \citet{hwangbo2015power,yagi2018SshapeUltraPassum} or in the deterministic setting see \citet{olesen2014maintaining}. In a particular application it is often not clear what should be the maintained axioms. In such cases testing the axioms is appropriate. \citet{yagi2018shape} provide a test for monotonicity and concavity and its statistical properties. Production functions are often used a component of larger economic model. Often some axiomatic conditions need to be imposed to make the production function estimation results useful to the larger model. For example mark-up models often require positive marginal products or monotonically decreasing benefits of additional inputs, \citet{de2016prices}. This would make the use of fully nonparametric estimators impossible.

\subsection{Convex Adaptively Partitioned Nonparametric Least Squares}
In this paper, we consider nonparametric approximation of $f(\bm{X})$ with several piecewise linear estimators. These estimators can consistently describe a general concave function allowing the concavity constraints in Equation \ref{eq:prod3} to be written as a system of linear inequalities. The first estimator we consider, the CNLS estimator, is a sieve estimator consistent with the functional description in Equations \ref{eq:prod1} -- \ref{eq:prod3}, \citet{kuosmanen2008representation}. CNLS is also the most flexible piecewise linear estimator we consider because it allows and has the most piecewise linear segments or hyperplanes. There are two limitations, however. The estimator imposes condition in Equation \ref{eq:prod3} by a set of numerous pairwise constraints, which requires significant computational enhancements to be applied on moderate datasets (see \citet{lee2013more} and \citet{mazumder2015Computational}). It also results in a parameter-intensive representation of $f(\bm{X})$, since it allows for potentially $N$ distinct hyperplanes. Thus, the highly detailed sample-specific fit limits the estimator's ability to predict the performance of unobserved establishments from the same industry. From an economics perspective, allowing for such a large number of distinct hyperplanes is an issue, because individual establishment observations can specify their own vector of marginal products i.e., they can place zero weight on some set of inputs and exclude them from the analysis of that establishment's production function. This implies that even if the establishment uses the inputs intensively, it can ignore the inputs recorded in the data when evaluating its performance. 

\citet{hannah2013multivariate} propose CAP, a convex regression method also consistent with Equation \ref{eq:prod1} -- \ref{eq:prod3}. The CAP algorithm partitions the dataset into input-space defined subsets (hereafter, Basis Regions) and estimate one hyperplane per basis region. CAP explores proposals for basis regions and greedily selects models with incrementally better fits as the number of hyperplanes increases/decreases/refit. CAP transition from simpler (initially linear) to more detailed models of the concave function and select the model that results in the best tradeoff between model fit and the number of parameters used.

We will now introduce Convex Adaptively Partitioned Nonparametric Least Squares (CAPNLS) which combines the advantages of both CNLS and CAP. We let $k$  and $[i]$ both be indices for basis regions to which observation $i$ is assigned for a given input set partitioning proposal. Let $\mathcal{K}_t$ be the set of basis regions defined by a partition at iteration $t$ and $K_t$ be the number of basis regions. Then, we approximate concave function $f(\bm{X})$ at input vector $\bm{X}_i$  with the estimator 
\begin{equation}
    \label{eq:prod4}
    \hat{f}_K(\bm{X}_i) = \beta_{0[i]}^* + \bm{\beta}_{-0[i]}^{*T} \bm{X}_i
\end{equation}
where
\begin{equation}
\label{eq:prod5}
    \begin{aligned}
    (\beta_{0k}^* + \bm{\beta}_{-0k}^*)_{k=1}^K &= \argmin_{(\beta_{0k} , \bm{\beta}_{-0k})_{k=1}^K} \sum_{i=1}^N (\beta_{0[i]} + \bm{\beta}_{-0[i]}^{T} \bm{X}_i - Y_i)^2 \\
    s.t. \beta_{0[i]} + \bm{\beta}_{-0[i]}^{T} \bm{X}_i \leq &\beta_{0k} + \bm{\beta}_{-0k}^{T} \bm{X}_i, \forall [i] = 1, \cdots, K, k = 1, \cdots, K \\
    &\bm{\beta}_{-0k} \geq \bm{0} \; \forall 1, \cdots, K.
    \end{aligned}
\end{equation}
The $k$th basis region is fit by a hyperplane with parameters $\bm{\beta}_{k} = (\beta_{0k}^*,  \bm{\beta}_{-0k}^{*})$. Note that like CNLS we are optimally fitting a piece-wise linear function; however, our estimator has a fixed number of hyperplanes, $K$, thus the total number of Afriat inequality constraints is $K(k-1)$ as opposed to the $N(N-1)$ constraints. Further, note that (\ref{eq:prod4}) estimates $f(\bm{X})$ conditionally on an input-space partition. Thus, to obtain an unconditional estimator of $f(\bm{X})$, we need to explore different input-space partitions as in CAP.

We estimate the function $f(\bm{X})$ by iteratively solving (\ref{eq:prod5}) inside of the partitioning proposal strategy. At each iteration, the strategy evaluates $KML$ partition-splitting proposals, where $M$, a tunable parameter, is the number of random input-space location proposals for a new knot at each iteration, $L = d$ is the number of randomly proposed directions, given the current basis regions and a proposed new knot location, that will define the new dataset partition, and $K$ is the current number of partitions at the current iteration.  The full estimation algorithm which nests (\ref{eq:prod5}) in the adaptive partitioning strategy is summaized in Algorithm 1 below.  
\algrenewcommand\algorithmicindent{2.0em}%
\begin{algorithm}
		\caption{CAPNLS Estimator}         
		\label{alg1}                          
		\begin{algorithmic}[1]                  
            \State Set $K=1$
            \State $n_{kml}=n$
            \While{$\exists n_{kml}\geq n_0/2$}
            \State Set $MSE_K=\infty$
            \For {$k=1,\ldots,K$}
            \For {$m=1,\ldots,M$ and $l=1,\ldots,L$}
                \If{$n_{kml}\geq n_0/2$}
                \State Fit (\ref{eq:prod4})
                \State Compute $MSE_{kml}$
                \If {$MSE_{kml}\leq MSE_K$}
                   \State $MSE_K = MSE_{kml}$ 
                    \State Store corresponding $\bm{\beta}_K$
                \EndIf
                \EndIf
            \EndFor
            \EndFor
            \State $K= K+1$
            \For {$k=1,\ldots,K$}
            \State Propose random split:
            \State \hspace{0.2in}$m=1,\ldots,M$ (knot location) 
            \State \hspace{0.2in}$l=1,\ldots,L$ (direction)
            \State Obtain $n_{kml}$.
            \EndFor
            \EndWhile
            \State $K_{max} = K$\\
            \Return $\hat{\bm{\beta}} = \argmin_{\bm{\beta}_k}\{k\}_{k=1}^{K_{max}}|MSE_k \leq MSE_K *1.01$
  \end{algorithmic}
\end{algorithm}

To ensure model parsimony, we select the smallest model (in terms of $K$) from the collection of models for which MSE is within a pre-specified tolerance of the MSE. This tolerance is calculated based on the largest $K$ considered available in such collection\footnote{The tolerance is set to 1\% and CNLS is used as the large model in all of our examples. Initially, we do not use the Generalized Cross Validation (GCV) score approximation used by \citet{hannah2013multivariate}, because they assert that GCV's predictive results are only comparable with full cross validation strategies for problems with $n \geq 5,000$, which are larger than the datasets we consider in this paper.}. The parameter $n_0$ is used in defining the minimum number of observations per hyperplane and is a tunable parameter bounded below by $2(d+1)$. CAPNLS has one-to-many hyperplane to observations mapping and requires at least $2(d+1)$ observations per partition to fit each hyperplane like CAP. This property is the key for superior out of sample performance. 

While CAPNLS was developed by combining CAP and CNLS, it has similarities to other estimators in the literature. Recently, \citet{OlesenRuggiero2017} propose the hinge function estimator (\citet{Breiman93}) as a piecewise linear (hyperplane) approximation of a monotone and concave production function in the case there is only a single regressor. Similar to CNLS, this estimator allows concavity to be imposed in a single-stage programming problem and avoids the Afriat inequalities by explicitly defining the location of the intersections of the hyperplanes (the hinges). Specifying the hinges has significant computational benefits at the cost of a reduction in flexibility. \citet{OlesenRuggiero2017} show that the hinge function estimator compares favorable to the CNLS estimator, which they refer to as the Afriat-Diewert-Parkan estimator. Critical to the performance of the hinge function estimator, is the implemention of the \textit{bagging} algorithm. Bagging is the process of subsampling the data set with replacement taking subsamples of size $n$ repeated $M$ times, \citet{breiman1996bagging}. Each subsample is used to create a new estimate and then the $M$ estimates are averaged.  \citet{hannah2012ensemble} had originally proposed this for the CAP estimator and all estimators discussed in the Monte Carlo simulation section could benefit from bagging.

 Though CAP and CAPNLS use the same partitioning strategy, there are three main differences between the estimators. First, CAPNLS imposes concavity via the Afriat Inequalities rather than a minimum-of-hyperplanes construction. Second, CAPNLS requires solving a global optimization problem rather than multiple localized optimization problems. As we will observe in Sections \ref{sec3:simulations} and \ref{sec4:ApplicationChilean}, the additional structure results in increased rates of convergence and improved robustness. 
 Third, CAPNLS does not require a refitting step, because it retains the observation-to-basis region correspondence before and after fitting problem (\ref{eq:prod4}).

\subsection{CAPNLS as a series of Quadratic Programs}
Taking advantage of the linearly-constrained quadratic programming structure of CAPNLS is essential to achieve computational feasibility. Therefore, we write Problem (\ref{eq:prod5}) in the standard form
\begin{equation}
\label{eq:prod6}
    \begin{aligned}
    \min_{\bm{\beta}} \frac{1}{2} \bm{\beta}^T H \bm{\beta} + \bm{\beta}^T g \\
    s.t. \; A \bm{\beta} \leq 0, \; \bm{\beta} \geq \bm{l}
    \end{aligned}
\end{equation}
Starting with the objective function from (\ref{eq:prod4}), we let $\tilde{\bm{X}} = (\bm{1}, \bm{X})$ and write
\begin{equation}
\label{eq:prod7}
    \begin{aligned}
    \min_{(\beta_{0k} , \bm{\beta}_{-0k})_{k=1}^K} \sum_{i=1}^N &(\beta_{0[i]} + \bm{\beta}_{-0[i]}^{T} \bm{X}_i - Y_i)^2 = \min_{(\bm{\beta})_{k=1}^K} \sum_{i=1}^N (\bm{\beta}_{[i]}^{T} \tilde{\bm{X}}_i - Y_i)^2 = \cdots \\
    &= \min_{(\bm{\beta})_{k=1}^K} \frac{1}{2} \sum_{i=1}^N (\bm{\beta}_{[i]}^{T} \tilde{\bm{X}}_i)^2 - \sum_{i=1}^N (\bm{\beta}_{[i]}^{T} \tilde{\bm{X}}_i Y_i)
    \end{aligned}
\end{equation}
where we have dropped constant $\sum_{i=1}^N Y_i^2$ and multiplied times one half. To write the last expression in Equation (\ref{eq:prod7}) in standard form, we first write $\sum_{i=1}^N (\bm{\beta}_{[i]}^{T} \tilde{\bm{X}}_i)^2$ using matrix operations. We define observation-to-hyperplane $n(d+1) \times K(d+1)$-dimensional mapping matrix $P$, with elements $P((i-1)(d+1)+j, ([i]-1)(d+1)+i) = \tilde{X}_{ij}, \; i=1, \cdots, n, \; j=1, \cdots, d+1$ and all other elements equal to zero. Similarly, we define $n \times n(d+1)$-dimensional observation-specific vector product matrix $S$, with elements  $S(i, (i-1)(d+1)+l) = 1$ for $l = 1, \cdots,3, i=1, \cdots, n, j=1, \cdots, d+1$. Then, we concatenate vectors $(\bm{\beta}_k)_{(k=1)}^K$ in $K(d+1) \times 1$-dimensional vector $\bm{\beta}$. It follows that 
\begin{equation}
    \label{eq:prod8}
    \sum_{i=1}^N (\bm{\beta}_{[i]}^{T} \tilde{\bm{X}}_i)^2 = \bm{\beta}^T P^T (S^T S) P \bm{\beta}  \text{ and } \sum_{i=1}^N (\bm{\beta}_{[i]}^{T} \tilde{\bm{X}}_i Y_i) = \bm{\beta}^T P^T S^T \bm{Y},
\end{equation}
from which we have that $H = P^T (S^T S) P$ and $g = -P^T S^T \bm{Y}$.

To write in the Afriat Inequality constraints as $nK \times K(d+1)$ - dimensional matrix $A$, we let elements $A \big( K(i-1)+k, \; j+(d+1)([i]-1) \big)= \tilde{X}_{ij}, \; i = 1, \cdots, n, \; j = 1, \cdots, d+1, \; k = 1, \cdots, K,$ and let all other elements equal zero. Finally, we define $K(d+1)$ – dimensional vector $\bm{l}$ to have elements $\bm{l} \big( (k-1)(d+1)+1 \big) = 0, \; k = 1, \cdots,K$, and all other elements be equal to negative infinity.

\section{Experiments on Simulated Data}
\label{sec3:simulations}
While a primary objective of this paper is to introduce an alternative to Monte Carlo simulations using application data, in this section and in the Appendix \ref{App:AppendixA}, we present Monte Carlo simulation results to allow a comparison between the Monte Carlo simulation approach and the application data results in Section \ref{sec4:ApplicationChilean}. We compare four estimators via Monte Carlo simulations: the proposed CAPNLS estimator, a correctly specified parametric estimator, 
a monotonically-constrained version of CAP estimator, and the CNLS estimator. Our analysis of simulated data is similar to the comparison of methods in published studies that propose classical frontier production function estimators discussed in the introduction. We run our experiments on a computer with Intel Core2 Quad CPU 3.00 GHz and 8GB RAM using MATLAB R2016b and the solvers \texttt{quadprog} and \texttt{lsqnonlin}. 

We consider Data Generation Processes (DGP) based on a Cobb-Douglas function and calculate our estimates for the expected \textit{in-sample error} of the production function estimators against the true DGP, $E(Err_{ISf})$, where the expectation is taken against all possible learning sets. We also estimate the following expected quantities: \textit{learning-to-testing set} or \textit{predictive error} against the true DGP, ($Err_f$), \textit{in-sample error} against observed output, $E(Err_{ISy})$, and \textit{predictive error} against observed output $E(Err_y)$ (\citet[p.~228-229]{hastie2009elements}). We note that in application datasets, the estimate of in-sample error against observed output is the most reliable fitting diagnostic when working with a \textit{census} or \textit{full set} of establishments. Conversely, the estimator's estimate compared to an additional sample drawn from the same DGP, which defines the expected \textit{predictive error}, is the primary diagnostic when assessing the fit of a functional estimator obtained from estimation on a learning set relative to a much larger population. Thus, assessing the fit of a functional estimator on a finite census from a non-exhaustive sample requires weighting the two errors by the relative sizes of the sets of observed and unobserved establishments. 

Thus, we estimate the expected in-sample error against the true DGP for a learning set of size $nLearn$, $E(Err_{ISf}^{nL})$, by $\widehat{E(Err_{ISf}^{nL})} = \bar{MSE}_{ISf}^{nL} = \sum_{v=1}^V \sum_{i=1}^{n_L} (\hat{f}_{vLi}^{nL} - f_{vLi})^2 / nV$ for each functional estimator\footnote{Note that the estimator ``hat'' character is over $E(Err_{ISf}^{nL})$ rather than $Err_{ISf}^{nL}$, the in-sample error for the particular learning set fitted with the production function.}, where $\hat{f}_{vL}^{nL}$ is the production function estimate obtained with the $v$th learning set and learning set of size $nL$, $f_{vLi}$ is the $i$th observation of the $v$th learning set, $n_L$ is the size of the learning set, and $V$ is the number of different learning sets considered. Analogously, we estimate the expected predictive error against the true DGP for a learning set of size $nLearn$, $E(Err_{f}^{nL})$, by computing the averaged MSE across the $V$ learning-testing set combinations of the same DGP, $\widehat{E(Err_{ISf}^{nL})} = \bar{MSE}_f^{nL} = \sum_{v=1}^V \sum_{i=1}^{n_T} (\hat{f}_{vLi}^{nL} - f_{vTi})^2 / nV$, where we choose the size of the testing set, $n_T = 1,000$, and $f_{vTi}$ is the $i$th output observation of the $v$th testing set. When estimating $E(Err_{ISy}^{nL})$, unlike when estimating $E(Err_{ISf}^{nL})$, we need to vary the random component of each observation of each learning set to avoid over-optimism (\citet[p.~228]{hastie2009elements}). Thus, we generate $W$ different sets of noise terms\footnote{Computing the in-sample error provides a more detailed measure of the quality of the production function on a full set than the \textit{learning error} $\bar{MSE}_{yIS}^{nL} = \sum_{v=1}^V \sum_{i=1}^{n_L} (\hat{f}_{vi} - Y_{vi})^2 / nv$, because it averages performance across many possible $\epsilon_i$ residual values for the learning set input vector.} for each learning set and estimate $\widehat{E(Err_{ISy}^{nL})} = \bar{MSE}_{yIS}^{nL} = \sum_(v=1)^V \sum_(w=1)^W \sum_(i=1)^(n_L) \big(\hat{f}_{vLi} - f(\bm{x}_{vLi}) + \epsilon_{wTi} \big)^2 / nVW$, where $\bm{x}_{vLi}$ is the $i$th input vector of the $v$th learning set, and $\epsilon_{wTi}$ is the ith residual of the $w$th testing set. Finally, we compute $\widehat{E(Err_y^{nL})} = \bar{MSE}_y^{nL} = \sum_{v=1}^V \sum_{i=1}^{n_T} (\hat{f}_{vLi} - Y_{vTi} )^2 / nV$ to estimate the predictive error against observed outputs, where $Y_{vTi}$ is the $i$th output observation of the $v$th testing set.

We consider results for \textit{full-sample} or \textit{census} scenarios, and \textit{learning set-to-full set} with finite full set scenarios. For the \textit{full-sample} scenarios, we report $\widehat{E(Err_{ISf}^n )}$ and $\widehat{E(Err_{ISy}^n )}$. For the \textit{learning set-to-full set} scenarios, we compute an estimator for the full set error
\begin{equation}
    \label{eq:prod9}
    \widehat{E(Err_{FS}^{nL})} = \bar{MSE}_{FS}^{nL} = (n_L / n) \widehat{E(Err_{IS\cdot}^{nL})} + ((n_F - n_L) / n) \widehat{E(Err_{\cdot}^{nL})}
\end{equation}
where $FS$ stands for full set, and either $f$ or $y$ replaces the dot operator. Note that $\widehat{E(Err_{ISy}^{nL})}$, $\widehat{E(Err_{y}^{nL})}$, and $\widehat{E(Err_{FSy}^{nL})}$ are also estimators for the noise level $\sigma^2$ of the DGP, we can use $\sigma^2$ as a benchmark for their estimations. Further, without our corrections for over-optimism, computing an estimator $\sigma^2$ will be complicated by the nonparametric nature of the regression methods used to fit the production functions\footnote{Specifically, if we intend to use the learning set's residual sum of squares, calculation of an estimator $\sigma^2$ would require knowledge of the functional estimator's effective number of parameters. However, effective parameters can be difficult to calculate for both nonparametric and sieve estimators.}. For our learning-to-testing scenarios, we compare the performance of the three methods on $100$ learning-testing set pairs, $V = 100$, using learning datasets of size $n_L= 30, 50, 80, 100, 150, 200, 240$, and $300$. For our full-sample scenarios, we consider $n_{Learn} = 100, 200, 300$. For all scenarios, we consider $30$ randomly drawn sets of noise testing vectors, $W= 30$, to compute the in-sample portion of Equation (\ref{eq:prod9}). We also estimate the correctly specified parametric estimator for the DGP. The true parametric form is never known in an application, i.e., we cannot select the correctly specified parametric estimator in practice, and our parametric estimation results are reported as a benchmark. 

 We present our estimates of expected full set errors measured against the true DGP, $\bar{MSE}_{FSf}^{nL}$, and expected fraction of unexplained variance on the full set, $\bar{MSE}_{FSy}^{nL} / var(Y_{FS})$, respectively. Also, note that the expected full set error is equal to expected In-Sample error for the full-sample scenarios. Due to the extensive nature of our results, we present them in graphical form for the two regressor case. The cases when $d = 3$ and $4$ are in Appendix \ref{App:AppendixA} along with the tabular results for all experiments. We record and report other relevant performance indicators, such as the number of hyperplanes fitted and the estimation time.

\subsection{Bivariate input Cobb-Douglas DGP}

We consider the DGP $Y_i = X_{i1}^{0.4} X_{i2}^{0.5} + \epsilon_i$, where $\epsilon_i \sim N(0,\sigma^2)$ and $\sigma = 0.01, 0.05, 0.1, 0.2, 0.3, 0.4$ for our six noise settings, which we split into low and high-noise settings. We assume $X_{ij} \sim Unif(0.1, 1)$ for $j = 1, 2$ and $i = 1, \cdots, n_L$. Our first observation from Figures \ref{fig:MCSim1} and \ref{fig2:MCSim_2var_lar} is that our estimated expected full set error results for all learning-to-testing set scenarios for CNLS exceed the scale of the $y$-axes (due to very high predictive error values); see Appendix \ref{App:AppendixA} for the values of CNLS for learning-to-testing set scenarios. The top set of graphs in Figure \ref{fig:MCSim1} shows that CAPNLS has similar to slightly better expected full set error values performance than both CAP and CNLS on full set scenarios, whereas CAPNLS clearly outperforms both methods on learning-to-testing set scenarios. The bottom panels of Figure \ref{fig:MCSim1} show that for these low-noise scenarios, $\sigma = 0.01, 0.05, 0.1$, where noise is  $0.3\%, 6.5\%$ and $22\%$ of the output variance respectively, the improvement of CAPNLS against CAP rarely exceeds $2\%$ of the variance of the full dataset. In other words, an $R$-squared measurement would differ by less than two percent. We observe that the difference between the correctly specified parametric estimator and CAPNLS is also within $2\%$ for almost all cases. In our higher noise settings, shown in Figure \ref{fig2:MCSim_2var_lar}, we observe a much larger performance gap  between CAP and CAPNLS, as the former becomes unstable in the full set context.

Table \ref{tab1:MCSim2D} lists the number of hyperplanes fitted for the Full Sample scenarios for the three nonparametric estimators. Larger values indicate a more complex production function using more hyperplanes to characterize the curvature. CNLS fits a much larger number of hyperplanes relative to CAPNLS, whereas CAP fits functions that are only slightly more complex than linear, by employing two hyperplanes in all estimates. Finally, although CAPNLS's runtimes are the highest, they are still small in absolute terms.

\renewcommand{\arraystretch}{0.65}
\begin{table}[t]
  \centering
  \caption{Number of Hyperplanes and Runtimes for Bivariate Input Cobb-Douglas DGP.}
    \begin{tabular}{rc|ccc|ccc|ccc}
    \toprule
    \multicolumn{1}{c}{} &       & \multicolumn{3}{c|}{CAPNLS} & \multicolumn{3}{c|}{CAP} & \multicolumn{3}{c}{CNLS} \\
        $\sigma$  & $n$ & 100   & 200   & 300   & 100   & 200   & 300   & 100   & 200   & 300 \\
    \hline
    \multicolumn{1}{c}{\multirow{2}[0]{*}{0.01}} & \textit{K} & 9     & 12    & 12    & 2     & 2     & 2     & 93    & 164   & 242 \\
    \multicolumn{1}{c}{} & Time (s) & 4     & 15    & 27    & 1     & 0.56  & 0.78  & 1     & 8     & 22 \\
    \hline
    \multicolumn{1}{c}{\multirow{2}[0]{*}{0.05}} & \textit{K} & 8     & 10    & 12    & 2     & 2     & 2     & 80    & 135   & 198 \\
    \multicolumn{1}{c}{} & Time (s) & 5     & 12    & 30    & 0.42  & 0.6   & 0.77  & 1     & 6     & 23 \\
    \hline
    \multicolumn{1}{c}{\multirow{2}[0]{*}{0.1}} & \textit{K} & 9     & 10    & 11    & 2     & 2     & 2     & 60    & 148   & 172 \\
    \multicolumn{1}{c}{} & Time (s) & 4     & 17    & 40    & 0.42  & 0.54  & 0.75  & 1     & 8     & 26 \\
    \hline
    \multicolumn{1}{c}{\multirow{2}[0]{*}{0.2}} & \textit{K} & 9     & 10    & 11    & 2     & 2     & 2     & 54    & 101   & 157 \\
    \multicolumn{1}{c}{} & Time (s) & 5     & 16    & 32    & 0.47  & 0.66  & 0.79  & 1     & 8     & 23 \\
    \hline
    \multicolumn{1}{c}{\multirow{2}[0]{*}{0.3}} & \textit{K} & 8     & 10    & 11    & 2     & 2     & 2     & 50    & 98    & 147 \\
    \multicolumn{1}{c}{} & Time (s) & 5     & 15    & 28    & 0.41  & 0.58  & 0.71  & 1     & 8     & 22 \\
    \hline
    \multicolumn{1}{c}{\multirow{2}[0]{*}{0.4}} & \textit{K} & 8     & 10    & 11    & 2     & 2     & 2     & 47    & 90    & 135 \\
    \multicolumn{1}{c}{} & Time (s) & 4     & 15    & 29    & 0.46  & 0.62  & 0.74  & 1     & 7     & 22 \\
    \bottomrule
    \end{tabular}%
  \label{tab1:MCSim2D}%
\end{table}%

\begin{landscape}
    \begin{figure}[ht]
        \includegraphics[width=\linewidth]{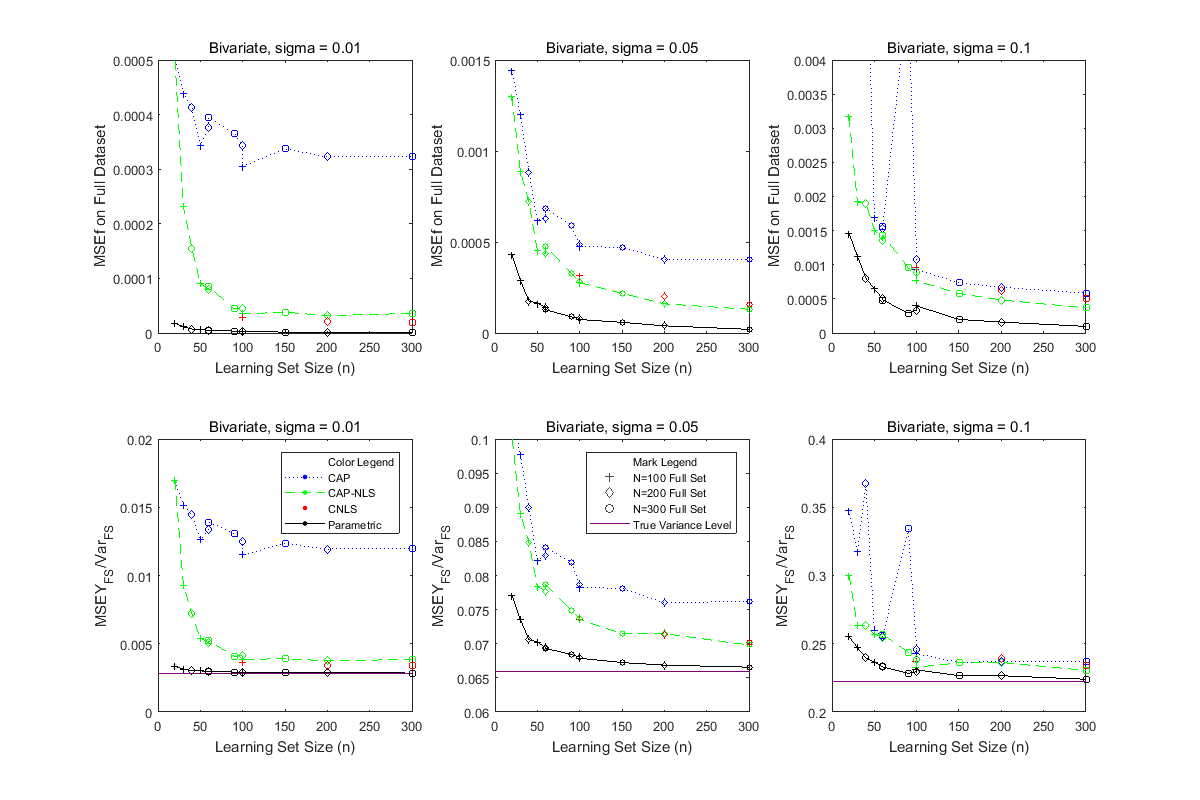}
        \centering
        \caption{Bivariate Input Cobb-Douglas DGP results for small noise settings.}
        \label{fig:MCSim1}
    \end{figure}
    \begin{figure}[ht]
        \includegraphics[width=\linewidth]{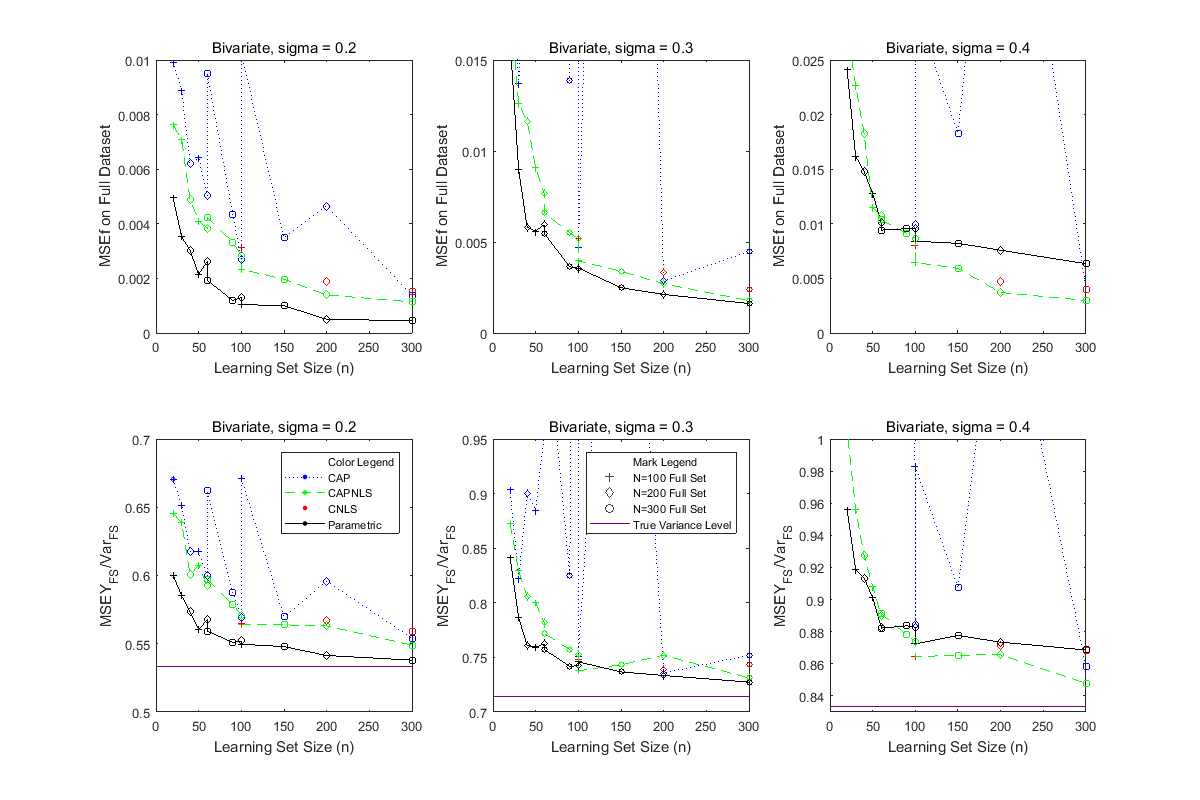}
        \centering
        \caption{Bivariate Input Cobb-Douglas DGP results for Large noise settings.}
        \label{fig2:MCSim_2var_lar}
    \end{figure}
\end{landscape}

\subsection{Other Monte Carlo Simulation Results}
The implications for the Monte Carlo simulations in Appendix \ref{App:AppendixA} are summarized as follows. For the cases when the number of inputs is 3 or 4, $(d= 3, 4)$, and our learning datasets of sizes $n_L= 30, 50, 80, 100, 150, 200, 240$, and $300$, we are asking a lot of a relatively small amount of data and one might suspect that the curse of dimensionality is a significant issue. We would expect the parametric estimators to be more robust to the curse of dimensionality for out of sample performance; however, the shape constraints improve the finite sample performance of the estimators considered (over general nonparametric methods) so the simulations in Appendix \ref{App:AppendixA} are important to understand the severity of the curse of dimensionality and the finite sample benefits of shape constraints. Focusing on $\sigma = 0.1$, $n=100$, the metric $MSEY_{FS} / Var_{FS}$ and the nonparametric shape constrained estimators, we see a significant drop in performance as we move from the bivariate to the trivariate to the tetravariate cases. Specifically, $MSEY_{FS} / Var_{FS}$ increase with dimensionality, $0.24$, $0.31$, and $0.39$, indicating a $\sim 30$\% drop in performance with a one unit increase in dimensionality. Notably, CAPNLS is the only functional estimator which performs robustly on a learning-to-full set basis across all dimensionalities and noise levels, while also being the nonparametric estimator with the lowest in-sample error on nearly all of the full set scenarios. Even though CNLS's overfitting of the learning set (as observed by the large number of hyperplanes it fitted), it has a severe detrimental effect on the expected full set error due to low predictive power, i.e., high expected predictive error, the overfitting has little effect on the in-sample performance (as observed through its expected full set error in the full set scenarios). Therefore, CNLS is a robust candidate estimator for analyzing full census datasets. CAP performs well on both full set and learning-to-full set scenarios for small noise settings at all dimensionalities, but its learning-to-full set performance deteriorates as the level of noise increases.  

Expected full set error on the full set scenarios is similar for all nonparametric methods, with the exception of CAP in the high noise settings with 3 or 4 inputs, when its performance deteriorates. CAPNLS and CNLS perform similarly in the full set scenarios in all cases. Runtimes for CAPNLS are the only ones to deteriorate significantly with dimensionality and they are the largest of the three nonparametric methods in all cases. Its runtimes, however, are still small in relative terms, i.e., no larger than 2 minutes for any fitted dataset. Finally, while dimensionality of production functions is typically low and therefore CAPNLS's scalability in dimensionality is not a concern, it implies that scalability in $N$ could be an issue to fit large production datasets (see Appendix \ref{App:AppendixB} for a modification to CAPNLS to address this potential issue). 

The performance of the parametric Cobb-Douglas in multiplicative form with an additive error term deteriorates in the 3 and 4 input scenarios relative to the nonparametric estimator. In fact the nonparametric estimator generally outperform the parametric estimator which might be surprising because typically the correctly specified parametric estimator will perform best in Monte Carlo simulation. Recall what is being graphed in this figure, we are graphing our metric for performance the full set error,  $\widehat{E(Err_{FS}^{nL})}$, which is a weighting of the expected in-sample error and the expected predictive error. While the parametric estimator should have the best performance in terms of expected predictive error, it is not the case for expected in-sample error. The parametric estimator is generally not able to fit the observed data with noise as well as the nonparametric estimators because of the parametric estimator's lack of flexibility. Or stated differently the nonparametric estimators benefit from being able to overfit the observed sample. We see the effects of lack of flexibility for the parametric estimator relative to the nonparametric estimator is pronounced in higher dimensional low noise cases. 

Finally a computational comment, the parametric Cobb-Douglas in multiplicative form with an additive error term estimator requires solving a nonlinear constrained optimization problem for which no polynomial time algorithm currently exists\footnote{As a reviewer pointed out, the parametric Cobb-Douglas in multiplicative form with an additive error term is not only challenging in cross-sectional data, but also in a panel setting due to the incidental parameters issue. Specifically analyzing the asymptotic performance, the variance of the noise term is a function of both $T$ and $N$, and unless both of these values approach infinite, the estimate of the noise variance will not converge to the true value.}. We uses the \texttt{lsqnonlin} solve in Matlab with the true value as the starting point of the solver. We also estimate the standard log-linear Cobb-Douglas with an additive error estimator and find both parametric estimators preform very similarly, see Appendix \ref{sec:APPG.CobbDouglas}. The next section presents the dataset. 

\section{Chilean Annual National Industrial Survey}
\label{sec4:ApplicationChilean}
\subsection{Dataset and considerations}
The Chilean Annual Industrial Survey (ENIA, by its initials in Spanish) is an annual census of all industrial establishments with 10 or more employees which are located inside the Chilean territory.  The census's main goal is to characterize Chile's manufacturing activity in terms of input usage, manufactured products, and means of production utilized in the diverse transformation processes. We focus on the five largest 4-digit industries in terms of sample size and only remove observations with non-positive value added or input values for any of the input variables used. In this paper, we refer to the learning sets as the survey subsamples and to the full sets as the survey full sample or census.

Our objective is to illustrate three key points largely overlooked in the production function estimation literature on working with national survey data for manufacturing. First, production data is highly clustered around particular scale sizes and input ratios, the data lacks the more complex curvature of data simulated from monotonic and concave DGPs. In view of this difference, the performance of estimators and their resulting rankings can vary significantly between Monte Carlo simulation experiments and the estimators' performance on survey data. Therefore, we assess the ability of the estimators discussed in Section 3 to fit industry-specific data from the ENIA dataset on a subsample to full sample setting. Second, we illustrate the replicate-specific performance of the selected functional estimators. Third, we graphically explore the increase in explanatory capability of our fitted production functions as a function of the relative size of survey subsample to survey full sample. The section concludes with a discussion of the implications of practical survey sample sizes.

\subsection{Methodology to compare functional estimator performance on real data}
We begin by comparing the additive error formulations of CAPNLS, CAP, and CNLS. We consider the additive-error Cobb-Douglas formulation of the form,  $Y=X_1^{\alpha_1} X_2^{\alpha_2} X_3^{\alpha_3} X_4^{\alpha_4} + \epsilon$, which we label CDA. As theory would imply, we restrict all input powers to be nonnegative for the Cobb-Douglas functional estimator. In total, we compare four different functional estimators. Our comparison focuses on the estimated expected error on the full survey set of establishments, given a survey subset size $\widehat{E({Err}_{FSy}^{nL})}$, but reports the scale-invariant quantity $R_{FS}^2 = \max (1-\widehat{E({Err}_{FSy})} / Var(Y_{FS}), 0)$, where $Var(Y_{FS})$ is the sample variance of the output on the full industry dataset\footnote{Note that the definition of $R_{FS}^2$ implies that if the evaluated estimator fails to explain more variability than the simply taking the mean of the output variable over the full sample, we will instead use the mean as our estimator.}.

As discussed in Section \ref{sec3:simulations}, Monte Carlo simulations, to compute $\widehat{E({Err}_{FSy}^{nL})}$ we rely on separate estimations of the expected predictive $E(Err_y^{nL})$ and in-sample $E(Err_{ISy}^{nL})$ errors, which we later weight by the relative size of the observed and unobserved establishment sets. Unlike in Section \ref{sec3:simulations}, we cannot generate more data from the same DGP as that of the observed dataset, or vectors of residuals with the same level of noise as the DGP, and thus we cannot compute the error estimators $(\bar{MSE}_y^{nL})$ and $(\bar{MSE}_{ISy}^{nL})$. To circumvent these issues, we estimate $E(Err_y^{nL})$ via an Repeated Learning Testing (RLT) procedure and we estimate $E(Err_{ISy}^{nL})$ by summing the learning error $MSE_{yL}^{nL}$ for a $n_L$-sized learning set and a parametric bootstrap covariance penalty estimator $\widehat{E(\omega^{nL})}$ for expected in-sample optimism $E(\omega^{nL})$ \citep{efron2004estimation}. For the RLT procedure, we consider $20$\%, $30$\%, $40$\%, and $50$\% learning subsets and $V = 100$ replicates to understand the predictive power of subsample-fitted functional estimators when inferring the industry-level production function as the subsample size increases\footnote{We reemphasize that unlike cross validation procedures in which the goal is to estimate $E(Err_y^n)$, our goal is to estimate $E(Err_y^{nL})$.}. For the bootstrap procedure, we consider $B=500$ parametric bootstrap replicates.

We compute our expected predictive error estimate given by RLT,  
\begin{equation}
    \label{eq:prod9a}
    MSE_{RLT}^{nL} = \sum_{\alpha = 1}^V \frac{n_L^\alpha}{n} \sum_{i \notin \{\alpha\}} (\hat{f}_i^{\alpha} - Y_i)^2 / n_T^{\alpha},
\end{equation}
where $\{\alpha\}$ is the index set of the $\alpha$th learning set, $\hat{f}_i^{\alpha}$ are the estimated functional values obtained from the $\alpha$th learning set, and testing set size $n_T^{\alpha} = n - n_L^{\alpha}$, where $n_L^{\alpha}$ is the size of the $\alpha$th learning set. Given that we only want to estimate the expected predictive error for a set equal in size to our learning set, our RLT estimator does not have the bias described by \citet{burman1989comparative} when estimating the usual cross-validation objective, which is the expected predictive error for a set equal in size to our full set. \citet{burman1989comparative} shows that the variance of RLT can be partially controlled with the number of replicates $V$. Finally, we acknowledge that independent of $V$, the variance of our RLT expected predictive error estimate could increase with the learning set size as the testing set size decreases, given our finite full survey. However, we do not observe an increase in variance in our estimates, as we explain in Section \ref{sec4.3:Comparison}.

To compute the estimator for expected\footnote{Again, expectations and averages over the error and optimism metrics discussed are done over all possible learning sets of a given size.}  in-sample error $\widehat{E({Err}_{ISy}^{nL})}$, we add the learning error $MSE_{yL}^{nL}$ and a covariance penalty term $E(\hat{\omega}^{nL})$ to account for expected optimism $E(\omega^{nL})$. If we consider an arbitrary estimator $\hat{Y}_i$ and a uniformly weighted squared loss function, i.e., $g(\hat{Y}_i, Y_i) = (\hat{Y}_i - Y_i)^2$ in our notation, \citet{efron2004estimation} shows that
\begin{equation}
    \label{eq:prod10}
    E(\hat{\omega}^{nL}) = \frac{2}{n_L} \sum_{i=1}^{n_L} cov (\hat{Y}_i, Y_i).
\end{equation}
We note that if all of the functional estimators being considered were in the linear smoother form $\hat{\bm{Y}} = S \bm{Y}$, we would write the penalty in terms of the trace. Clearly, the Cobb-Douglas functional estimator is not, and so we use the parametric bootstrap algorithm by \citet{efron2004estimation}, which directly estimates $cov (\hat{Y}_i, Y_i)$ (see Appendix \ref{App:AppendixC} for the details about this algorithm). Thus, our full expression for $\widehat{E({Err}_{FSy}^{nL})}$ for learning set sizes of size $nL$ is
\begin{equation}
    \label{eq:prod11}
    \widehat{E({Err}_{FSy}^{nL})} = \frac{n_T}{n} MSE_{RLT}^{nL} + \frac{n_L}{n} \big( MSE_{yL}^{nL} + E(\hat{\omega}^{nL}) \big).
\end{equation}
We assume the data is homoscedastic and use an error measure that is uniformly-weighted over observations, such as (\ref{eq:prod10}). Thus, if we intended to use multiplicative or other residual assumptions, our error estimators would need to reflect a similar residual-weighting scheme. 
To define the inputs and output for our production function, we follow the KLEMS framework and fit a Value-Added production function
\begin{equation}
    \label{eq:prod12}
    VA = Y - M = f(KLES)  ,
\end{equation}
where $VA$ is value added, $Y$ is output, $M$ is intermediate goods, $K$ is capital stock, $L$ are labor man-hours, $E$ is energy, and $S$ is service expenditures, respectively. The variables are readily found in the Chilean manufacturing dataset, except for Energy, for which we also add the fuel expenditures costs. All variables except for $L$ are measured in thousands of Chilean pesos.

\subsection{Functional Estimator Comparison Results}
\label{sec4.3:Comparison}
In Table \ref{tab2:SumStats}, the Best Method field lists the functional estimator with the highest $R_{FS}^2$ for each subset size, considering ties for functional estimators with $R_{FS}^2$ values within $2$\% of the best estimator. Table \ref{tab2:SumStats} also shows a field for $K_{nL}^{CAPNLS}$, the average number of CAPNLS hyperplanes fitted to either the learning sets in the case of $20$, $30$, $40$, and $50$ percent subset sizes, or the bootstrapped sets used to compute $\widehat{E(\omega^n)}$ in the case of the full set.  The average number of CAPNLS hyperplanes fitted allows us to compare the complexity of the estimated production functions relative the estimates in the Monte Carlo simulation section. As expected due to the simpler curvature and more concentrated nature of real manufacturing survey data relative to Monte Carlo simulated data, the number of CAPNLS hyperplanes fitted for data sets with $100$ or $200$ observations is generally smaller than those fitted to similar sample sizes in Appendix \ref{App:AppendixA}, in which the production function also has a four-dimensional input space. 
Further exploring our results, we observe both similarities and discrepancies regarding the insights obtained from testing estimators with Census data. The clearest similarity to all our low noise settings\footnote{The maximum attainable, i.e., using the full set as the learning set, noise-to-total variance levels of our real datasets are very similar to those of our low noise settings. Compare $(1 - \bar{MSE}_{FSy}^{nL}) / var(Y_{FS})$ in our low noise settings against the $R_{FS}^2$ results of the $100$\% survey real datasets.}  is the multiple ties across functional estimators in terms of $R_{FS}^2$, meaning that several of the estimators describe the production function with the same accuracy. Discrepancies include better CDA performance for larger datasets. Surprisingly, CDA's performance is remarkably good, especially if we consider that now the true DGP is unknown. Table \ref{tab3:CDAPerformance} presents the capabilities of the CDA parametric estimator against the best estimate achieved for each subset size.  In general, the CDA estimator describes nearly as much variance as the best estimator. Further in Appendix \ref{App:AppendixD}, we include equivalent results to those of Table \ref{tab3:CDAPerformance} including estimates from the classical multiplicative error assumption for Cobb-Douglas (labeled CDM)\footnote{Recall we use the Cobb-Douglas function with an additive error term is used to maintain consistency of the error structure across estimators.}. The results for CDM show that a multiplicative error assumption when fitting the Cobb-Douglas model is a significantly better assumption for the other metal products and wood industries (industry codes 2899 and 2010) (even if tested in terms of (\ref{eq:prod12}) with uniform error weighting) and a significantly worse assumption for bakeries (1541). These results show that common characteristics of manufacturing survey data, such as a high concentration of establishments around popular scale sizes or the economically efficient input ratios, sparse data on large establishments and simpler curvature, reduce the performance gap between estimators such as CAP or Cobb-Douglas and our proposed estimator.

Table \ref{tab4:BestMethod} shows that the best estimator in the Chilean manufacturing dataset is perhaps more closely related to the learning set size regardless of the residual noise level. CAPNLS is dominant for very small learning set sizes (less than 50 observations). CAPNLS, CAP and CDA perform similarly for larger datasets. The additional structure of CAPNLS relative to CAP seems to lose its benefits as the learning set size increases for our application datasets, which is a much more direct statement than we could make from extrapolating the results across the three small noise settings in Appendix \ref{App:AppendixA}. Some insights obtained from evaluating estimators on the actual application dataset are not observed from those on the simulated data. For instance, our simulated data examples show potential problems when fitting the CDA model at high dimensionalities or high noise settings, yet for the application datasets considered, CDA is a reliable production function estimator for all learning set sizes.

\begin{table}[htbp]
  \centering
  \caption{Method comparison across the 5 largest sampled industries from the Chilean Annual National Industrial Survey, 2010.}
    \begin{tabular}{lccccl}
    \toprule
    \multicolumn{1}{c}{Industry Name (Code)} &    $n$   & Survey Size &   $R_{FS}^2$    & $K_{nL}^{CAPNLS}$  & \multicolumn{1}{c}{Best Method} \\
    \midrule
    \multirow{5}[0]{*}{Other Metal Products (2899)} & \multicolumn{1}{c}{\multirow{5}[0]{*}{144}} & 20\%  & \multicolumn{1}{c}{50\%} & \multicolumn{1}{c}{1} & CAPNLS, CDA \\
          & \multicolumn{1}{c}{} & 30\%  & \multicolumn{1}{c}{60\%} & \multicolumn{1}{c}{2} & CAPNLS, CDA \\
          & \multicolumn{1}{c}{} & 40\%  & \multicolumn{1}{c}{64\%} & \multicolumn{1}{c}{2} & CAPNLS, CDA \\
          & \multicolumn{1}{c}{} & 50\%  & \multicolumn{1}{c}{72\%} & \multicolumn{1}{c}{3} & CAPNLS \\
          & \multicolumn{1}{c}{} & 100\% & \multicolumn{1}{c}{88\%} & \multicolumn{1}{c}{7} & CAPNLS \\
    \midrule
    \multirow{5}[0]{*}{Wood (2010)} & \multicolumn{1}{c}{\multirow{5}[0]{*}{150}} & 20\%  & \multicolumn{1}{c}{35\%} & \multicolumn{1}{c}{1} & CDA \\
          & \multicolumn{1}{c}{} & 30\%  & \multicolumn{1}{c}{40\%} & \multicolumn{1}{c}{1} & CAPNLS, CDA \\
          & \multicolumn{1}{c}{} & 40\%  & \multicolumn{1}{c}{47\%} & \multicolumn{1}{c}{2} & CAPNLS, CDA \\
          & \multicolumn{1}{c}{} & 50\%  & \multicolumn{1}{c}{52\%} & \multicolumn{1}{c}{3} & CAPNLS, CDA \\
          & \multicolumn{1}{c}{} & 100\% & \multicolumn{1}{c}{66\%} & \multicolumn{1}{c}{6} & CAPNLS \\
    \midrule
    \multirow{5}[0]{*}{Structural Use Metal (2811)} & \multicolumn{1}{c}{\multirow{5}[0]{*}{161}} & 20\%  & \multicolumn{1}{c}{77\%} & \multicolumn{1}{c}{1} & CAPNLS, CAP \\
          & \multicolumn{1}{c}{} & 30\%  & \multicolumn{1}{c}{82\%} & \multicolumn{1}{c}{2} & CAPNLS \\
          & \multicolumn{1}{c}{} & 40\%  & \multicolumn{1}{c}{87\%} & \multicolumn{1}{c}{3} & CAPNLS, CAP \\
          & \multicolumn{1}{c}{} & 50\%  & \multicolumn{1}{c}{90\%} & \multicolumn{1}{c}{4} & CAPNLS \\
          & \multicolumn{1}{c}{} & 100\% & \multicolumn{1}{c}{95\%} & \multicolumn{1}{c}{9} & CAPNLS, CAP \\
    \midrule
    \multirow{5}[0]{*}{Plastics (2520)} & \multicolumn{1}{c}{\multirow{5}[0]{*}{249}} & 20\%  & \multicolumn{1}{c}{54\%} & \multicolumn{1}{c}{2} & CAPNLS, CAP, CDA \\
          & \multicolumn{1}{c}{} & 30\%  & \multicolumn{1}{c}{57\%} & \multicolumn{1}{c}{3} & CDA \\
          & \multicolumn{1}{c}{} & 40\%  & \multicolumn{1}{c}{57\%} & \multicolumn{1}{c}{5} & CAPNLS, CAP, CDA \\
          & \multicolumn{1}{c}{} & 50\%  & \multicolumn{1}{c}{60\%} & \multicolumn{1}{c}{7} & CAPNLS, CAP, CDA \\
          & \multicolumn{1}{c}{} & 100\% & \multicolumn{1}{c}{64\%} & \multicolumn{1}{c}{11}    & CAPNLS, CAP, CDA \\
    \midrule
    \multirow{5}[0]{*}{Bakeries (1541)} & \multicolumn{1}{c}{\multirow{5}[0]{*}{250}} & 20\%  & \multicolumn{1}{c}{72\%} & \multicolumn{1}{c}{3} & CAP \\
          & \multicolumn{1}{c}{} & 30\%  & \multicolumn{1}{c}{77\%} & \multicolumn{1}{c}{3} & CAP \\
          & \multicolumn{1}{c}{} & 40\%  & \multicolumn{1}{c}{78\%} & \multicolumn{1}{c}{4} & CAP, CDA \\
          & \multicolumn{1}{c}{} & 50\%  & \multicolumn{1}{c}{85\%} & \multicolumn{1}{c}{4} & CAP \\
          & \multicolumn{1}{c}{} & 100\% & \multicolumn{1}{c}{99\%} & \multicolumn{1}{c}{5} & CAPNLS, CAP, CDA \\
    \bottomrule
    \end{tabular}%
  \label{tab2:SumStats}%
\end{table}%

\begin{table}[htbp]
  \centering
  \caption{Ratio of CDA to Best Model performance.}
    \begin{tabular}{lccccl}
    \toprule
    \multicolumn{1}{c}{Industry Name (Code)} & $n$ & Survey Size & $R_{FS}^2$ & $R_{CDA}^2$ & \multicolumn{1}{c}{Ratio vs. Best Method} \\
    \midrule
    \multirow{5}[0]{*}{Other Metal Products (2899)} & \multicolumn{1}{c}{\multirow{5}[0]{*}{144}} & 20\%  & \multicolumn{1}{c}{50\%} & \multicolumn{1}{c}{49\%} & CDA ties for Best Method \\
          & \multicolumn{1}{c}{} & 30\%  & \multicolumn{1}{c}{60\%} & \multicolumn{1}{c}{59\%} & CDA ties for Best Method \\
          & \multicolumn{1}{c}{} & 40\%  & \multicolumn{1}{c}{64\%} & \multicolumn{1}{c}{64\%} & CDA ties for Best Method \\
          & \multicolumn{1}{c}{} & 50\%  & \multicolumn{1}{c}{72\%} & \multicolumn{1}{c}{60\%} & 0.83 vs. CAPNLS \\
          & \multicolumn{1}{c}{} & 100\% & \multicolumn{1}{c}{88\%} & \multicolumn{1}{c}{79\%} & 0.90 vs. CAPNLS \\
    \midrule
    \multirow{5}[0]{*}{Wood (2010)} & \multicolumn{1}{c}{\multirow{5}[0]{*}{150}} & 20\%  & \multicolumn{1}{c}{35\%} & \multicolumn{1}{c}{35\%} & CDA ties for Best Method \\
          & \multicolumn{1}{c}{} & 30\%  & \multicolumn{1}{c}{40\%} & \multicolumn{1}{c}{40\%} & CDA ties for Best Method \\
          & \multicolumn{1}{c}{} & 40\%  & \multicolumn{1}{c}{47\%} & \multicolumn{1}{c}{47\%} & CDA ties for Best Method \\
          & \multicolumn{1}{c}{} & 50\%  & \multicolumn{1}{c}{52\%} & \multicolumn{1}{c}{51\%} & CDA ties for Best Method \\
          & \multicolumn{1}{c}{} & 100\% & \multicolumn{1}{c}{66\%} & \multicolumn{1}{c}{62\%} & 0.94 vs. CAPNLS \\
    \midrule
    \multirow{5}[0]{*}{Structural Use Metal (2811)} & \multicolumn{1}{c}{\multirow{5}[0]{*}{161}} & 20\%  & \multicolumn{1}{c}{77\%} & \multicolumn{1}{c}{69\%} & 0.90 vs. CAPNLS \\
          & \multicolumn{1}{c}{} & 30\%  & \multicolumn{1}{c}{82\%} & \multicolumn{1}{c}{76\%} & 0.93 vs. CAPNLS \\
          & \multicolumn{1}{c}{} & 40\%  & \multicolumn{1}{c}{87\%} & \multicolumn{1}{c}{81\%} & 0.93 vs. CAPNLS \\
          & \multicolumn{1}{c}{} & 50\%  & \multicolumn{1}{c}{90\%} & \multicolumn{1}{c}{87\%} & 0.97 vs. CAPNLS \\
          & \multicolumn{1}{c}{} & 100\% & \multicolumn{1}{c}{95\%} & \multicolumn{1}{c}{91\%} & 0.96 vs. CAPNLS \\
   \midrule
        \multirow{5}[0]{*}{Plastics (2520)} & \multicolumn{1}{c}{\multirow{5}[0]{*}{249}} & 20\%  & \multicolumn{1}{c}{54\%} & \multicolumn{1}{c}{53\%} & CDA ties for Best Method \\
          & \multicolumn{1}{c}{} & 30\%  & \multicolumn{1}{c}{57\%} & \multicolumn{1}{c}{57\%} & CDA ties for Best Method \\
          & \multicolumn{1}{c}{} & 40\%  & \multicolumn{1}{c}{57\%} & \multicolumn{1}{c}{57\%} & CDA ties for Best Method \\
          & \multicolumn{1}{c}{} & 50\%  & \multicolumn{1}{c}{60\%} & \multicolumn{1}{c}{60\%} & CDA ties for Best Method \\
          & \multicolumn{1}{c}{} & 100\% & \multicolumn{1}{c}{64\%} & \multicolumn{1}{c}{64\%} & CDA ties for Best Method \\
    \midrule
    \multirow{5}[0]{*}{Bakeries (1541)} & \multicolumn{1}{c}{\multirow{5}[0]{*}{250}} & 20\%  & \multicolumn{1}{c}{72\%} & \multicolumn{1}{c}{61\%} & 0.85 vs. CAP \\
          & \multicolumn{1}{c}{} & 30\%  & \multicolumn{1}{c}{77\%} & \multicolumn{1}{c}{71\%} & 0.92 vs. CAP \\
          & \multicolumn{1}{c}{} & 40\%  & \multicolumn{1}{c}{78\%} & \multicolumn{1}{c}{78\%} & CDA ties for Best Method \\
          & \multicolumn{1}{c}{} & 50\%  & \multicolumn{1}{c}{85\%} & \multicolumn{1}{c}{82\%} & 0.96 vs. CAP \\
          & \multicolumn{1}{c}{} & 100\% & \multicolumn{1}{c}{99\%} & \multicolumn{1}{c}{99\%} & CDA ties for Best Method \\
    \bottomrule
    \end{tabular}%
  \label{tab3:CDAPerformance}%
\end{table}%

\begin{table}[htbp]
  \centering
  \caption{Most frequently selected Best Method for different sample size ranges.}
    \begin{tabular}{ccccc}
    \toprule
    \multirow{2}[0]{*}{Times selected as ``Best Method''} & \multicolumn{4}{c}{Learning Set Size} \\
          & 29 - 50 & 51 - 80 & 81 - 149 & 150+ \\
    \midrule
   \multicolumn{1}{c}{CAPNLS} & 7     & 5     & 3     & 4 \\
    \multicolumn{1}{c}{CAP} & 3     & 3     & 4     & 3 \\
    \multicolumn{1}{c}{CDA} & 5     & 2     & 2     & 2 \\
    \bottomrule
    \end{tabular}%
  \label{tab4:BestMethod}%
\end{table}%

An important feature of production functions are characteristics such as most productive scale size (MPSS), marginal products, and marginal rates of technical substitution. A significant benefit of nonparametric shape constrained estimators over parametric methods is the flexibility to estimate these characteristics over the input domain. For example all of the shape constrained nonparametric estimators considered are nonhomothetic estimators\footnote{Within the class of monotonically increasing and concave production function, the additional assumption of homotheticity can be imposed. A homothetic production function $f(\bm{X})$ can be written as $g(h(\bm{X}))$ where $g$ is a monotonically increasing function and $h$ is homogeneous of degree 1. Practically what nonhomothetic implies is that while input isoquants are convex for any output level, the shape of the input isoquant can vary across output levels.} and therefore the level of the MPSS may vary with the input mix. Table \ref{tab:MPSS} reports the MPSS estimates for CAP, CAPNLS, and CNLS for several different capital to labor ratios while holding all other inputs at their median levels. Notice the significant variance over capital to labor ratios, but relatively consistent performance across estimators. CAPNLS provides the lowest estimates of MPSS for all capital to labor combinations and all industry except for Structural Use Metal (2811) in which CAP has smaller MPSS for all capital to labor combinations in which the capital percentile is greater than or equal the labor percentile. This can be attributed to CAPNLS' more complicated hyperplane structure relative to CAP and thus estimates more hyperplanes at higher output levels with shallower slopes (lower marginal products). For a more extensive set of results, see Appendix \ref{sec:APPF.AppComp}, Table \ref{tabA:MPSS}. Note for the parametric Cobb-Douglas estimators the MPSS can only be $0$ or $\infty$. For all five Chilean manufacturing industries, the Cobb-Douglas production functions with a multiplicative residual had parameter estimates indicating decreasing returns-to-scale and thus a MPSS of zero. 

Tables \ref{tab:MP_K} and \ref{tabA:MP_L} report the marginal product estimates for the three methods for several different capital to labor ratios while holding all other inputs and output at their median levels. Here we observe that CNLS has larger marginal product estimates in most cases relative to either CAP or CAPNLS. This indicates that slope of the function estimated is steeper with CNLS which is consistent with using more hyperplanes to approximate the technology. Table \ref{tab:MRTS_KL} reports the marginal rates of technical substitution (MRTS). Here again, CNLS's use of more hyperplanes leads to a higher variation in the MRTS across different capital to labor ratios. Both CAP and CAPNLS have explicit criteria to assure parsimonious models, for example CAPNLS requires $n_0/2$ observations per hyperplane whereas CNLS can have just a single observation per hyperplane. This leads to CNLS using more hyperplanes and having more curvature in the estimated isoquants.

\afterpage{
\begin{landscape}

\begin{table}[t]
  \centering
  \caption{Most Product Scale Size ($y$) conditional on intermediate goods $M$, energy $E$, and service expenditures $S$ held at the $50$th percentile}
    \begin{tabular}{cc|ccccc|ccccc|ccccc}
    \toprule
    \multicolumn{2}{c|}{Percentile} & \multicolumn{5}{c|}{CAP}              & \multicolumn{5}{c|}{CAPNLS}            & \multicolumn{5}{c}{CNLS} \\
    \multicolumn{1}{c}{$K$} & \multicolumn{1}{c|}{$L$} & \multicolumn{1}{c}{2899} & \multicolumn{1}{c}{2010} & \multicolumn{1}{c}{2811} & \multicolumn{1}{c}{2520} & \multicolumn{1}{c|}{1541} & \multicolumn{1}{c}{2899} & \multicolumn{1}{c}{2010} & \multicolumn{1}{c}{2811} & \multicolumn{1}{c}{2520} & \multicolumn{1}{c|}{1541} & \multicolumn{1}{c}{2899} & \multicolumn{1}{c}{2010} & \multicolumn{1}{c}{2811} & \multicolumn{1}{c}{2520} & \multicolumn{1}{c}{1541} \\
    \hline
   25    & 25    & 5.7   & 7.6   & 6.4   & 17.5  & 8.9   & 4.4   & 5.1   & 5.8   & 9.1   & 3.9   & 5.6   & 8.8   & 14.3  & 18.3  & 10.4 \\
    50    & 25    & 6.9   & 11.4  & 5.7   & 10.8  & 9.5   & 5.1   & 7.6   & 6.2   & 17.8  & 4.1   & 7.9   & 20.2  & 9.3   & 29.6  & 9.6 \\
    75    & 25    & 4.3   & 12.5  & 2.5   & 48.2  & 10.8  & 7.2   & 15.2  & 6.4   & 36.9  & 4.7   & 9.3   & 30.0  & 7.4   & 39.3  & 8.5 \\
    25    & 50    & 4.7   & 5.1   & 10.1  & 7.3   & 6.6   & 3.5   & 3.8   & 7.9   & 8.3   & 3.0   & 3.3   & 3.1   & 12.9  & 9.6   & 8.1 \\
    50    & 50    & 5.9   & 5.9   & 6.5   & 7.8   & 7.0   & 3.8   & 4.5   & 8.0   & 12.5  & 3.1   & 5.3   & 9.2   & 14.1  & 18.9  & 10.9 \\
    75    & 50    & 5.0   & 12.8  & 3.3   & 38.4  & 8.0   & 4.7   & 5.8   & 6.2   & 14.7  & 3.2   & 6.7   & 20.3  & 10.5  & 17.7  & 8.2 \\
    25    & 75    & 3.9   & 3.0   & 13.6  & 4.6   & 4.7   & 2.6   & 2.4   & 5.5   & 7.1   & 2.3   & 2.2   & 2.6   & 14.7  & 4.5   & 7.3 \\
    50    & 75    & 4.4   & 3.5   & 12.0  & 5.3   & 4.9   & 2.7   & 2.4   & 5.5   & 8.1   & 2.3   & 2.1   & 5.6   & 15.5  & 5.8   & 9.4 \\
    75    & 75    & 5.9   & 7.0   & 4.9   & 7.4   & 5.4   & 3.2   & 2.6   & 5.1   & 9.1   & 2.4   & 3.2   & 10.8  & 13.6  & 11.9  & 10.1 \\
    \bottomrule
    \end{tabular}%
  \label{tab:MPSS}%
\end{table}%

\begin{table}[t]
  \centering
  \caption{Marginal Product of Capital}
    \begin{tabular}{cc|ccccc|ccccc|ccccc}
    \toprule
    \multicolumn{2}{c|}{Percentile} & \multicolumn{5}{c|}{CAP}              & \multicolumn{5}{c|}{CAPNLS}            & \multicolumn{5}{c}{CNLS} \\
    \multicolumn{1}{c}{$K$} & \multicolumn{1}{c|}{$L$} & \multicolumn{1}{c}{2899} & \multicolumn{1}{c}{2010} & \multicolumn{1}{c}{2811} & \multicolumn{1}{c}{2520} & \multicolumn{1}{c|}{1541} & \multicolumn{1}{c}{2899} & \multicolumn{1}{c}{2010} & \multicolumn{1}{c}{2811} & \multicolumn{1}{c}{2520} & \multicolumn{1}{c|}{1541} & \multicolumn{1}{c}{2899} & \multicolumn{1}{c}{2010} & \multicolumn{1}{c}{2811} & \multicolumn{1}{c}{2520} & \multicolumn{1}{c}{1541} \\
    \hline
    25    & 25    & 0.09  & 0.38  & 0.53  & 0.19  & 0.45  & 0.18  & 0.32  & 0.37  & 0.42  & 0.30  & 0.77  & 1.81  & 1.26  & 1.47  & 2.80 \\
    50    & 25    & 0.09  & 0.30  & 0.44  & 0.08  & 0.45  & 0.12  & 0.29  & 0.34  & 0.25  & 0.28  & 0.17  & 1.42  & 0.56  & 0.35  & 1.43 \\
    75    & 25    & 0.09  & 0.30  & 0.20  & 0.04 & 0.15  & 0.03  & 0.10  & 0.26  & 0.07  & 0.24  & 0.00  & 0.01  & 0.02  & 0.01  & 0.04 \\
    25    & 50    & 0.09  & 0.39  & 0.52  & 0.34  & 0.45  & 0.28  & 0.34  & 0.38  & 0.46  & 0.33  & 1.40  & 2.11  & 2.39  & 1.69  & 2.90 \\
    50    & 50    & 0.09  & 0.39  & 0.52  & 0.15  & 0.45  & 0.24  & 0.30  & 0.35  & 0.42  & 0.27  & 0.35  & 1.62  & 0.61  & 0.73  & 2.70 \\
    75    & 50    & 0.09  & 0.29  & 0.22  & 0.00  & 0.15  & 0.12  & 0.10  & 0.24  & 0.14  & 0.24  & 0.00  & 0.00  & 0.07  & 0.05  & 0.08 \\
    25    & 75    & 0.29  & 0.40  & 0.45  & 0.37  & 0.59  & 0.27  & 0.32  & 0.31  & 0.47  & 0.32  & 2.17  & 2.52  & 2.86  & 2.58  & 3.13 \\
    50    & 75    & 0.29  & 0.39  & 0.41  & 0.24  & 0.59  & 0.26  & 0.32  & 0.28  & 0.44  & 0.31  & 0.66  & 1.51  & 1.31  & 0.66  & 2.96 \\
    75    & 75    & 0.13  & 0.39  & 0.36  & 0.12  & 0.29  & 0.20  & 0.19  & 0.18  & 0.29  & 0.25  & 0.15  & 0.03  & 0.08  & 0.02  & 2.16 \\
    \bottomrule
    \end{tabular}%
  \label{tab:MP_K}%
\end{table}%

\begin{table}[t]
  \centering
  \caption{Marginal Product of Labor}
     \begin{tabular}{cc|ccccc|ccccc|ccccc}
    \toprule
    \multicolumn{2}{c|}{Percentile} & \multicolumn{5}{c|}{CAP}              & \multicolumn{5}{c|}{CAPNLS}            & \multicolumn{5}{c}{CNLS} \\
    \multicolumn{1}{c}{$K$} & \multicolumn{1}{c|}{$L$} & \multicolumn{1}{c}{2899} & \multicolumn{1}{c}{2010} & \multicolumn{1}{c}{2811} & \multicolumn{1}{c}{2520} & \multicolumn{1}{c|}{1541} & \multicolumn{1}{c}{2899} & \multicolumn{1}{c}{2010} & \multicolumn{1}{c}{2811} & \multicolumn{1}{c}{2520} & \multicolumn{1}{c|}{1541} & \multicolumn{1}{c}{2899} & \multicolumn{1}{c}{2010} & \multicolumn{1}{c}{2811} & \multicolumn{1}{c}{2520} & \multicolumn{1}{c}{1541} \\
    \hline
    25    & 25    & 0.14  & 0.05  & 0.10  & 0.27  & 0.07  & 0.22  & 0.10  & 0.13  & 0.20  & 0.16  & 0.19  & 0.09  & 0.17  & 0.16  & 0.07 \\
    50    & 25    & 0.14  & 0.08  & 0.10  & 0.27  & 0.07  & 0.23  & 0.11  & 0.13  & 0.25  & 0.16  & 0.20  & 0.10  & 0.17  & 0.29  & 0.15 \\
    75    & 25    & 0.14  & 0.08  & 0.10  & 0.28  & 0.06  & 0.27  & 0.10  & 0.13  & 0.27  & 0.16  & 0.21  & 0.10  & 0.19  & 0.41  & 0.24 \\
    25    & 50    & 0.14  & 0.04  & 0.09  & 0.16  & 0.07  & 0.16  & 0.06  & 0.12  & 0.16  & 0.13  & 0.12  & 0.03  & 0.07  & 0.10  & 0.05 \\
    50    & 50    & 0.14  & 0.04  & 0.09  & 0.21  & 0.07  & 0.15  & 0.06  & 0.11  & 0.16  & 0.13  & 0.15  & 0.03  & 0.11  & 0.12  & 0.06 \\
    75    & 50    & 0.14  & 0.07  & 0.09  & 0.22  & 0.06  & 0.18  & 0.08  & 0.12  & 0.21  & 0.13  & 0.17  & 0.04  & 0.12  & 0.12  & 0.21 \\
    25    & 75    & 0.09  & 0.01  & 0.08  & 0.14  & 0.05  & 0.08  & 0.03  & 0.10  & 0.13  & 0.08  & 0.01  & 0.00  & 0.04  & 0.00  & 0.01 \\
    50    & 75    & 0.09  & 0.02  & 0.08  & 0.14  & 0.05  & 0.08  & 0.03  & 0.09  & 0.13  & 0.08  & 0.01  & 0.00  & 0.05  & 0.01  & 0.01 \\
    75    & 75    & 0.12  & 0.02  & 0.08  & 0.14  & 0.04  & 0.09  & 0.03  & 0.08  & 0.13  & 0.09  & 0.03  & 0.00  & 0.06  & 0.03  & 0.03 \\
        \bottomrule
    \end{tabular}%
  \label{tab:MP_L}%
\end{table}%



\begin{table}[t]
  \centering
  \caption{Marginal Rate of Technical Substitution ($K/L$)}
     \begin{tabular}{cc|ccccc|ccccc|ccccc}
    \toprule
    \multicolumn{2}{c|}{Percentile} & \multicolumn{5}{c|}{CAP}              & \multicolumn{5}{c|}{CAPNLS}            & \multicolumn{5}{c}{CNLS} \\
    \multicolumn{1}{c}{$K$} & \multicolumn{1}{c|}{$L$} & \multicolumn{1}{c}{2899} & \multicolumn{1}{c}{2010} & \multicolumn{1}{c}{2811} & \multicolumn{1}{c}{2520} & \multicolumn{1}{c|}{1541} & \multicolumn{1}{c}{2899} & \multicolumn{1}{c}{2010} & \multicolumn{1}{c}{2811} & \multicolumn{1}{c}{2520} & \multicolumn{1}{c|}{1541} & \multicolumn{1}{c}{2899} & \multicolumn{1}{c}{2010} & \multicolumn{1}{c}{2811} & \multicolumn{1}{c}{2520} & \multicolumn{1}{c}{1541} \\
    \hline
    25    & 25    & 0.67  & 7.01  & 5.26  & 0.70  & 6.50  & 0.83  & 3.03  & 2.86  & 2.15  & 1.91  & 4.13  & 19.62 & 7.38  & 9.28  & 37.99 \\
    50    & 25    & 0.67  & 3.75  & 4.39  & 0.30  & 6.50  & 0.51  & 2.69  & 2.58  & 1.00  & 1.80  & 0.85  & 13.76 & 3.33  & 1.20  & 9.50 \\
    75    & 25    & 0.67  & 3.75  & 1.91  & 0.15 & 2.41  & 0.11  & 1.00  & 2.00  & 0.27  & 1.48  & 0.01  & 0.05  & 0.10  & 0.02  & 0.15 \\
    25    & 50    & 0.67  & 8.85  & 6.01  & 2.10  & 6.50  & 1.77  & 5.70  & 3.21  & 2.87  & 2.47  & 12.17 & 70.90 & 32.68 & 16.12 & 54.18 \\
    50    & 50    & 0.67  & 8.85  & 6.01  & 0.68  & 6.50  & 1.60  & 4.87  & 3.10  & 2.56  & 2.04  & 2.33  & 47.52 & 5.34  & 6.15  & 42.42 \\
    75    & 50    & 0.67  & 4.22  & 2.45  & 0.01  & 2.41  & 0.64  & 1.30  & 2.04  & 0.67  & 1.83  & 0.02  & 0.06  & 0.56  & 0.43  & 0.38 \\
    25    & 75    & 3.24  & 34.64 & 5.55  & 2.60  & 12.51 & 3.25  & 10.29 & 3.26  & 3.56  & 4.05  & 251.26 & 1989.27 & 63.89 & 775.82 & 241.40 \\
    50    & 75    & 3.24  & 21.11 & 5.18  & 1.75  & 12.51 & 3.23  & 10.28 & 3.02  & 3.43  & 3.97  & 59.83 & 1168.04 & 24.68 & 68.81 & 199.46 \\
    75    & 75    & 1.09  & 21.11 & 4.60  & 0.81  & 7.10  & 2.08  & 5.54  & 2.15  & 2.14  & 2.85  & 5.56  & 9.40  & 1.47  & 0.84  & 76.52 \\
    \bottomrule
    \end{tabular}%
  \label{tab:MRTS_KL}%
\end{table}%

\end{landscape}
}

\subsection{Estimator performance measures as a function of subsample size and surveying implications}
\begin{figure}[ht]
    \includegraphics[width=0.7\textwidth]{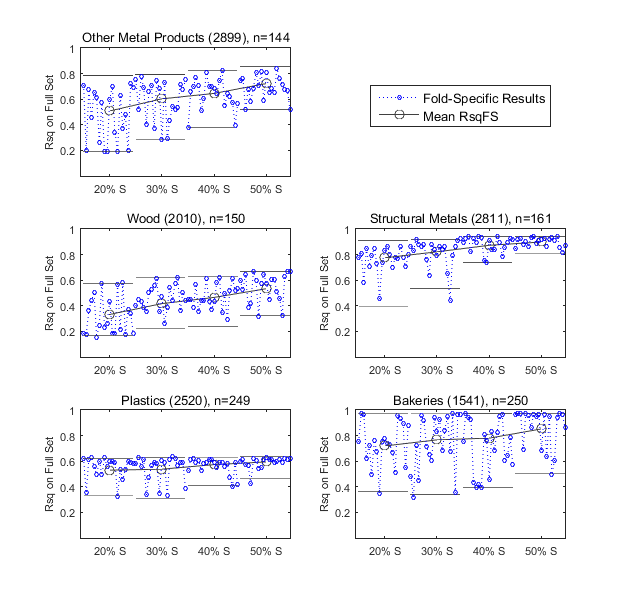}
    \centering
    \caption{Best Method's fit on the full census, $R_{Full}^2$, as a function of relative subset size for selected industries. CAPNLS is the Best Method for industry codes 2899, 2010, 2811, and 2520, whereas CAP is the Best Method for industry code 1541.}
    \label{fig2:IndPerform}
\end{figure}

We apply the results from our framework to make recommendations about the minimal size that a randomly-sampled production survey needs to represent a census. We compute simulation-based confidence intervals on $R_{FS}^2$ across the replicates of our RLT results. As mentioned, increased testing set variance as the learning set size increases does not seem to be large enough to affect the variance of our estimates across the different learning and testing set sizes considered. Based on Table \ref{tab2:SumStats}, we label  CAPNLS as the Best Method across the different survey sizes for all industries, except Bakeries, for which CAP is identified as the Best Method. Figure \ref{fig2:IndPerform} shows the learning subset-specific results for the Best Method in terms of goodness-of-fit, $R_{FS}^2$, for the industries. We note that the variance of $R_{FS}^2$ and overall predictive power is significantly enhanced by the inclusion of the in-sample component of the expected full set error. In Appendix \ref{App:AppendixE}, we further explore the sensitivity of the results shown in Figure \ref{fig2:IndPerform} to our assumption of a finite population of firms and discuss the consequences of considering an infinite amount of unobserved firms when assessing the predictive capabilities of our estimators, thus only evaluating estimator performance in terms of predictive error.

The mean goodness-of-fit increases in survey subsample size for all industries with different degrees of diminishing returns. The results are of significant practical importance for countries and organizations that do not conduct annual censuses. Although the goodness-of-fit results we obtain are specific to the particular census data sets. To use the data from the census year to inform the sample size needed in the following (non-census) years, requires assuming that both the set of establishments within an industry and the complexity of the production function do not change significantly over the time period.  For example based on the Chilean 2010 census data, if production functions with $75$\% of the predictive power of a census-fitted production function are desired in 2011, the relative survey sample size needs to be approximately $40$\%, $45$\%, $< 20$\%, $< 20$\%,and $25$\% for industry codes 2899, 2010, 2811, 2520, and 1541, respectively. 

\section{Conclusions}
This paper has two significant contributions to the production function estimation literature. Firstly and most importantly, we constructed a framework to test the adequateness of a production function estimator on real data. Specifically, we established a procedure based on repeated learning-testing and parametric bootstrapping that is able to assess the quality of subsample-fitted production functions to fit full survey (census) samples. Further, this procedure estimates the relative quality of the subsample-fitted production function to that of one fitted with a full sample. Using our framework, we demonstrated for our application that unlike for simulated data, CAPNLS, CAP and a Cobb-Douglas specification performed similarly. Our functional estimator selection procedure is widely applicable, and thus should be routinely used for model selection of econometrically-estimated production functions. Finally, we discovered that the commonly-used Cobb Douglas production function results in very competitive approximations on the Chilean manufacturing dataset at all learning set sizes if an additive residual is used. 

Secondly, we introduced CAPNLS, a nonparametric estimator, which imposes global optimization and no refitting relative to CAP, and additional smoothing relative to CNLS. We formulated a homoscedastic version of CAPNLS as a series of quadratic programs, which improves computational performance. We demonstrated that CAPNLS' additional structure relative to CAP and parsimonious structure relative to CNLS translates into superior performance, smaller sensitivity to noise and input vector dimensionality,  increased robustness in learning-to-full estimation and a faster empirical rate of convergence on simulated data when the noise level is high relative to the full variance of the output. When the noise level is relatively low to the full variance of the output, CAPNLS's performance is similar to CAP and better than CNLS. 


Further work can be done in applying our estimator selection framework to a broader array of datasets, as we have restricted this exposition to the largest industries in the Chilean manufacturing dataset. Theoretical research related to CAPNLS, such as proving consistency and setting bounds on CAPNLS' fast rate of convergence remain open. Incorporation of smoothing strategies to CAPNLS, such as the one presented in \citet{mazumder2015Computational}, also are outstanding future lines of work. Finally, we have focused on standard axioms of monotonicity and concavity, \citet{shephard1970theory}; however this excludes the phenomenon of increasing returns-to-scale. Extending this study to include estimators with alternative assumptions to concave, i.e. \citet{hwangbo2015power,yagi2018SshapeUltraPassum}, also seems promising.

\clearpage

		\begin{center}
			{\large\bf Appendix}
		\end{center}

    This appendix includes:
    \begin{itemize}
    	\item Results for the case with 3 and 4 regressors; More extensive simulated dataset results (Appendix \ref{App:AppendixA}),
    	\item Scalability of CAPNLS to larger datasets (Appendix \ref{App:AppendixB}).
    	\item Parametric Bootstrap algorithm to calculate expected optimism (Appendix \ref{App:AppendixC}).
    	\item Comparison between different error structures for the parametric Cobb-Douglas function (Appendix \ref{sec:APPG.CobbDouglas}).
    	\item Cobb-Douglas results with multiplicative residual assumption for Chilean manufacturing data (Appendix \ref{App:AppendixD}).
    	\item Application Results for Infinite Populations (Appendix \ref{App:AppendixE}).
    	\item Comprehensive Application Results: MPSS, MP, and MRTS (Appendix \ref{sec:APPF.AppComp}).
    \end{itemize}

	\numberwithin{equation}{section}
	\numberwithin{table}{section}
	\numberwithin{figure}{section}
	\clearpage
	\appendix	
	\section{Results for the case with 3 and 4 regressors; More extensive simulated dataset results} \label{App:AppendixA}
	In this section, we provide additional Monte Carlo simulation results for the case when there are 3 and 4 regressors. The performance of all estimators deteriorates significantly due to the curse of dimensionality. The results reported are for the performance metric full set error,  $\widehat{E(Err_{FS}^{nL})}$, which is a weighting of the expected in-sample error and the expected predictive error. While the parametric estimator should have the best performance in terms of expected predictive error, it is not the case for expected in-sample error. The parametric estimator is generally not able to fit the observed data with noise in it as well as the nonparametric estimators because of its lack of flexibility. Or stated differently, the nonparametric estimators benefit from being able to overfit the observed sample. We see the effects of lack of flexibility for the parametric estimator relative to the nonparametric estimator is pronounced in higher dimensional low noise cases.

	\subsection{Tri-variate input Cobb-Douglas DGP}
	\label{App:Extensions}
	We consider the data generation process (DGP) $Y_i = X_{i1}^{0.4} X_{i2}^{0.3} X_{i3}^{0.2} + \epsilon_i$, where $\epsilon_i \sim N(0, \sigma^2)$, $\sigma = \{0.01, 0.05, 0.1, 0.2, 0.3, 0.4\}$ for our six noise settings having the same small and large noise split as the previous example, and $X_{ij} \sim Unif(0.1, 1)$ for $j = 1, 2, 3, \; i = 1, \ldots, n_L$. Again, CNLS's expected full set errors exceed the displayed range for the learning-to-full set scenarios regardless of noise level, due to the poor predictive error values, which are partly linked to the higher proportion of ill-defined hyperplanes\footnote{These are hyperplanes which have zero coefficients on some input dimensions, implying it is possible to obtain output without the zero-coefficient inputs.}  CNLS fits. Compared to bi-variate case, the parametric estimator deteriorates relative to the nonparametric estimators, i.e., the higher errors for the parametric estimator exceed the small scale of most panels in Figure \ref{fig:1.Trivariat}. 
	
	Figure \ref{fig:2.Trivariatelarge}, however, shows that the errors given by the parametric estimator are lower than the errors for CNLS in learning-to-full settings. Further, CAP's expected performance deteriorates relative to the bi-variate example. As in the bi-variate example, as the learning set grows, the expected full set errors gap between CAPNLS and the correctly specified parametric estimator either favors CAPNLS at every learning set size or becomes more favorable for CAPNLS as the learning set size increases. Finally, Figure \ref{fig:2.Trivariatelarge} shows that CAPNLS can accurately recover a production function even when noise composes nearly 85\% of the variance, as shown when $\sigma = 0.4$.
	
	In Table \ref{tab:A1.Trivariate}, we observe that all methods fit slightly more hyperplanes than for the bi-variate example. The increase in the number of hyperplanes with increased dimensionality is moderate for both CAPNLS and CAP at all settings. For CNLS, while the number of hyperplanes does not significantly increase for $n=100$, it significantly increases for the two larger datasets. The runtimes for all methods are also higher than in the previous example, i.e., CAPNLS's times nearly double, although staying below one minute for all scenarios.
	
	\begin{landscape}
		\begin{figure}[ht]
			\includegraphics[width=\linewidth]{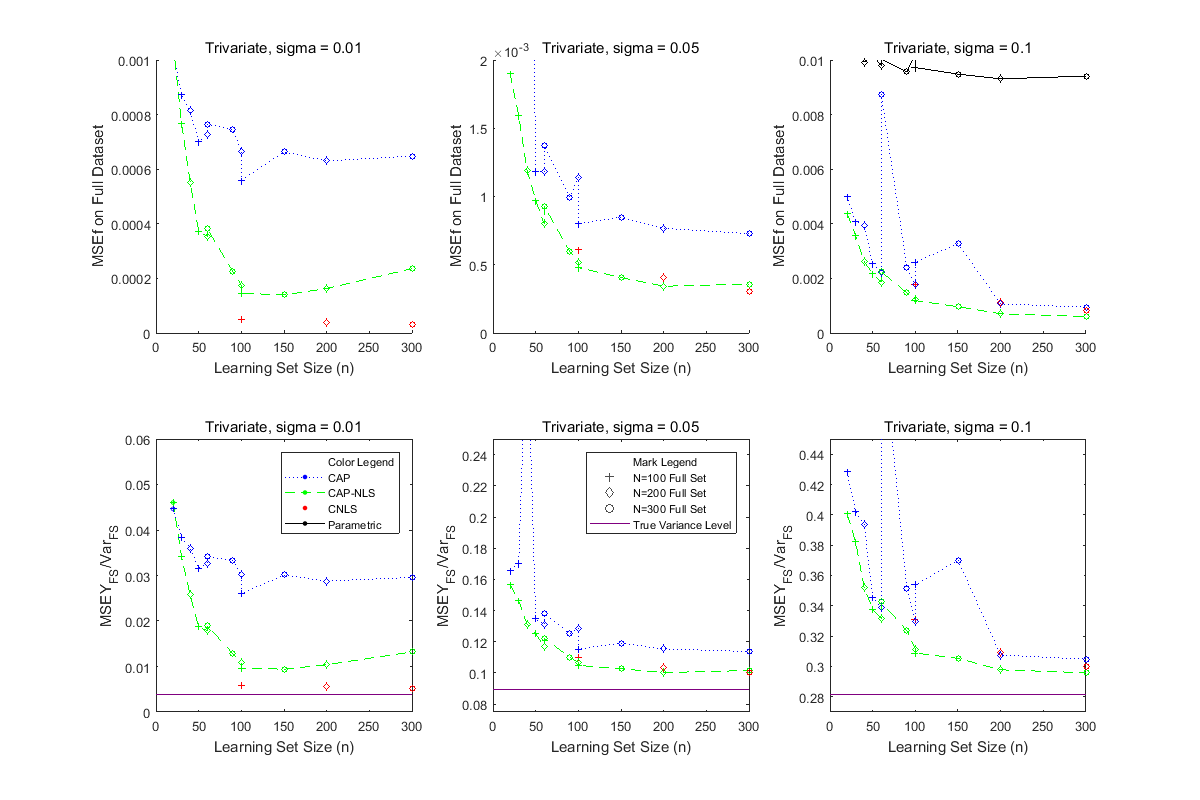}
			\centering
			\caption{Trivariate Input Cobb-Douglas DGP results for small noise settings. \label{fig:1.Trivariat}}
		\end{figure}
		\begin{figure}[!ht]
			\begin{center}
				\includegraphics[width=\linewidth]{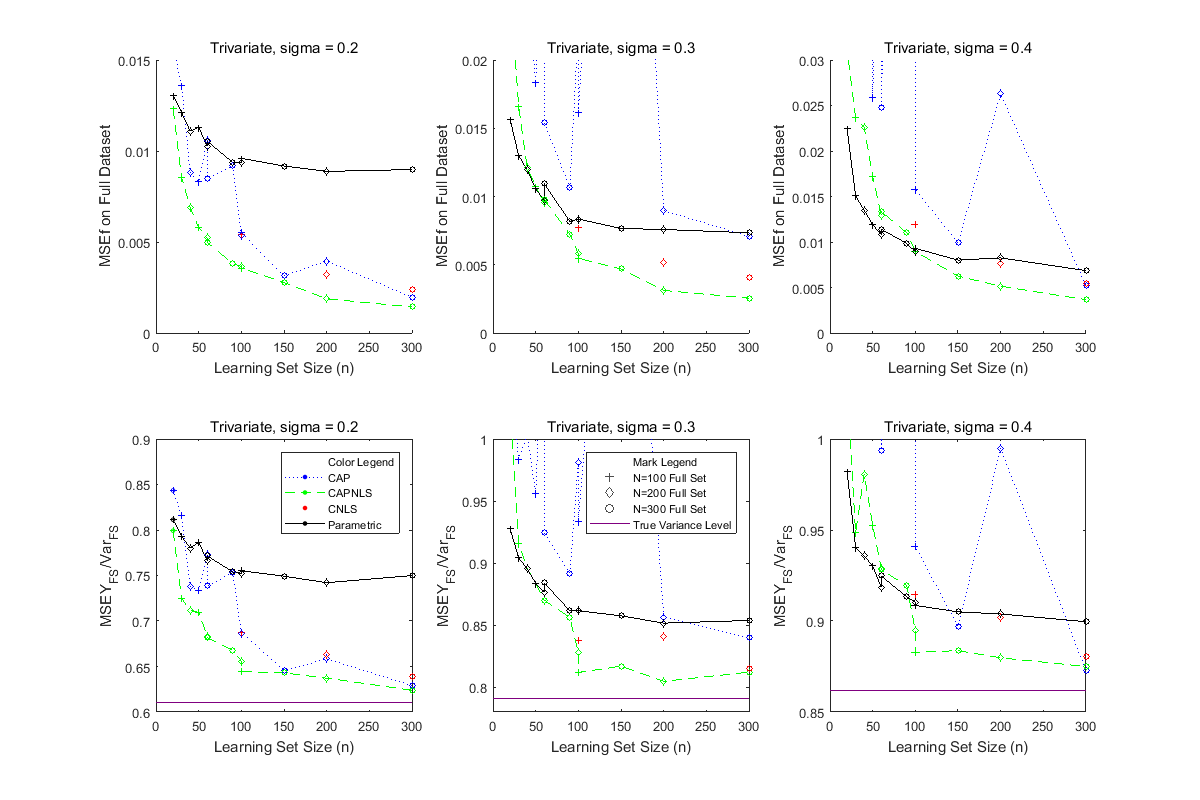}
			\end{center}
			\caption{Trivariate Input Cobb-Douglas DGP results for large noise settings. \label{fig:2.Trivariatelarge}}
		\end{figure}
	\end{landscape}
	
	\renewcommand{\arraystretch}{0.65}
	\begin{table}[t]
		\centering
		\caption{Number of Hyperplanes and Runtimes for Trivariate Input Cobb-Douglas DGP.}
		\begin{tabular}{rc|ccc|ccc|ccc}
			\toprule
			\multicolumn{1}{c}{} &       & \multicolumn{3}{c|}{CAPNLS} & \multicolumn{3}{c|}{CAP} & \multicolumn{3}{c}{CNLS} \\
			$\sigma$  & $n$ & 100   & 200   & 300   & 100   & 200   & 300   & 100   & 200   & 300 \\
			\hline
			
			\multirow{2}[0]{*}{0.01} & \textit{K} & 9     & 12    & 13    & 2     & 2     & 2     & 96    & 193   & 294 \\
			& Time (s) & 5     & 28    & 51    & 0.59  & 0.78  & 1     & 1     & 9     & 27 \\
			\hline
			\multirow{2}[0]{*}{0.05} & \textit{K} & 10    & 11    & 12    & 2     & 2     & 3     & 80    & 169   & 235 \\
			& Time (s) & 5     & 28    & 45    & 0.53  & 0.83  & 1     & 1     & 9     & 28 \\
			\hline
			\multirow{2}[0]{*}{0.1} & \textit{K} & 8     & 12    & 14    & 2     & 2     & 3     & 75    & 136   & 199 \\
			& Time (s) & 6     & 24    & 54    & 0.54  & 1     & 1     & 1     & 10    & 24 \\
			\hline
			\multirow{2}[0]{*}{0.2} & \textit{K} & 8     & 11    & 12    & 2     & 3     & 2     & 61    & 126   & 193 \\
			& Time (s) & 8     & 23    & 50    & 0.52  & 0.93  & 1     & 1     & 9     & 28 \\
			\hline  
			\multirow{2}[0]{*}{0.3} & \textit{K} & 8     & 11    & 12    & 2     & 3     & 3     & 57    & 123   & 184 \\
			& Time (s) & 5     & 23    & 49    & 0.53  & 0.92  & 1     & 1     & 9     & 28 \\
			\hline  
			\multirow{2}[0]{*}{0.4} & \textit{K} & 8     & 11    & 12    & 2     & 3     & 3     & 54    & 115   & 179 \\
			& Time (s) & 5     & 25    & 49    & 0.52  & 0.92  & 1     & 1     & 9     & 29 \\
			
			\bottomrule
		\end{tabular}%
		\label{tab:A1.Trivariate}%
	\end{table}%

	\subsection{Tetravariate input Cobb-Douglas DGP}
	\label{sec:A.2.Four-variate}	
	We consider the DGP $Y_i = X_{i1}^{0.3} X_{i2}^{0.25} X_{i3}^{0.25} X_{i4}^{0.1} + \epsilon_i$, where $\epsilon_i \sim N(0, \sigma^2)$ and $X_{ij} \sim Unif(0.1, 1)$ for $j = 1, 2, 3, 4; \; i = 1, \ldots, n_L$ and $\sigma = \{0.01, 0.05, 0.1, 0.2, 0.3, 0.4\}$ for our six noise settings. The preformance of all estimators further deteriorates due the curse of dimensionality. Figures \ref{fig:3.Fourvariat} and \ref{fig:4.Fourvariatelarge} show that for this higher dimensional example, the parameters in the parametric estimator are increasingly difficult to learn, and thus the parametric estimator can only predict the true function up to a certain accuracy, namely $\bar{MSE}_{FSf}^{nL} = 0.015$, and then tends to plateau at this error level even as the learning set size increases. Moreover, the benefits of CAPNLS over the other nonparametric methods are similar to the tri-variate example for the small noise settings, but significantly larger for the large noise settings. Finally, the gap between CAPNLS and all the other functional estimators, parametric or nonparametric, favors CAPNLS for all noise settings and learning set sizes.
	
	Table \ref{tab:A2.Fourvariate} shows that the number of hyperplanes needed to fit the four-variate input production function does not significantly increase from the trivariate-input case for any of the methods. CAPNLS has 40-60\% longer runtimes compared to the trivariate-input case. The runtime increase with dimensionality; however, this is not a severe concern, because the input information to fit a production function (or output in the case of a cost function) rarely exceeds four variables. The maximum recorded runtime for CAPNLS is still below two minutes, i.e., it is not large in absolute terms. 
	
	\begin{landscape}
		\begin{figure}[!ht]
			\begin{center}
				\includegraphics[width=\linewidth]{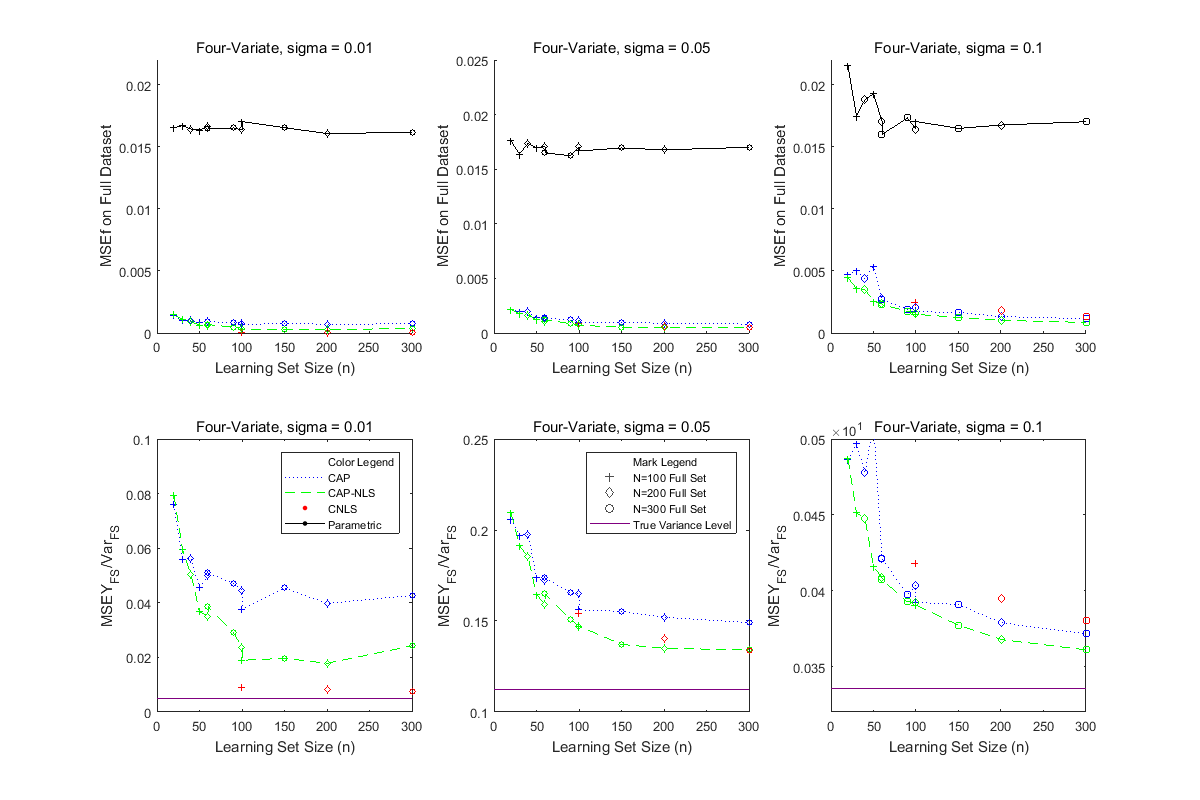}
			\end{center}
			\caption{Tetra-variate Input Cobb-Douglas DGP results for small noise settings. \label{fig:3.Fourvariat}}
		\end{figure}
		\begin{figure}[!ht]
			\begin{center}
				\includegraphics[width=\linewidth]{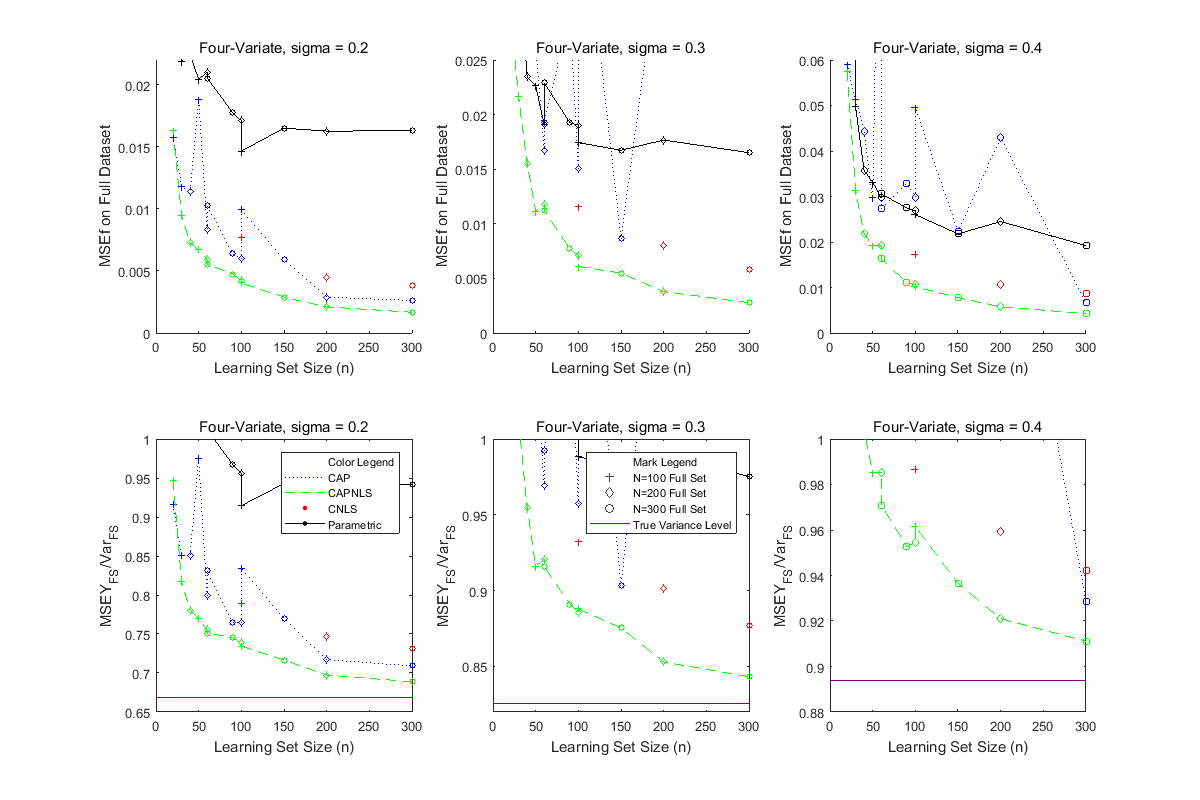}
			\end{center}
			\caption{Tetra-variate Input Cobb-Douglas DGP results for large noise settings. \label{fig:4.Fourvariatelarge}}
		\end{figure}	
	\end{landscape}
	
	\begin{table}[t]\
		\small
		\centering
		\caption{Number of Hyperplanes and Runtimes for Tetra-variate Input Cobb-Douglas DGP.}
		\begin{tabular}{rc|ccc|ccc|ccc}
			\toprule
			\multicolumn{1}{c}{} &       & \multicolumn{3}{c|}{CAPNLS} & \multicolumn{3}{c|}{CAP} & \multicolumn{3}{c}{CNLS} \\
			$\sigma$  & $n$ & 100   & 200   & 300   & 100   & 200   & 300   & 100   & 200   & 300 \\
			\hline
			\multirow{2}[0]{*}{0.01} & \textit{K} & 7     & 11    & 12    & 2     & 2     & 2     & 98    & 194   & 234 \\
			& Time (s) & 5     & 29    & 70    & 0.42  & 1     & 2     & 1     & 8     & 24 \\
			\hline
			\multirow{2}[0]{*}{0.05} & \textit{K} & 7     & 12    & 13    & 2     & 2     & 2     & 87    & 170   & 215 \\
			& Time (s) & 4     & 33    & 65    & 0.39  & 1     & 1     & 1     & 9     & 27 \\
			\hline
			\multirow{2}[0]{*}{0.1} & \textit{K} & 7     & 12    & 12    & 2     & 2     & 2     & 55    & 166   & 207 \\
			& Time (s) & 4     & 30    & 62    & 0.6   & 1     & 2     & 1     & 10    & 32 \\
			\hline
			\multirow{2}[0]{*}{0.2} & \textit{K} & 7     & 12    & 13    & 2     & 2     & 2     & 63    & 132   & 192 \\
			& Time (s) & 4     & 36    & 79    & 0.45  & 1     & 2     & 1     & 9     & 31 \\
			\hline
			\multirow{2}[0]{*}{0.3} & \textit{K} & 7     & 12    & 12    & 2     & 2     & 2     & 59    & 122   & 192 \\
			& Time (s) & 4     & 36    & 74    & 0.46  & 1     & 2     & 1     & 10    & 32 \\
			\hline
			\multirow{2}[0]{*}{0.4} & \textit{K} & 7     & 12    & 12    & 2     & 2     & 2     & 57    & 122   & 186 \\
			& Time (s) & 4     & 36    & 75    & 0.48  & 1     & 2     & 1     & 10    & 33 \\
			\bottomrule
		\end{tabular}%
		\label{tab:A2.Fourvariate}%
	\end{table}%
	
	\subsection{More extensive simulated dataset results}
	Tables \ref{tab:A3}-\ref{tab:A20} provide more detail results. This information was show in Figures \ref{fig:MCSim1} and \ref{fig2:MCSim_2var_lar} in the main text and Figures \ref{fig:1.Trivariat}-\ref{fig:4.Fourvariatelarge} in the Appendix. 
	
	\begin{landscape}
		
		\begin{table}[ht]\
			\small
			\centering
			\caption{$MSE_f$, $MSE_{ISf}$, $MSE_{FSf}$, $MSE_y$, $MSE_{ISy}$, $MSE_{FSy}$, time, and $K$ results for $d = 2$, $\sigma = 0.2$.}
			
\begin{tabular}{rlrrrrrrrrr}
				\toprule
				\multicolumn{2}{c}{$d$ = 2, $\sigma$ = 0.2} & \multicolumn{3}{c}{nFull = 100} & \multicolumn{3}{c}{nFull = 200} & \multicolumn{3}{c}{nFull = 300} \\
				\multicolumn{1}{l}{nL/nFull} & Metric & \multicolumn{1}{c}{CAPNLS} & \multicolumn{1}{c}{CAP} & \multicolumn{1}{c}{CNLS} & \multicolumn{1}{c}{CAPNLS} & \multicolumn{1}{c}{CAP} & \multicolumn{1}{c}{CNLS} & \multicolumn{1}{c}{CAPNLS} & \multicolumn{1}{c}{CAP} & \multicolumn{1}{c}{CNLS} \\
				\midrule
				100\% & MSEFInSamp & 0.0025 & 0.0069 & 0.003 & 0.0014 & 0.002 & 0.0017 & 0.0012 & 0.0019 & 0.0015 \\
				100\% & MSEFTesting & 0.0028 & 0.0073 & 292.9882 & 0.0014 & 0.002 & 93.3874 & 0.0012 & 0.0019 & 168.9078 \\
				100\% & MSEFCensus & 0.0025 & 0.0069 & 0.003 & 0.0014 & 0.002 & 0.0017 & 0.0012 & 0.0019 & 0.0015 \\
				\midrule
				80\%  & MSEFInSamp & 0.0045 & 0.016 & 0.0054 & 0.0024 & 0.0053 & 0.0029 & 0.0019 & 0.0273 & 0.0023 \\
				80\%  & MSEFTesting & 0.0047 & 0.0173 & 56.2393 & 0.0026 & 0.0055 & 414.9028 & 0.0019 & 0.0249 & 148.8055 \\
				80\%  & MSEFCensus & 0.0046 & 0.0166 & 28.1224 & 0.0025 & 0.0054 & 207.4528 & 0.0019 & 0.0261 & 74.4039 \\
\bottomrule
			\end{tabular}%
			
			\label{tab:A3}%
		\end{table}%
		\begin{table}[ht]\
			\small
			\centering
			\caption{$MSE_f$, $MSE_{ISf}$, $MSE_{FSf}$, $MSE_y$, $MSE_{ISy}$, $MSE_{FSy}$, time, and $K$ results for $d = 2$, $\sigma = 0.3$.}
			\begin{tabular}{rlrrrrrrrrr}
				\toprule
				\multicolumn{2}{c}{$d$ = 2, $\sigma$ = 0.3} & \multicolumn{3}{c}{nFull = 100} & \multicolumn{3}{c}{nFull = 200} & \multicolumn{3}{c}{nFull = 300} \\
				\multicolumn{1}{l}{nL/nF} & Metric & \multicolumn{1}{c}{CAPNLS} & \multicolumn{1}{c}{CAP} & \multicolumn{1}{c}{CNLS} & \multicolumn{1}{c}{CAPNLS} & \multicolumn{1}{c}{CAP} & \multicolumn{1}{c}{CNLS} & \multicolumn{1}{c}{CAPNLS} & \multicolumn{1}{c}{CAP} & \multicolumn{1}{c}{CNLS} \\
				\midrule
				100\% & MSEFInSamp & 0.0051 & 0.0061 & 0.0057 & 0.0027 & 0.003 & 0.0033 & 0.0019 & 0.0029 & 0.0024 \\
				100\% & MSEFTesting & 0.0056 & 0.0067 & 308.9863 & 0.0028 & 0.0031 & 102.7481 & 0.0019 & 0.0029 & 198.8724 \\
				100\% & MSEFCensus & 0.0051 & 0.0061 & 0.0057 & 0.0027 & 0.003 & 0.0033 & 0.0019 & 0.0029 & 0.0024 \\
				\midrule
				80\%  & MSEFInSamp & 0.0075 & 0.0122 & 0.0084 & 0.0048 & 0.0111 & 0.0057 & 0.0035 & 0.015 & 0.0042 \\
				80\%  & MSEFTesting & 0.0084 & 0.0144 & 10.9325 & 0.0052 & 0.0139 & 217.2453 & 0.0035 & 0.015 & 116.0226 \\
				80\%  & MSEFCensus & 0.0079 & 0.0133 & 5.4704 & 0.005 & 0.0125 & 108.6255 & 0.0035 & 0.015 & 58.0134 \\
\bottomrule
			\end{tabular}%
			\label{tab:A4}%
		\end{table}%
		\begin{table}[ht]\
			\small
			\centering
			\caption{$MSE_f$, $MSE_{ISf}$, $MSE_{FSf}$, $MSE_y$, $MSE_{ISy}$, $MSE_{FSy}$, time, and $K$ results for $d = 2$, $\sigma = 0.4$.}
			\begin{tabular}{rlrrrrrrrrr}
				\toprule
				\multicolumn{2}{c}{$d$ = 2, $\sigma$ = 0.4} & \multicolumn{3}{c}{nFull = 100} & \multicolumn{3}{c}{nFull = 200} & \multicolumn{3}{c}{nFull = 300} \\
				\multicolumn{1}{l}{nL/nF} & Metric & \multicolumn{1}{c}{CAPNLS} & \multicolumn{1}{c}{CAP} & \multicolumn{1}{c}{CNLS} & \multicolumn{1}{c}{CAPNLS} & \multicolumn{1}{c}{CAP} & \multicolumn{1}{c}{CNLS} & \multicolumn{1}{c}{CAPNLS} & \multicolumn{1}{c}{CAP} & \multicolumn{1}{c}{CNLS} \\
				\midrule
				100\% & MSEFInSamp & 0.0076 & 0.0099 & 0.0087 & 0.0041 & 0.0059 & 0.0048 & 0.0027 & 0.1022 & 0.0033 \\
				100\% & MSEFTesting & 0.0081 & 0.0105 & 281.3251 & 0.0043 & 0.0064 & 198.6378 & 0.0027 & 0.1093 & 89.5596 \\
				100\% & MSEFCensus & 0.0076 & 0.0099 & 0.0087 & 0.0041 & 0.0059 & 0.0048 & 0.0027 & 0.1022 & 0.0033 \\
				\midrule
				80\%  & MSEFInSamp & 0.0124 & 0.0408 & 0.0136 & 0.0062 & 0.1578 & 0.0073 & 0.0054 & 0.0112 & 0.0063 \\
				80\%  & MSEFTesting & 0.0125 & 0.0398 & 10.2707 & 0.0064 & 0.1723 & 118.2487 & 0.0055 & 0.0115 & 198.9407 \\
				80\%  & MSEFCensus & 0.0125 & 0.0403 & 5.1422 & 0.0063 & 0.165 & 59.128 & 0.0054 & 0.0113 & 99.4735 \\
\bottomrule
			\end{tabular}%
			\label{tab:A5}%
		\end{table}%
		\begin{table}[ht]\
			\small
			\centering
			\caption{$MSE_f$, $MSE_{ISf}$, $MSE_{FSf}$, $MSE_y$, $MSE_{ISy}$, $MSE_{FSy}$, time, and $K$ results for $d = 3$, $\sigma = 0.2$.}
			\begin{tabular}{rlrrrrrrrrr}
				\toprule
				\multicolumn{2}{c}{$d$ = 3, $\sigma$ = 0.2} & \multicolumn{3}{c}{nFull = 100} & \multicolumn{3}{c}{nFull = 200} & \multicolumn{3}{c}{nFull = 300} \\
				\multicolumn{1}{l}{nL/nF} & Metric & \multicolumn{1}{c}{CAPNLS} & \multicolumn{1}{c}{CAP} & \multicolumn{1}{c}{CNLS} & \multicolumn{1}{c}{CAPNLS} & \multicolumn{1}{c}{CAP} & \multicolumn{1}{c}{CNLS} & \multicolumn{1}{c}{CAPNLS} & \multicolumn{1}{c}{CAP} & \multicolumn{1}{c}{CNLS} \\
				\midrule
				100\% & MSEFInSamp & 0.0031 & 0.0173 & 0.0048 & 0.0019 & 0.0026 & 0.0031 & 0.0014 & 0.0042 & 0.0024 \\
				100\% & MSEFTesting & 0.0034 & 0.0205 & 591.2154 & 0.002 & 0.0028 & 1240.058 & 0.0014 & 0.0045 & 1174.255 \\
				100\% & MSEFCensus & 0.0031 & 0.0173 & 0.0048 & 0.0019 & 0.0026 & 0.0031 & 0.0014 & 0.0042 & 0.0024 \\
				\midrule
				80\%  & MSEFInSamp & 0.0047 & 0.0099 & 0.0074 & 0.0033 & 0.0046 & 0.005 & 0.0025 & 0.0038 & 0.004 \\
				80\%  & MSEFTesting & 0.0053 & 0.0114 & 53.6331 & 0.0035 & 0.0051 & 357.0193 & 0.0027 & 0.0041 & 907.1555 \\
				80\%  & MSEFCensus & 0.005 & 0.0106 & 26.8203 & 0.0034 & 0.0048 & 178.5121 & 0.0026 & 0.0039 & 453.5797 \\
\bottomrule
			\end{tabular}%
			\label{tab:A6}%
		\end{table}%
		\begin{table}[ht]\
			\small
			\centering
			\caption{$MSE_f$, $MSE_{ISf}$, $MSE_{FSf}$, $MSE_y$, $MSE_{ISy}$, $MSE_{FSy}$, time, and $K$ results for $d = 3$, $\sigma = 0.3$.}
			\begin{tabular}{rlrrrrrrrrr}
				\toprule
				\multicolumn{2}{c}{$d$ = 3, $\sigma$ = 0.3} & \multicolumn{3}{c}{nFull = 100} & \multicolumn{3}{c}{nFull = 200} & \multicolumn{3}{c}{nFull = 300} \\
				\multicolumn{1}{l}{nL/nF} & Metric & \multicolumn{1}{c}{CAPNLS} & \multicolumn{1}{c}{CAP} & \multicolumn{1}{c}{CNLS} & \multicolumn{1}{c}{CAPNLS} & \multicolumn{1}{c}{CAP} & \multicolumn{1}{c}{CNLS} & \multicolumn{1}{c}{CAPNLS} & \multicolumn{1}{c}{CAP} & \multicolumn{1}{c}{CNLS} \\
				\midrule
				100\% & MSEFInSamp & 0.0052 & 0.0699 & 0.0078 & 0.0034 & 0.0094 & 0.0053 & 0.0026 & 0.0053 & 0.0044 \\
				100\% & MSEFTesting & 0.0057 & 0.0753 & 782.2253 & 0.0035 & 0.0095 & 717.8571 & 0.0026 & 0.0055 & 1235.091 \\
				100\% & MSEFCensus & 0.0052 & 0.0699 & 0.0078 & 0.0034 & 0.0094 & 0.0053 & 0.0026 & 0.0053 & 0.0044 \\
				\midrule
				80\%  & MSEFInSamp & 0.0095 & 0.0429 & 0.0144 & 0.0058 & 0.0303 & 0.0088 & 0.0041 & 0.1537 & 0.0064 \\
				80\%  & MSEFTesting & 0.0104 & 0.0491 & 119.1442 & 0.0064 & 0.0312 & 768.1003 & 0.0043 & 0.1663 & 926.5316 \\
				80\%  & MSEFCensus & 0.0099 & 0.046 & 59.5793 & 0.0061 & 0.0307 & 384.0546 & 0.0042 & 0.16  & 463.269 \\
\bottomrule
			\end{tabular}%
			\label{tab:A7}%
		\end{table}%
		\begin{table}[ht]\
			\small
			\centering
			\caption{$MSE_f$, $MSE_{ISf}$, $MSE_{FSf}$, $MSE_y$, $MSE_{ISy}$, $MSE_{FSy}$, time, and $K$ results for $d = 3$, $\sigma = 0.4$.}
			\begin{tabular}{rlrrrrrrrrr}
				\toprule
				\multicolumn{2}{c}{$d$ = 3, $\sigma$ = 0.4} & \multicolumn{3}{c}{nFull = 100} & \multicolumn{3}{c}{nFull = 200} & \multicolumn{3}{c}{nFull = 300} \\
				\multicolumn{1}{l}{nL/nF} & Metric & \multicolumn{1}{c}{CAPNLS} & \multicolumn{1}{c}{CAP} & \multicolumn{1}{c}{CNLS} & \multicolumn{1}{c}{CAPNLS} & \multicolumn{1}{c}{CAP} & \multicolumn{1}{c}{CNLS} & \multicolumn{1}{c}{CAPNLS} & \multicolumn{1}{c}{CAP} & \multicolumn{1}{c}{CNLS} \\
				\midrule
				100\% & MSEFInSamp & 0.0096 & 0.347 & 0.0126 & 0.0051 & 0.0437 & 0.0078 & 0.0039 & 0.0836 & 0.0061 \\
				100\% & MSEFTesting & 0.0104 & 0.318 & 940.716 & 0.0052 & 0.0429 & 1025.042 & 0.004 & 0.0848 & 1404.57 \\
				100\% & MSEFCensus & 0.0096 & 0.347 & 0.0126 & 0.0051 & 0.0437 & 0.0078 & 0.0039 & 0.0836 & 0.0061 \\
				\midrule
				80\%  & MSEFInSamp & 0.0154 & 0.034 & 0.0195 & 0.0089 & 0.0247 & 0.0138 & 0.0056 & 0.024 & 0.0086 \\
				80\%  & MSEFTesting & 0.0181 & 0.0397 & 75.3542 & 0.0092 & 0.0258 & 669.2816 & 0.0061 & 0.0267 & 1050.676 \\
				80\%  & MSEFCensus & 0.0168 & 0.0369 & 37.6869 & 0.009 & 0.0252 & 334.6477 & 0.0058 & 0.0253 & 525.3422 \\
\bottomrule
			\end{tabular}%
			\label{tab:A8}%
		\end{table}%
		\begin{table}[ht]\
			\small
			\centering
			\caption{$MSE_f$, $MSE_{ISf}$, $MSE_{FSf}$, $MSE_y$, $MSE_{ISy}$, $MSE_{FSy}$, time, and $K$ results for $d = 4$, $\sigma = 0.2$.}
			\begin{tabular}{rlrrrrrrrrr}
				\toprule
				\multicolumn{2}{c}{$d$ = 4, $\sigma$ = 0.2} & \multicolumn{3}{c}{nFull = 100} & \multicolumn{3}{c}{nFull = 200} & \multicolumn{3}{c}{nFull = 300} \\
				\multicolumn{1}{l}{nL/nF} & Metric & \multicolumn{1}{c}{CAPNLS} & \multicolumn{1}{c}{CAP} & \multicolumn{1}{c}{CNLS} & \multicolumn{1}{c}{CAPNLS} & \multicolumn{1}{c}{CAP} & \multicolumn{1}{c}{CNLS} & \multicolumn{1}{c}{CAPNLS} & \multicolumn{1}{c}{CAP} & \multicolumn{1}{c}{CNLS} \\
				\midrule
				100\% & MSEFInSamp & 0.0038 & 0.0082 & 0.0072 & 0.0023 & 0.0031 & 0.0048 & 0.0017 & 0.0022 & 0.004 \\
				100\% & MSEFTesting & 0.0041 & 0.0084 & 366.5847 & 0.0023 & 0.0034 & 1201.953 & 0.0018 & 0.0023 & 2096.697 \\
				100\% & MSEFCensus & 0.0038 & 0.0082 & 0.0072 & 0.0023 & 0.0031 & 0.0048 & 0.0017 & 0.0022 & 0.004 \\
				\midrule
				80\%  & MSEFInSamp & 0.0056 & 0.0101 & 0.0105 & 0.0036 & 0.0049 & 0.0071 & 0.0026 & 0.0055 & 0.0054 \\
				80\%  & MSEFTesting & 0.0067 & 0.012 & 13.227 & 0.0038 & 0.0054 & 569.4433 & 0.0028 & 0.0063 & 1140.87 \\
				80\%  & MSEFCensus & 0.0061 & 0.011 & 6.6187 & 0.0037 & 0.0052 & 284.7252 & 0.0027 & 0.0059 & 570.4375 \\
\bottomrule
			\end{tabular}%
			\label{tab:A9}%
		\end{table}%
		\begin{table}[ht]\
			\small
			\centering
			\caption{$MSE_f$, $MSE_{ISf}$, $MSE_{FSf}$, $MSE_y$, $MSE_{ISy}$, $MSE_{FSy}$, time, and $K$ results for $d = 4$, $\sigma = 0.3$.}
			\begin{tabular}{rlrrrrrrrrr}
				\toprule
				\multicolumn{2}{c}{$d$ = 4, $\sigma$ = 0.3} & \multicolumn{3}{c}{nFull = 100} & \multicolumn{3}{c}{nFull = 200} & \multicolumn{3}{c}{nFull = 300} \\
				\multicolumn{1}{l}{nL/nF} & Metric & \multicolumn{1}{c}{CAPNLS} & \multicolumn{1}{c}{CAP} & \multicolumn{1}{c}{CNLS} & \multicolumn{1}{c}{CAPNLS} & \multicolumn{1}{c}{CAP} & \multicolumn{1}{c}{CNLS} & \multicolumn{1}{c}{CAPNLS} & \multicolumn{1}{c}{CAP} & \multicolumn{1}{c}{CNLS} \\
				\midrule
				100\% & MSEFInSamp & 0.0068 & 0.0138 & 0.0126 & 0.0042 & 0.0267 & 0.0084 & 0.0028 & 0.0042 & 0.0061 \\
				100\% & MSEFTesting & 0.0074 & 0.0157 & 330.2075 & 0.0043 & 0.0294 & 956.4055 & 0.0028 & 0.0043 & 2061.143 \\
				100\% & MSEFCensus & 0.0068 & 0.0138 & 0.0126 & 0.0042 & 0.0267 & 0.0084 & 0.0028 & 0.0042 & 0.0061 \\
				\midrule
				80\%  & MSEFInSamp & 0.0113 & 0.0198 & 0.0199 & 0.0063 & 0.1558 & 0.0127 & 0.0051 & 0.0505 & 0.0098 \\
				80\%  & MSEFTesting & 0.0123 & 0.0227 & 6.9686 & 0.0061 & 0.1795 & 231.3875 & 0.0054 & 0.0542 & 846.5812 \\
				80\%  & MSEFCensus & 0.0118 & 0.0213 & 3.4943 & 0.0062 & 0.1677 & 115.7001 & 0.0053 & 0.0523 & 423.2955 \\
\bottomrule
			\end{tabular}%
			\label{tab:A10}%
		\end{table}%
		\begin{table}[ht]\
			\small
			\centering
			\caption{$MSE_f$, $MSE_{ISf}$, $MSE_{FSf}$, $MSE_y$, $MSE_{ISy}$, $MSE_{FSy}$, time, and $K$ results for $d = 4$, $\sigma = 0.4$.}
			\begin{tabular}{rlrrrrrrrrr}
				\toprule
				\multicolumn{2}{c}{$d$ = 4, $\sigma$ = 0.4} & \multicolumn{3}{c}{nFull = 100} & \multicolumn{3}{c}{nFull = 200} & \multicolumn{3}{c}{nFull = 300} \\
				\multicolumn{1}{l}{nL/nF} & Metric & \multicolumn{1}{c}{CAPNLS} & \multicolumn{1}{c}{CAP} & \multicolumn{1}{c}{CNLS} & \multicolumn{1}{c}{CAPNLS} & \multicolumn{1}{c}{CAP} & \multicolumn{1}{c}{CNLS} & \multicolumn{1}{c}{CAPNLS} & \multicolumn{1}{c}{CAP} & \multicolumn{1}{c}{CNLS} \\
				\midrule
				100\% & MSEFInSamp & 0.0091 & 0.0333 & 0.0162 & 0.0061 & 0.0109 & 0.012 & 0.0044 & 0.0151 & 0.0088 \\
				100\% & MSEFTesting & 0.0104 & 0.0336 & 269.5879 & 0.0063 & 0.0123 & 1809.438 & 0.0044 & 0.0149 & 2638.173 \\
				100\% & MSEFCensus & 0.0091 & 0.0333 & 0.0162 & 0.0061 & 0.0109 & 0.012 & 0.0044 & 0.0151 & 0.0088 \\
				\midrule
				80\%  & MSEFInSamp & 0.0183 & 0.0637 & 0.0296 & 0.011 & 0.0194 & 0.0177 & 0.0078 & 0.027 & 0.0141 \\
				80\%  & MSEFTesting & 0.02  & 0.0755 & 14.6548 & 0.012 & 0.0206 & 378.636 & 0.0084 & 0.0279 & 671.0586 \\
				80\%  & MSEFCensus & 0.0191 & 0.0696 & 7.3422 & 0.0115 & 0.02  & 189.3269 & 0.0081 & 0.0274 & 335.5364 \\
\bottomrule
			\end{tabular}%
			\label{tab:A11}%
		\end{table}%
		\begin{table}[ht]\
			\small
			\centering
			\caption{$MSE_f$, $MSE_{ISf}$, $MSE_{FSf}$, $MSE_y$, $MSE_{ISy}$, $MSE_{FSy}$, time, and $K$ results for $d = 2$, $\sigma = 0.01$.}
			\begin{tabular}{rlrrrrrrrrr}
				\toprule
				\multicolumn{2}{c}{$d$ = 2, $\sigma$ = 0.01} & \multicolumn{3}{c}{nFull = 100} & \multicolumn{3}{c}{nFull = 200} & \multicolumn{3}{c}{nFull = 300} \\
				\multicolumn{1}{l}{nL/nF} & Metric & \multicolumn{1}{c}{CAPNLS} & \multicolumn{1}{c}{CAP} & \multicolumn{1}{c}{CNLS} & \multicolumn{1}{c}{CAPNLS} & \multicolumn{1}{c}{CAP} & \multicolumn{1}{c}{CNLS} & \multicolumn{1}{c}{CAPNLS} & \multicolumn{1}{c}{CAP} & \multicolumn{1}{c}{CNLS} \\
				\midrule
				100\% & MSEFInSamp & 0     & 0.0003 & 0     & 0     & 0.0003 & 0     & 0     & 0.0003 & 0 \\
				100\% & MSEFTesting & 0     & 0.0003 & 87.91 & 0     & 0.0003 & 76.2429 & 0     & 0.0003 & 55.3655 \\
				100\% & MSEFCensus & 0     & 0.0003 & 0     & 0     & 0.0003 & 0     & 0     & 0.0003 & 0 \\
				\midrule
				80\%  & MSEFInSamp & 0.0001 & 0.0002 & 0     & 0     & 0.0003 & 0     & 0     & 0.0003 & 0 \\
				80\%  & MSEFTesting & 0.0001 & 0.0004 & 69.4946 & 0     & 0.0003 & 138.3418 & 0     & 0.0003 & 252.9521 \\
				80\%  & MSEFCensus & 0.0001 & 0.0003 & 34.7473 & 0     & 0.0003 & 69.1709 & 0     & 0.0003 & 126.4761 \\
\bottomrule
			\end{tabular}%
			\label{tab:A12}%
		\end{table}%
		\begin{table}[ht]\
			\small
			\centering
			\caption{$MSE_f$, $MSE_{ISf}$, $MSE_{FSf}$, $MSE_y$, $MSE_{ISy}$, $MSE_{FSy}$, time, and $K$ results for $d = 2$, $\sigma = 0.05$.}
			\begin{tabular}{rlrrrrrrrrr}
				\toprule
				\multicolumn{2}{c}{$d$ = 2, $\sigma$ = 0.05} & \multicolumn{3}{c}{nFull = 100} & \multicolumn{3}{c}{nFull = 200} & \multicolumn{3}{c}{nFull = 300} \\
				\multicolumn{1}{l}{nL/nF} & Metric & \multicolumn{1}{c}{CAPNLS} & \multicolumn{1}{c}{CAP} & \multicolumn{1}{c}{CNLS} & \multicolumn{1}{c}{CAPNLS} & \multicolumn{1}{c}{CAP} & \multicolumn{1}{c}{CNLS} & \multicolumn{1}{c}{CAPNLS} & \multicolumn{1}{c}{CAP} & \multicolumn{1}{c}{CNLS} \\
				\midrule
				100\% & MSEFInSamp & 0.0003 & 0.0004 & 0.0004 & 0.0002 & 0.0004 & 0.0002 & 0.0001 & 0.0003 & 0.0002 \\
				100\% & MSEFTesting & 0.0003 & 0.0005 & 344.2265 & 0.0002 & 0.0004 & 104.3058 & 0.0001 & 0.0004 & 208.427 \\
				100\% & MSEFCensus & 0.0003 & 0.0004 & 0.0004 & 0.0002 & 0.0004 & 0.0002 & 0.0001 & 0.0003 & 0.0002 \\
				\midrule
				80\%  & MSEFInSamp & 0.0004 & 0.0005 & 0.0005 & 0.0003 & 0.0005 & 0.0004 & 0.0002 & 0.0004 & 0.0002 \\
				80\%  & MSEFTesting & 0.0006 & 0.0007 & 56.796 & 0.0003 & 0.0005 & 163.3753 & 0.0002 & 0.0004 & 276.1076 \\
				80\%  & MSEFCensus & 0.0005 & 0.0006 & 28.3982 & 0.0003 & 0.0005 & 81.6878 & 0.0002 & 0.0004 & 138.0539 \\
\bottomrule
			\end{tabular}%
			\label{tab:A13}%
		\end{table}%
		\begin{table}[ht]\
			\small
			\centering
			\caption{$MSE_f$, $MSE_{ISf}$, $MSE_{FSf}$, $MSE_y$, $MSE_{ISy}$, $MSE_{FSy}$, time, and $K$ results for $d = 2$, $\sigma = 0.1$.}
			\begin{tabular}{rlrrrrrrrrr}
				\toprule
				\multicolumn{2}{c}{$d$ = 2, $\sigma$ = 0.1} & \multicolumn{3}{c}{nFull = 100} & \multicolumn{3}{c}{nFull = 200} & \multicolumn{3}{c}{nFull = 300} \\
				\multicolumn{1}{l}{nL/nF} & Metric & \multicolumn{1}{c}{CAPNLS} & \multicolumn{1}{c}{CAP} & \multicolumn{1}{c}{CNLS} & \multicolumn{1}{c}{CAPNLS} & \multicolumn{1}{c}{CAP} & \multicolumn{1}{c}{CNLS} & \multicolumn{1}{c}{CAPNLS} & \multicolumn{1}{c}{CAP} & \multicolumn{1}{c}{CNLS} \\
				\midrule
				100\% & MSEFInSamp & 0.0008 & 0.0009 & 0.001 & 0.0005 & 0.0006 & 0.0007 & 0.0003 & 0.0005 & 0.0004 \\
				100\% & MSEFTesting & 0.0009 & 0.001 & 370.924 & 0.0005 & 0.0007 & 218.0245 & 0.0003 & 0.0005 & 141.735 \\
				100\% & MSEFCensus & 0.0008 & 0.0009 & 0.001 & 0.0005 & 0.0006 & 0.0007 & 0.0003 & 0.0005 & 0.0004 \\
				\midrule
				80\%  & MSEFInSamp & 0.0015 & 0.0017 & 0.0018 & 0.0008 & 0.001 & 0.0011 & 0.0006 & 0.0007 & 0.0007 \\
				80\%  & MSEFTesting & 0.0017 & 0.002 & 130.458 & 0.0009 & 0.0011 & 219.5168 & 0.0007 & 0.0008 & 329.2754 \\
				80\%  & MSEFCensus & 0.0016 & 0.0018 & 65.2299 & 0.0009 & 0.0011 & 109.7589 & 0.0006 & 0.0007 & 164.6381 \\
\bottomrule
			\end{tabular}%
			\label{tab:A14}%
		\end{table}%
		\begin{table}[ht]\
			\small
			\centering
			\caption{$MSE_f$, $MSE_{ISf}$, $MSE_{FSf}$, $MSE_y$, $MSE_{ISy}$, $MSE_{FSy}$, time, and $K$ results for $d = 3$, $\sigma = 0.01$.}
			\begin{tabular}{rlrrrrrrrrr}
				\toprule
				\multicolumn{2}{c}{$d$ = 3, $\sigma$ = 0.01} & \multicolumn{3}{c}{nFull = 100} & \multicolumn{3}{c}{nFull = 200} & \multicolumn{3}{c}{nFull = 300} \\
				\multicolumn{1}{l}{nL/nF} & Metric & \multicolumn{1}{c}{CAPNLS} & \multicolumn{1}{c}{CAP} & \multicolumn{1}{c}{CNLS} & \multicolumn{1}{c}{CAPNLS} & \multicolumn{1}{c}{CAP} & \multicolumn{1}{c}{CNLS} & \multicolumn{1}{c}{CAPNLS} & \multicolumn{1}{c}{CAP} & \multicolumn{1}{c}{CNLS} \\
				\midrule
				100\% & MSEFInSamp & 0.0001 & 0.0006 & 0.0001 & 0.0002 & 0.0006 & 0     & 0.0003 & 0.0007 & 0 \\
				100\% & MSEFTesting & 0.0002 & 0.0008 & 413.5722 & 0.0002 & 0.0007 & 423.1706 & 0.0003 & 0.0007 & 714.7379 \\
				100\% & MSEFCensus & 0.0001 & 0.0006 & 0.0001 & 0.0002 & 0.0006 & 0     & 0.0003 & 0.0007 & 0 \\
				\midrule
				80\%  & MSEFInSamp & 0.0003 & 0.0005 & 0.0001 & 0.0001 & 0.0006 & 0.0001 & 0.0001 & 0.0006 & 0 \\
				80\%  & MSEFTesting & 0.0005 & 0.0009 & 32.9737 & 0.0002 & 0.0008 & 187.1054 & 0.0002 & 0.0007 & 429.8912 \\
				80\%  & MSEFCensus & 0.0004 & 0.0007 & 16.4869 & 0.0002 & 0.0007 & 93.5527 & 0.0001 & 0.0007 & 214.9456 \\
\bottomrule
			\end{tabular}%
			\label{tab:A15}%
		\end{table}%
		\begin{table}[ht]\
			\small
			\centering
			\caption{$MSE_f$, $MSE_{ISf}$, $MSE_{FSf}$, $MSE_y$, $MSE_{ISy}$, $MSE_{FSy}$, time, and $K$ results for $d = 3$, $\sigma = 0.05$.}
			\begin{tabular}{rlrrrrrrrrr}
				\toprule
				\multicolumn{2}{c}{$d$ = 3, $\sigma$ = 0.05} & \multicolumn{3}{c}{nFull = 100} & \multicolumn{3}{c}{nFull = 200} & \multicolumn{3}{c}{nFull = 300} \\
				\multicolumn{1}{l}{nL/nF} & Metric & \multicolumn{1}{c}{CAPNLS} & \multicolumn{1}{c}{CAP} & \multicolumn{1}{c}{CNLS} & \multicolumn{1}{c}{CAPNLS} & \multicolumn{1}{c}{CAP} & \multicolumn{1}{c}{CNLS} & \multicolumn{1}{c}{CAPNLS} & \multicolumn{1}{c}{CAP} & \multicolumn{1}{c}{CNLS} \\
				\midrule
				100\% & MSEFInSamp & 0.0005 & 0.0012 & 0.0006 & 0.0003 & 0.0008 & 0.0004 & 0.0003 & 0.0007 & 0.0003 \\
				100\% & MSEFTesting & 0.0006 & 0.0014 & 762.8861 & 0.0004 & 0.0008 & 975.1576 & 0.0004 & 0.0008 & 823.8718 \\
				100\% & MSEFCensus & 0.0005 & 0.0012 & 0.0006 & 0.0003 & 0.0008 & 0.0004 & 0.0003 & 0.0007 & 0.0003 \\
				\midrule
				80\%  & MSEFInSamp & 0.0008 & 0.001 & 0.0009 & 0.0004 & 0.0008 & 0.0006 & 0.0004 & 0.0009 & 0.0005 \\
				80\%  & MSEFTesting & 0.0011 & 0.0013 & 144.0639 & 0.0005 & 0.001 & 601.6849 & 0.0004 & 0.001 & 608.9285 \\
				80\%  & MSEFCensus & 0.0009 & 0.0012 & 72.0324 & 0.0005 & 0.0009 & 300.8427 & 0.0004 & 0.001 & 304.4645 \\
\bottomrule
			\end{tabular}%
			\label{tab:A16}%
		\end{table}%
		\begin{table}[ht]\
			\small
			\centering
			\caption{$MSE_f$, $MSE_{ISf}$, $MSE_{FSf}$, $MSE_y$, $MSE_{ISy}$, $MSE_{FSy}$, time, and $K$ results for $d = 3$, $\sigma = 0.1$.}
			\begin{tabular}{rlrrrrrrrrr}
				\toprule
				\multicolumn{2}{c}{$d$ = 3, $\sigma$ = 0.1} & \multicolumn{3}{c}{nFull = 100} & \multicolumn{3}{c}{nFull = 200} & \multicolumn{3}{c}{nFull = 300} \\
				\multicolumn{1}{l}{nL/nF} & Metric & \multicolumn{1}{c}{CAPNLS} & \multicolumn{1}{c}{CAP} & \multicolumn{1}{c}{CNLS} & \multicolumn{1}{c}{CAPNLS} & \multicolumn{1}{c}{CAP} & \multicolumn{1}{c}{CNLS} & \multicolumn{1}{c}{CAPNLS} & \multicolumn{1}{c}{CAP} & \multicolumn{1}{c}{CNLS} \\
				\midrule
				100\% & MSEFInSamp & 0.0012 & 0.0019 & 0.0018 & 0.0008 & 0.0018 & 0.0011 & 0.0006 & 0.001 & 0.0008 \\
				100\% & MSEFTesting & 0.0014 & 0.0022 & 978.6187 & 0.0009 & 0.0019 & 1562.132 & 0.0006 & 0.001 & 1432.728 \\
				100\% & MSEFCensus & 0.0012 & 0.0019 & 0.0018 & 0.0008 & 0.0018 & 0.0011 & 0.0006 & 0.001 & 0.0008 \\
				\midrule
				80\%  & MSEFInSamp & 0.0019 & 0.0062 & 0.0027 & 0.0012 & 0.0021 & 0.0018 & 0.001 & 0.0013 & 0.0014 \\
				80\%  & MSEFTesting & 0.0026 & 0.0068 & 83.7592 & 0.0014 & 0.0023 & 545.8418 & 0.0011 & 0.0015 & 711.7752 \\
				80\%  & MSEFCensus & 0.0023 & 0.0065 & 41.8809 & 0.0013 & 0.0022 & 272.9218 & 0.001 & 0.0014 & 355.8883 \\
\bottomrule
			\end{tabular}%
			\label{tab:A17}%
		\end{table}%
		\begin{table}[ht]\
			\small
			\centering
			\caption{$MSE_f$, $MSE_{ISf}$, $MSE_{FSf}$, $MSE_y$, $MSE_{ISy}$, $MSE_{FSy}$, time, and $K$ results for $d = 4$, $\sigma = 0.01$.}
			\begin{tabular}{rlrrrrrrrrr}
				\toprule
				\multicolumn{2}{c}{$d$ = 4, $\sigma$ = 0.01} & \multicolumn{3}{c}{nFull = 100} & \multicolumn{3}{c}{nFull = 200} & \multicolumn{3}{c}{nFull = 300} \\
				\multicolumn{1}{l}{nL/nF} & Metric & \multicolumn{1}{c}{CAPNLS} & \multicolumn{1}{c}{CAP} & \multicolumn{1}{c}{CNLS} & \multicolumn{1}{c}{CAPNLS} & \multicolumn{1}{c}{CAP} & \multicolumn{1}{c}{CNLS} & \multicolumn{1}{c}{CAPNLS} & \multicolumn{1}{c}{CAP} & \multicolumn{1}{c}{CNLS} \\
				\midrule
				100\% & MSEFInSamp & 0.0003 & 0.0007 & 0.0001 & 0.0003 & 0.0007 & 0.0001 & 0.0004 & 0.0007 & 0.0001 \\
				100\% & MSEFTesting & 0.0005 & 0.001 & 161.6089 & 0.0003 & 0.0009 & 365.6785 & 0.0005 & 0.0008 & 584.3493 \\
				100\% & MSEFCensus & 0.0003 & 0.0007 & 0.0001 & 0.0003 & 0.0007 & 0.0001 & 0.0004 & 0.0007 & 0.0001 \\
				\midrule
				80\%  & MSEFInSamp & 0.0004 & 0.0005 & 0.0001 & 0.0003 & 0.0007 & 0.0001 & 0.0003 & 0.0007 & 0.0001 \\
				80\%  & MSEFTesting & 0.0008 & 0.0011 & 8.6689 & 0.0005 & 0.0009 & 204.995 & 0.0004 & 0.0009 & 263.8065 \\
				80\%  & MSEFCensus & 0.0006 & 0.0008 & 4.3345 & 0.0004 & 0.0008 & 102.4976 & 0.0003 & 0.0008 & 131.9033 \\
\bottomrule
			\end{tabular}%
			\label{tab:A18}%
		\end{table}%
		\begin{table}[ht]\
			\small
			\centering
			\caption{$MSE_f$, $MSE_{ISf}$, $MSE_{FSf}$, $MSE_y$, $MSE_{ISy}$, $MSE_{FSy}$, time, and $K$ results for $d = 4$, $\sigma = 0.05$.}
			\begin{tabular}{rlrrrrrrrrr}
				\toprule
				\multicolumn{2}{c}{$d$ = 4, $\sigma$ = 0.05} & \multicolumn{3}{c}{nFull = 100} & \multicolumn{3}{c}{nFull = 200} & \multicolumn{3}{c}{nFull = 300} \\
				\multicolumn{1}{l}{nL/nF} & Metric & \multicolumn{1}{c}{CAPNLS} & \multicolumn{1}{c}{CAP} & \multicolumn{1}{c}{CNLS} & \multicolumn{1}{c}{CAPNLS} & \multicolumn{1}{c}{CAP} & \multicolumn{1}{c}{CNLS} & \multicolumn{1}{c}{CAPNLS} & \multicolumn{1}{c}{CAP} & \multicolumn{1}{c}{CNLS} \\
				\midrule
				100\% & MSEFInSamp & 0.0007 & 0.001 & 0.0009 & 0.0005 & 0.0009 & 0.0007 & 0.0005 & 0.0008 & 0.0005 \\
				100\% & MSEFTesting & 0.0009 & 0.0012 & 440.3271 & 0.0006 & 0.001 & 1105.787 & 0.0006 & 0.0009 & 630.5477 \\
				100\% & MSEFCensus & 0.0007 & 0.001 & 0.0009 & 0.0005 & 0.0009 & 0.0007 & 0.0005 & 0.0008 & 0.0005 \\
				\midrule
				80\%  & MSEFInSamp & 0.0009 & 0.0024 & 0.0013 & 0.0007 & 0.001 & 0.0009 & 0.0005 & 0.0009 & 0.0007 \\
				80\%  & MSEFTesting & 0.0014 & 0.0029 & 20.5826 & 0.0009 & 0.0012 & 701.4585 & 0.0007 & 0.0011 & 1027.998 \\
				80\%  & MSEFCensus & 0.0011 & 0.0026 & 10.2919 & 0.0008 & 0.0011 & 350.7297 & 0.0006 & 0.001 & 513.9992 \\
\bottomrule
			\end{tabular}%
			\label{tab:A19}%
		\end{table}%
		\begin{table}[ht]\
			\small
			\centering
			\caption{$MSE_f$, $MSE_{ISf}$, $MSE_{FSf}$, $MSE_y$, $MSE_{ISy}$, $MSE_{FSy}$, time, and $K$ results for $d = 4$, $\sigma = 0.1$.}
			\begin{tabular}{rlrrrrrrrrr}
				\toprule
				\multicolumn{2}{c}{$d$ = 4, $\sigma$ = 0.1} & \multicolumn{3}{c}{nFull = 100} & \multicolumn{3}{c}{nFull = 200} & \multicolumn{3}{c}{nFull = 300} \\
				\multicolumn{1}{l}{nL/nF} & Metric & \multicolumn{1}{c}{CAPNLS} & \multicolumn{1}{c}{CAP} & \multicolumn{1}{c}{CNLS} & \multicolumn{1}{c}{CAPNLS} & \multicolumn{1}{c}{CAP} & \multicolumn{1}{c}{CNLS} & \multicolumn{1}{c}{CAPNLS} & \multicolumn{1}{c}{CAP} & \multicolumn{1}{c}{CNLS} \\
				\midrule
				100\% & MSEFInSamp & 0.0015 & 0.0018 & 0.0025 & 0.001 & 0.0013 & 0.0017 & 0.0008 & 0.0012 & 0.0014 \\
				100\% & MSEFTesting & 0.0017 & 0.0022 & 322.4462 & 0.001 & 0.0015 & 1624.469 & 0.0009 & 0.0013 & 2548.58 \\
				100\% & MSEFCensus & 0.0015 & 0.0018 & 0.0025 & 0.001 & 0.0013 & 0.0017 & 0.0008 & 0.0012 & 0.0014 \\
				\midrule
				80\%  & MSEFInSamp & 0.0025 & 0.003 & 0.0043 & 0.0015 & 0.0017 & 0.0025 & 0.0012 & 0.0015 & 0.0019 \\
				80\%  & MSEFTesting & 0.0031 & 0.0043 & 20.1828 & 0.0017 & 0.002 & 398.8514 & 0.0013 & 0.0017 & 1202.019 \\
				80\%  & MSEFCensus & 0.0028 & 0.0036 & 10.0935 & 0.0016 & 0.0019 & 199.427 & 0.0012 & 0.0016 & 601.0107 \\
\bottomrule
			\end{tabular}%
			\label{tab:A20}%
		\end{table}%
		
	\end{landscape}
	
	\section{Scalability of CAPNLS to larger datasets}
	\label{App:AppendixB}
	
	To demonstrate the performance of CAPNLS in large data sets, we revisit the DGP used in the Tri-variate case, specifically, $Y_i = X_{i1}^{0.4} X_{i2}^{0.3} X_{i3}^{0.2} + \epsilon_i$, where $\epsilon_i \sim N(0, \sigma^2)$, $\sigma = 0.1$ and $X_{ij} \sim Unif(0.1, 1)$ for $j = 1, 2, 3, \; i = 1, \cdots, n$, and $n = 500, 1000, 2000, 3000$, and $5000$. Table \ref{tab:B1} reports the estimator's performance\footnote{Due to the increased computational burden of using larger datasets, we present results for a single replicate of the DGP for each sample size and only include learning set results. For this section we report RMSE results rather than MSE results because the latter are small and the differences are indistinguishable across settings.}. 
	\renewcommand{\arraystretch}{1}
	\begin{table}[ht]\
		\small
		\centering
		\caption{Number of Hyperplanes and Runtimes for Trivariate Input Cobb-Douglas DGP on Larger Datasets.}
		\begin{tabular}{rccccc}
			\toprule
			\multicolumn{1}{c}{\textit{n}} & 500   & 1000  & 2000  & 3000  & 5000 \\
			\midrule
			$RMSEf_{Learn}^{CAPNLS}$  & 0.025 & 0.022 & 0.025 & 0.025 & 0.024 \\
			$RMSEf_{Learn}^{CAPNLSF}$  & 0.023 & 0.025 & 0.026 & 0.027 & 0.028 \\
			$MSEy_{Learn}^{CAPNLS}$  & 0.009 & 0.01  & 0.011 & 0.011 & 0.011 \\
			$MSEy_{Learn}^{CAPNLSF}$  & 0.01  & 0.011 & 0.01  & 0.011 & 0.011 \\
			$K^{CAPNLS}$  & 5     & 5     & 7     & 5     & 5 \\
			$K^{CAPNLSF}$  & 4     & 4     & 5     & 4     & 4 \\
			$Time(min)^{CAPNLS}$  & 3     & 8     & 43    & 114   & 367 \\
			$Time(min)^{CAPNLSF}$  & 3     & 5     & 10    & 11    & 41 \\
			\bottomrule
		\end{tabular}%
		\label{tab:B1}%
	\end{table}%
	We first conduct standard CAPNLS analysis and report learning errors, number of fitted hyperplanes and runtime results.  Runtimes for datasets up to $2,000$ observations are well below the one hour threshold, but there are significant scalability challenges for datasets larger than $2,000$ observations. Thus, we apply the Fast CAP stopping criterion in \citet{hannah2013multivariate}, which measures the GCV score improvement by the addition of one more hyperplane and stops the algorithm if no improvement has been achieved in two consecutive additions. Unlike Fast CAP, however, we apply it directly to the learning error against observations. We denote the results for those runs with the CAPNLSF superscript and observe that differences are minimal compared to following our standard partitioning strategy. This alternative stopping rule results in a highly scalable algorithm which can fit datasets up to $5,000$ observations in $\sim 40$ minutes. 
	
	\section{Parametric Bootstrap algorithm to calculate expected optimism}
	\label{App:AppendixC}
	
	We apply the following algorithm from \citet{efron2004estimation} to compute in-sample optimism. First, we assume a Gaussian density $p(\bm{Y}) = N (\hat{\bm{Y}}, \hat{\sigma}^2 \bm{I})$, where $\bm{Y}$ is the vector of estimated output values of the estimator for which we are assessing the in-sample optimism. We use $\bm{I}$ to indicate an identity matrix. We obtain $\sigma^2$ from the residuals of a ``big'' model presumed to have negligible bias. Given CNLS's high flexibility and complex description (many hyperplanes), we choose it as our ``big'' model. Although obtaining an unbiased estimate for $\sigma^2$ from CNLS's residuals is complicated, i.e., there are no formal results regarding the effective number of parameters CNLS uses, using $MSE_{yLearn}^{CNLS}$ as $\sigma^2$ results in a downward biased estimator of $\sigma^2$. This downward bias in fact results in improved efficiency for the parametric bootstrap algorithm and is an example of a ``little'' bootstrap, see (\citet{breiman1992little}. Thus, we let $\hat{\sigma} = MSE_{yLearn}^{CNLS}$. \citet{efron2004estimation} then suggests to run a large number $B$ of simulated observations $\bm{Y}^*$ from $p(\bm{Y})$, fit them to obtain estimates $\hat{\bm{Y}}^*$, and estimate $cov_i = cov(\hat{Y}_i, Y_i)$ computing
	\begin{equation}
	\label{eqC1}
	\hat{cov}_i = \sum_{b=1}^B \hat{Y}_i^{*b} ({Y}_i^{*b} - {Y}_i^{*}) / (B - 1); Y_i^* = \sum_{b=1}^B \hat{Y}_i^{*b} / B.
	\end{equation}
	We select $B = 500$ for all our experiments based on observed convergence of the $\sum_{i =1}^N \hat{cov}_i$ quantity.
	Further, we note that if the researcher is not comfortable with the assumption made about the size of $MSE_{yLearn}^{CNLS}$ relative to $\sigma^2$, sensitivity analysis (by adding a multiplier $c > 1$, such that $p(\bm{Y}) = N (\hat{\bm{Y}}, c \hat{\sigma}^2 \bm{I})$) can be performed. Finally, we also note that non-Gaussian distributions can be used to draw the bootstrapped $\bm{Y}^*$ vectors. This is especially useful when considering inefficiency, because it can imply a skewed distribution.
	
	\section{Comparison between different error structures for the parametric Cobb-Douglas function}
	\label{sec:APPG.CobbDouglas}
	This Appendix compares the performance of the standard log-linear Cobb-Douglas with an additive error estimator and the multiplicative Cobb-Douglas with an additive error estimator. Figures \ref{fig:F1.CDACDM_Low2D}-\ref{fig:F6.CDACDM_High4D} consider the bivariate, trivariate, and tetravariate case with both a low and high noise setting. We find both parametric estimators preform very similarly with the correctly specified multiplicative Cobb-Douglas with an additive error estimator performing slightly better.
	
	\begin{landscape}
		\begin{figure}[!ht]
			\begin{center}			\includegraphics[width=7in]{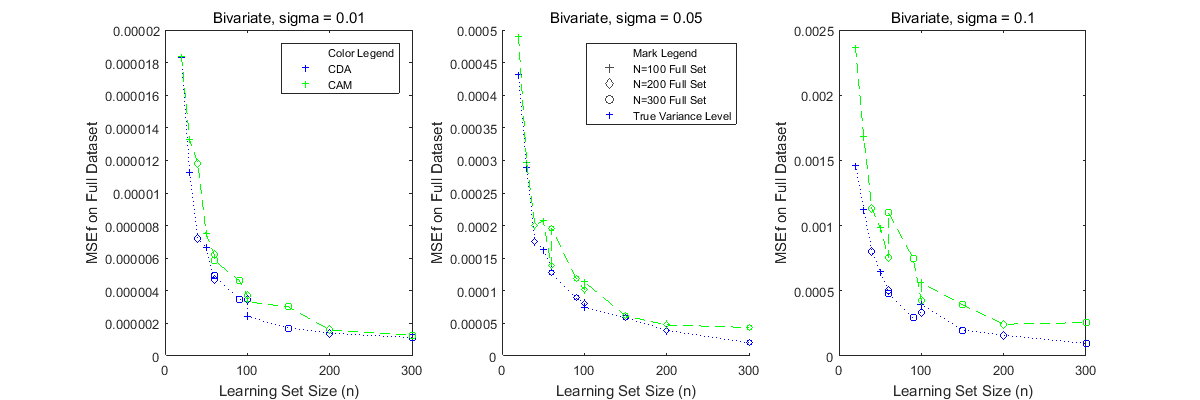}
			\end{center}
			\caption{A comparison of Cobb-Douglas multiplicative function with an additive error term (CDA) vs. Cobb-Douglas log-linear function with an additive error term (CDM) when the noise level is low and the model includes two inputs.  
				\label{fig:F1.CDACDM_Low2D}}
		\end{figure}
		\begin{figure}[!ht]
			\begin{center}
				\includegraphics[width=7in]{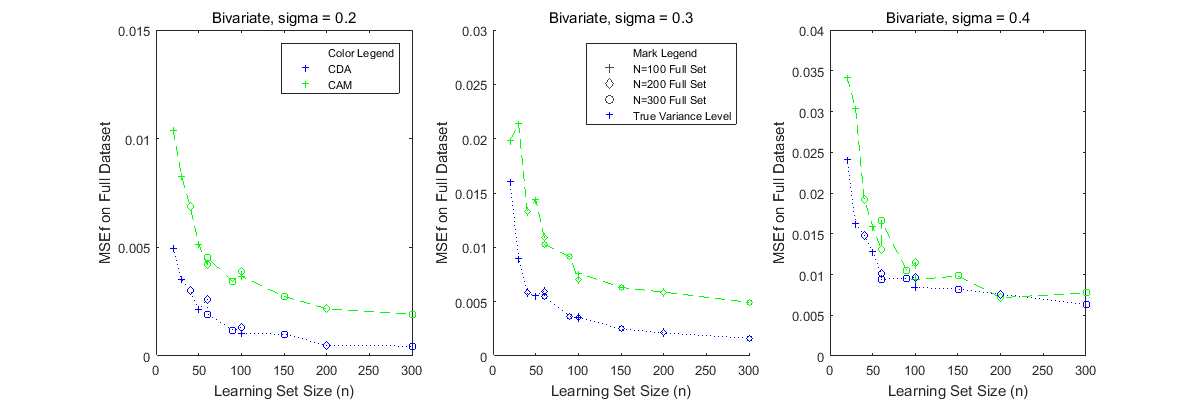}
			\end{center}
			\caption{A comparison of Cobb-Douglas multiplicative function with an additive error term (CDA) vs. Cobb-Douglas log-linear function with an additive error term (CDM) when the noise level is high and the model includes two inputs.  
				\label{fig:F2.CDACDM_high2D}}
		\end{figure}
		\begin{figure}[!ht]
			\begin{center}
				\includegraphics[width=7in]{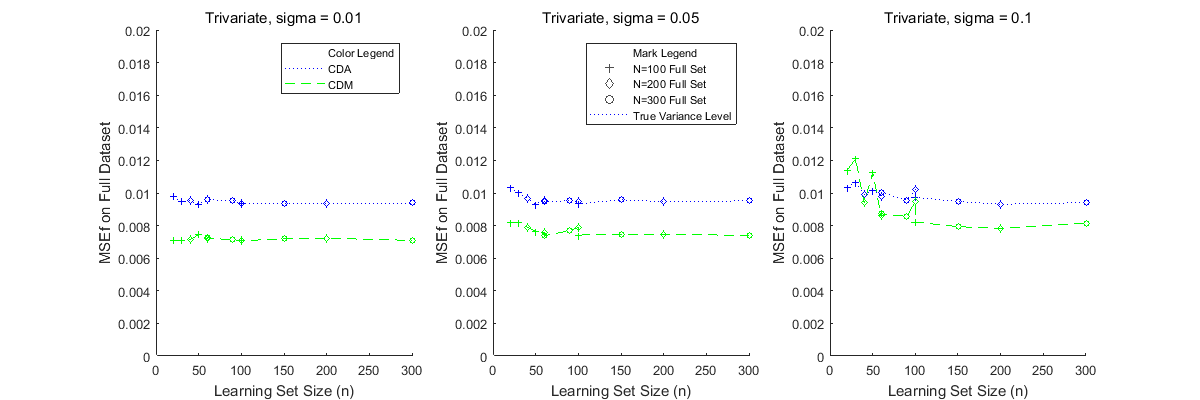}
			\end{center}
			\caption{A comparison of Cobb-Douglas multiplicative function with an additive error term (CDA) vs. Cobb-Douglas log-linear function with an additive error term (CDM) when the noise level is low and the model includes three inputs.  
				\label{fig:F3.CDACDM_Low3D}}
		\end{figure}
		\begin{figure}[!ht]
			\begin{center}
				\includegraphics[width=7in]{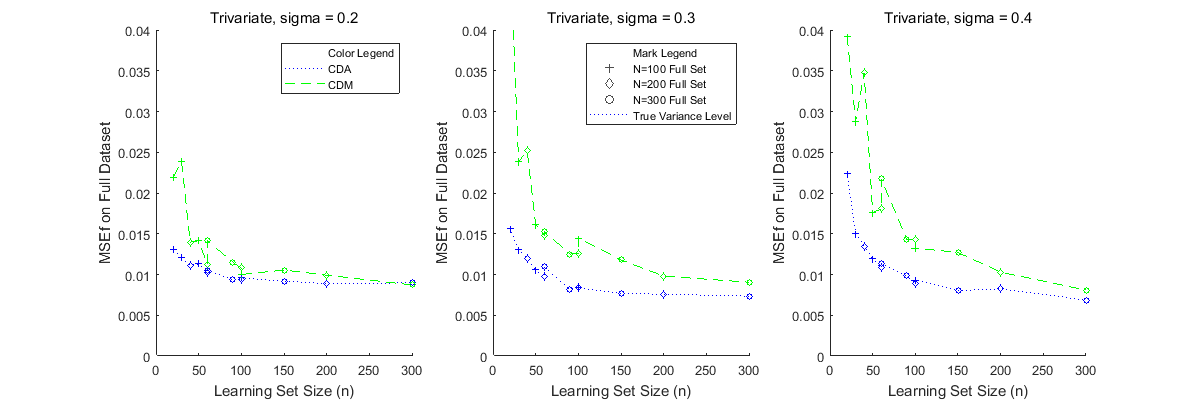}
			\end{center}
			\caption{A comparison of Cobb-Douglas multiplicative function with an additive error term (CDA) vs. Cobb-Douglas log-linear function with an additive error term (CDM) when the noise level is high and the model includes three inputs.  
				\label{fig:F4.CDACDM_High3D}}
		\end{figure}
		\begin{figure}[!ht]
			\begin{center}
				\includegraphics[width=7in]{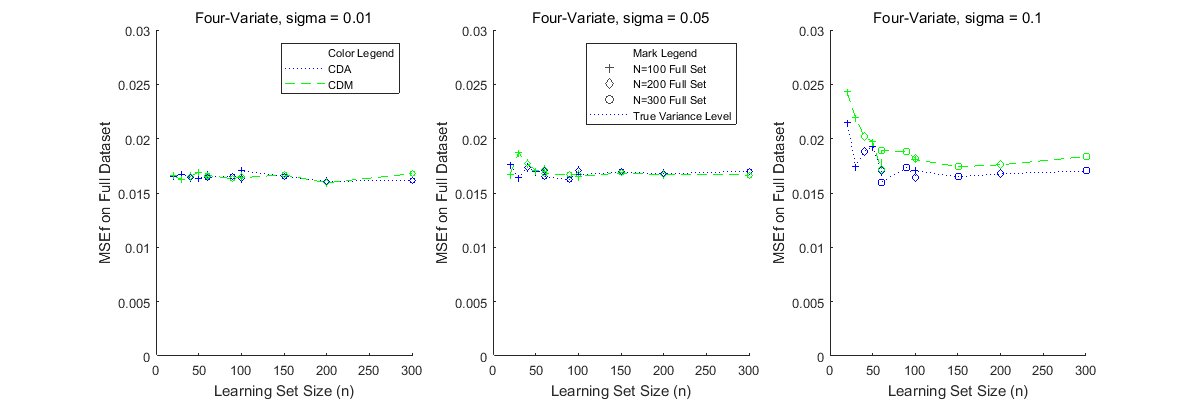}
			\end{center}
			\caption{A comparison of Cobb-Douglas multiplicative function with an additive error term (CDA) vs. Cobb-Douglas log-linear function with an additive error term (CDM) when the noise level is low and the model includes four inputs.  
				\label{fig:F5.CDACDM_Low4D}}
		\end{figure}
		\begin{figure}[!ht]
			\begin{center}
				\includegraphics[width=7in]{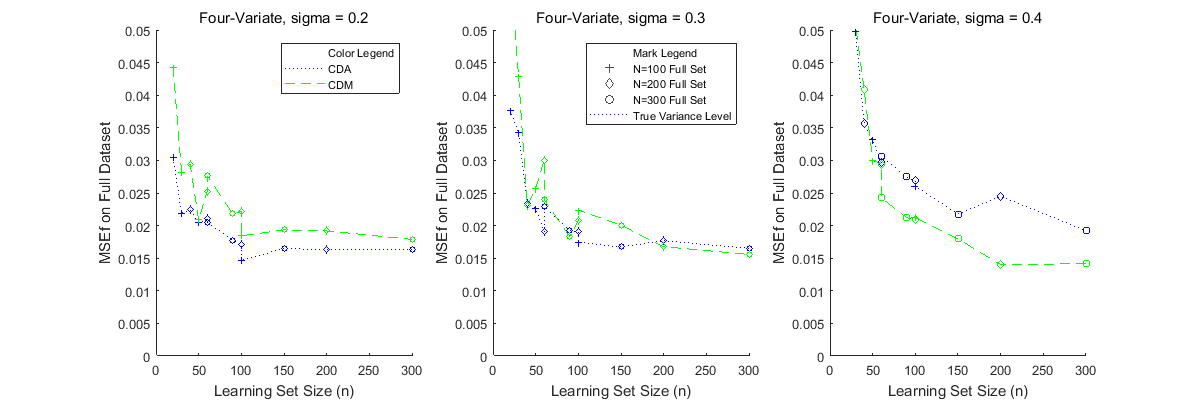}
			\end{center}
			\caption{A comparison of Cobb-Douglas multiplicative function with an additive error term (CDA) vs. Cobb-Douglas log-linear function with an additive error term (CDM) when the noise level is high and the model includes four inputs.  
				\label{fig:F6.CDACDM_High4D}}
		\end{figure}
	\end{landscape}
	
	\clearpage
	
	\section{Cobb-Douglas results with multiplicative residual assumption for Chilean manufacturing data} 
	\label{App:AppendixD}
	
	In this appendix we reconstruct Table \ref{tab3:CDAPerformance} from the main text, but consider the alternative parametric estimator, a log-linear Cobb-Douglas function with an additive error term (CDM).
	
	\renewcommand{\arraystretch}{0.65}
	\begin{table}[ht]\
		\small
		\centering
		\caption{Ratio of CDM to Best Model performance.}
		\begin{tabular}{lccccl}
			\toprule
			\multicolumn{1}{c}{Industry Name (Code)} &    $n$   & Survey Size &   $R_{FS}^2$    & $R_{CDM}^2$  & \multicolumn{1}{c}{Ratio vs. Best Method} \\
			\midrule
			\multicolumn{1}{r}{\multirow{5}[0]{*}{Other Metal Products (2899)}} & \multirow{5}[0]{*}{144} & 20\%  & 50\%  & 82\%  & CDM is Best Method \\
			&       & 30\%  & 60\%  & 85\%  & CDM is Best Method \\
			&       & 40\%  & 64\%  & 86\%  & CDM is Best Method \\
			&       & 50\%  & 72\%  & 86\%  & CDM is Best Method \\
			&       & 100\% & 88\%  & 87\%  & CDM ties for Best Method \\
			\midrule
			\multicolumn{1}{r}{\multirow{5}[0]{*}{Wood (2010)}} & \multirow{5}[0]{*}{150} & 20\%  & 35\%  & 45\%  & CDM is Best Method \\
			&       & 30\%  & 40\%  & 50\%  & CDM is Best Method \\
			&       & 40\%  & 47\%  & 51\%  & CDM is Best Method \\
			&       & 50\%  & 52\%  & 53\%  & CDA ties for Best Method \\
			&       & 100\% & 66\%  & 62\%  & 0.94 vs. CAPNLS \\
			\midrule
			\multicolumn{1}{r}{\multirow{5}[0]{*}{Structural Use Metal (2811)}} & \multirow{5}[0]{*}{161} & 20\%  & 77\%  & 79\%  & CDM ties for Best Method \\
			&       & 30\%  & 82\%  & 81\%  & CDM ties for Best Method \\
			&       & 40\%  & 87\%  & 84\%  & 0.97 vs. CAPNLS \\
			&       & 50\%  & 90\%  & 85\%  & 0.94 vs. CAPNLS \\
			&       & 100\% & 95\%  & 92\%  & 0.97 vs. CAPNLS \\
			\midrule
			\multicolumn{1}{r}{\multirow{5}[0]{*}{Plastics (2520)}} & \multirow{5}[0]{*}{249} & 20\%  & 54\%  & 56\%  & CDM ties for Best Method \\
			&       & 30\%  & 57\%  & 56\%  & CDM ties for Best Method \\
			&       & 40\%  & 57\%  & 57\%  & CDM ties for Best Method \\
			&       & 50\%  & 60\%  & 57\%  & CDM ties for Best Method \\
			&       & 100\% & 64\%  & 60\%  & 0.94 vs. CAPNLS \\
			\midrule
			\multicolumn{1}{r}{\multirow{5}[0]{*}{Bakeries (1541)}} & \multirow{5}[0]{*}{250} & 20\%  & 72\%  & 46\%  & 0.64 vs. CAP \\
			&       & 30\%  & 77\%  & 50\%  & 0.65 vs. CAP \\
			&       & 40\%  & 78\%  & 50\%  & 0.64 vs. CAP \\
			&       & 50\%  & 85\%  & 51\%  & 0.60 vs. CAP \\
			&       & 100\% & 99\%  & 58\%  & 0.59 vs. CAP \\
			\bottomrule
		\end{tabular}%
		\label{tab:D1}%
	\end{table}%
	
	\section{Application Results for Infinite Populations}
	\label{App:AppendixE}
	
\begin{figure}[!ht]
		\begin{center}
			\includegraphics[width=3.8in]{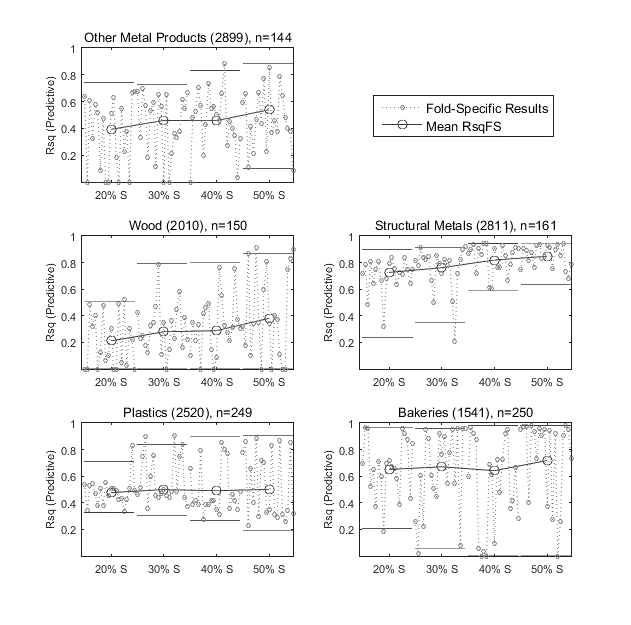}
		\end{center}
		\caption{Best Method's $R_{Pred}^2$ as function of relative subset size for selected industries. CAPNLS was chosen as Best Method for industry codes 2899, 2010, 2811 and 2520, while CDA was chosen for industry code 1541. 
			\label{fig:E1.CAPNLS}}
	\end{figure}

	In this Appendix we explore the sensitivity of our application insights in the case when predictive ability at any point of the production function is equally important. This represents a key departure from the assumptions made in the main body of the paper, as it translates into not weighting in-sample and predictive errors. Rather, since we are interested in the descriptive ability of the fitted production function on an infinite number of unobserved input vectors, we only consider the predictive error. To illustrate the consequences of this alternative assumptions in detail, we present our results in Figure \ref{fig:E1.CAPNLS} which is analogous to Figure \ref{fig2:IndPerform}. In Figure \ref{fig:E1.CAPNLS} we show replicate-specific as well as averaged $R^2$ values. In this case, rather than using $R_{FS}^2$ as our predictive power indicator, we use $R_{Pred}^2 = \max(1-(E(\hat{Err}_y) / Var (Y_{FS}), 0)$. 
	
	In Figure \ref{fig:E1.CAPNLS}, we observe that for the majority of studied industries, weighting the in-sample error with the predictive error, the direct consequence of our finite full population assumption, does not affect the diagnostic of the mean predictive power of our production function models. Using our notation, this means that for most industries the expected $R_{Pred}^2$ for each given subsample size did not differ greatly from $R_{FS}^2$. However, the $R_{Pred}^2$ figures have significantly higher variance than their $R_{FS}^2$ counterparts for each subsample size. For all industries except for industry code 2811, we obtain at least one replicate with negligible predictive power. For some industries, such as Industry Codes 2520 and 1541, this causes the predictive power bound to be very wide (although the upper bound and mean values increase monotonically in the subsample size). 
	
	\section{Comprehensive Application Results: MPSS, MP, and MRTS}
	\label{sec:APPF.AppComp}
	Tables \ref{tabA:MPSS}-\ref{tabA:MRTS} provide more detail results regarding MPSS, MP for both capital and labor, and MRTS for different percentiles of the four inputs (capital, labor, energy, and services). This information was summarized in Tables \ref{tab:MPSS}-\ref{tab:MRTS_KL} in the main text.

	\begin{table}[htbp]
		\tiny
		\centering
		\caption{Most Product Scaler Size ($y$)}
		\begin{tabular}{rrrr|rrrrr|rrrrr|rrrrr}
			\toprule
			\multicolumn{4}{c|}{Percentile} & \multicolumn{5}{c|}{CAP}               & \multicolumn{5}{c|}{CAPNLS}            & \multicolumn{5}{c}{CNLS} \\
			\multicolumn{1}{l}{$K$} & \multicolumn{1}{l}{$L$} & \multicolumn{1}{l}{$E$} & \multicolumn{1}{l|}{$S$} & \multicolumn{1}{l}{2899} & \multicolumn{1}{l}{2010} & \multicolumn{1}{l}{2811} & \multicolumn{1}{l}{2520} & \multicolumn{1}{l|}{1541} & \multicolumn{1}{l}{2899} & \multicolumn{1}{l}{2010} & \multicolumn{1}{l}{2811} & \multicolumn{1}{l}{2520} & \multicolumn{1}{l|}{1541} & \multicolumn{1}{l}{2899} & \multicolumn{1}{l}{2010} & \multicolumn{1}{l}{2811} & \multicolumn{1}{l}{2520} & \multicolumn{1}{l}{1541} \\
			\midrule
			25    & 25    & 25    & 25    & 5.0   & 5.6   & 15.2  & 6.7   & 6.0   & 3.1   & 3.7   & 6.7   & 10.2  & 2.6   & 3.8   & 6.6   & 15.6  & 12.0  & 9.1 \\
			50    & 25    & 25    & 25    & 6.1   & 7.8   & 3.4   & 8.3   & 6.6   & 3.6   & 4.0   & 5.7   & 12.4  & 2.7   & 5.9   & 15.1  & 12.0  & 13.2  & 8.6 \\
			75    & 25    & 25    & 25    & 3.4   & 9.5   & 1.7   & 28.8  & 7.8   & 5.2   & 6.0   & 4.9   & 20.0  & 2.9   & 5.7   & 14.6  & 9.1   & 13.5  & 8.7 \\
			25    & 50    & 25    & 25    & 4.2   & 3.3   & 10.7  & 4.8   & 4.7   & 2.6   & 2.5   & 4.8   & 7.5   & 2.2   & 1.8   & 4.2   & 19.1  & 4.9   & 8.2 \\
			50    & 50    & 25    & 25    & 5.0   & 4.7   & 4.7   & 5.2   & 5.1   & 2.9   & 2.5   & 4.5   & 8.7   & 2.3   & 2.2   & 6.8   & 13.3  & 7.3   & 8.8 \\
			75    & 50    & 25    & 25    & 4.3   & 12.1  & 1.9   & 13.5  & 6.1   & 4.0   & 4.7   & 4.5   & 9.6   & 2.4   & 3.3   & 9.1   & 9.4   & 11.2  & 11.7 \\
			25    & 75    & 25    & 25    & 3.6   & 1.4   & 8.5   & 3.8   & 3.6   & 2.4   & 1.7   & 4.4   & 5.8   & 1.8   & 1.5   & 3.5   & 16.4  & 4.1   & 7.6 \\
			50    & 75    & 25    & 25    & 4.0   & 1.8   & 5.7   & 4.4   & 3.8   & 2.4   & 1.8   & 4.2   & 6.3   & 2.0   & 1.2   & 6.1   & 16.6  & 4.7   & 8.6 \\
			75    & 75    & 25    & 25    & 5.5   & 4.2   & 3.1   & 5.7   & 4.4   & 2.7   & 2.1   & 3.9   & 7.1   & 2.1   & 1.6   & 5.5   & 11.3  & 7.0   & 13.3 \\
			25    & 25    & 50    & 25    & 4.6   & 8.8   & 7.1   & 6.1   & 7.4   & 3.5   & 4.9   & 13.4  & 8.8   & 3.3   & 4.5   & 8.2   & 14.5  & 20.2  & 9.1 \\
			50    & 25    & 50    & 25    & 5.7   & 15.3  & 6.3   & 12.1  & 8.2   & 4.0   & 5.0   & 10.4  & 11.1  & 3.4   & 6.2   & 14.8  & 9.2   & 28.0  & 8.7 \\
			75    & 25    & 50    & 25    & 3.4   & 10.1  & 2.7   & 23.7  & 9.6   & 5.6   & 15.7  & 7.7   & 21.3  & 3.9   & 6.3   & 16.0  & 7.0   & 39.1  & 8.6 \\
			25    & 50    & 50    & 25    & 4.0   & 5.3   & 10.7  & 4.4   & 5.8   & 2.8   & 2.8   & 8.1   & 6.7   & 2.6   & 2.3   & 4.6   & 12.8  & 4.2   & 6.7 \\
			50    & 50    & 50    & 25    & 4.7   & 6.8   & 8.8   & 5.9   & 6.4   & 3.0   & 3.0   & 7.1   & 7.9   & 2.7   & 3.5   & 9.4   & 9.3   & 15.3  & 9.4 \\
			75    & 50    & 50    & 25    & 4.2   & 13.0  & 3.6   & 11.8  & 7.4   & 4.1   & 5.2   & 6.1   & 8.4   & 2.8   & 4.3   & 10.6  & 5.6   & 17.3  & 8.8 \\
			25    & 75    & 50    & 25    & 3.7   & 1.9   & 16.2  & 3.7   & 4.3   & 2.4   & 1.9   & 5.2   & 5.5   & 2.0   & 1.6   & 2.5   & 20.8  & 3.3   & 6.3 \\
			50    & 75    & 50    & 25    & 3.9   & 2.3   & 13.4  & 4.2   & 4.6   & 2.5   & 2.0   & 5.1   & 6.4   & 2.1   & 1.6   & 3.7   & 8.8   & 6.7   & 8.7 \\
			75    & 75    & 50    & 25    & 5.4   & 5.2   & 5.8   & 5.5   & 5.2   & 3.0   & 2.4   & 4.7   & 7.0   & 2.2   & 2.1   & 5.4   & 6.1   & 8.3   & 8.8 \\
			25    & 25    & 75    & 25    & 3.6   & 4.3   & 2.4   & 5.5   & 9.0   & 4.4   & 5.7   & 12.0  & 9.4   & 4.9   & 3.6   & 8.6   & 14.5  & 21.0  & 9.3 \\
			50    & 25    & 75    & 25    & 4.1   & 5.3   & 1.6   & 14.5  & 10.0  & 4.7   & 6.7   & 13.9  & 13.1  & 5.9   & 8.1   & 16.0  & 8.7   & 29.8  & 8.7 \\
			75    & 25    & 75    & 25    & 3.5   & 7.1   & 2.0   & 14.9  & 13.0  & 9.0   & 9.0   & 18.4  & 17.6  & 9.1   & 11.3  & 15.7  & 6.2   & 18.2  & 8.5 \\
			25    & 50    & 75    & 25    & 3.5   & 4.7   & 2.2   & 3.8   & 7.8   & 3.4   & 5.2   & 14.7  & 6.9   & 3.9   & 2.3   & 5.2   & 10.0  & 4.2   & 6.6 \\
			50    & 50    & 75    & 25    & 3.9   & 6.3   & 2.4   & 5.3   & 8.4   & 3.8   & 5.5   & 20.0  & 8.0   & 4.0   & 4.6   & 7.8   & 8.7   & 15.3  & 8.5 \\
			75    & 50    & 75    & 25    & 4.1   & 9.8   & 2.7   & 16.4  & 10.0  & 5.4   & 7.4   & 15.8  & 8.7   & 4.8   & 6.6   & 8.5   & 4.1   & 18.3  & 8.9 \\
			25    & 75    & 75    & 25    & 3.4   & 4.9   & 4.2   & 3.3   & 5.8   & 2.7   & 3.7   & 9.2   & 5.3   & 2.7   & 2.0   & 3.4   & 20.8  & 2.5   & 5.1 \\
			50    & 75    & 75    & 25    & 3.6   & 6.0   & 4.2   & 3.7   & 6.2   & 2.8   & 4.1   & 9.1   & 5.5   & 2.7   & 2.1   & 4.6   & 7.3   & 5.3   & 7.3 \\
			75    & 75    & 75    & 25    & 4.9   & 12.1  & 4.2   & 5.2   & 7.0   & 3.3   & 5.4   & 8.4   & 6.0   & 2.9   & 4.1   & 4.6   & 3.1   & 10.5  & 8.4 \\
			25    & 25    & 25    & 50    & 5.3   & 8.2   & 4.6   & 4.9   & 6.7   & 4.1   & 4.4   & 4.0   & 8.8   & 3.2   & 4.5   & 6.2   & 8.1   & 7.9   & 10.5 \\
			50    & 25    & 25    & 50    & 6.1   & 10.6  & 3.2   & 9.1   & 7.3   & 4.9   & 6.3   & 4.4   & 17.7  & 3.3   & 6.7   & 19.0  & 6.2   & 11.0  & 9.3 \\
			75    & 25    & 25    & 50    & 4.4   & 12.0  & 1.9   & 37.1  & 8.9   & 6.8   & 9.9   & 4.1   & 20.1  & 3.7   & 7.6   & 34.8  & 5.4   & 19.8  & 9.5 \\
			25    & 50    & 25    & 50    & 4.6   & 4.5   & 5.5   & 5.2   & 5.2   & 3.2   & 3.3   & 5.5   & 7.5   & 2.6   & 2.3   & 3.2   & 8.9   & 6.1   & 8.6 \\
			50    & 50    & 25    & 50    & 5.7   & 5.3   & 3.4   & 4.8   & 5.5   & 3.4   & 3.3   & 5.7   & 11.1  & 2.6   & 3.0   & 9.8   & 9.8   & 10.1  & 8.8 \\
			75    & 50    & 25    & 50    & 5.1   & 10.1  & 2.3   & 44.3  & 6.5   & 4.5   & 4.5   & 5.2   & 15.3  & 2.8   & 3.9   & 23.5  & 9.5   & 12.2  & 12.3 \\
			25    & 75    & 25    & 50    & 3.9   & 2.4   & 6.2   & 4.8   & 3.9   & 2.5   & 2.2   & 5.1   & 6.5   & 2.0   & 1.9   & 2.8   & 11.3  & 3.8   & 7.5 \\
			50    & 75    & 25    & 50    & 4.4   & 2.8   & 4.6   & 5.1   & 4.1   & 2.6   & 2.2   & 4.7   & 7.3   & 2.1   & 1.6   & 8.2   & 10.7  & 5.2   & 8.8 \\
			75    & 75    & 25    & 50    & 6.1   & 5.6   & 3.0   & 7.1   & 4.6   & 3.1   & 2.4   & 4.4   & 8.7   & 2.2   & 2.0   & 17.6  & 12.2  & 7.7   & 11.1 \\
			25    & 25    & 50    & 50    & 5.7   & 7.6   & 6.4   & 17.5  & 8.9   & 4.4   & 5.1   & 5.8   & 9.1   & 3.9   & 5.6   & 8.8   & 14.3  & 18.3  & 10.4 \\
			50    & 25    & 50    & 50    & 6.9   & 11.4  & 5.7   & 10.8  & 9.5   & 5.1   & 7.6   & 6.2   & 17.8  & 4.1   & 7.9   & 20.2  & 9.3   & 29.6  & 9.6 \\
			75    & 25    & 50    & 50    & 4.3   & 12.5  & 2.5   & 48.2  & 10.8  & 7.2   & 15.2  & 6.4   & 36.9  & 4.7   & 9.3   & 30.0  & 7.4   & 39.3  & 8.5 \\
			25    & 50    & 50    & 50    & 4.7   & 5.1   & 10.1  & 7.3   & 6.6   & 3.5   & 3.8   & 7.9   & 8.3   & 3.0   & 3.3   & 3.1   & 12.9  & 9.6   & 8.1 \\
			50    & 50    & 50    & 50    & 5.9   & 5.9   & 6.5   & 7.8   & 7.0   & 3.8   & 4.5   & 8.0   & 12.5  & 3.1   & 5.3   & 9.2   & 14.1  & 18.9  & 10.9 \\
			75    & 50    & 50    & 50    & 5.0   & 12.8  & 3.3   & 38.4  & 8.0   & 4.7   & 5.8   & 6.2   & 14.7  & 3.2   & 6.7   & 20.3  & 10.5  & 17.7  & 8.2 \\
			25    & 75    & 50    & 50    & 3.9   & 3.0   & 13.6  & 4.6   & 4.7   & 2.6   & 2.4   & 5.5   & 7.1   & 2.3   & 2.2   & 2.6   & 14.7  & 4.5   & 7.3 \\
			50    & 75    & 50    & 50    & 4.4   & 3.5   & 12.0  & 5.3   & 4.9   & 2.7   & 2.4   & 5.5   & 8.1   & 2.3   & 2.1   & 5.6   & 15.5  & 5.8   & 9.4 \\
			75    & 75    & 50    & 50    & 5.9   & 7.0   & 4.9   & 7.4   & 5.4   & 3.2   & 2.6   & 5.1   & 9.1   & 2.4   & 3.2   & 10.8  & 13.6  & 11.9  & 10.1 \\
			25    & 25    & 75    & 50    & 4.5   & 5.6   & 5.8   & 7.3   & 10.3  & 5.0   & 7.3   & 8.7   & 9.7   & 5.4   & 4.5   & 10.5  & 16.3  & 15.4  & 10.4 \\
			50    & 25    & 75    & 50    & 5.2   & 6.6   & 5.6   & 17.1  & 11.4  & 5.8   & 8.2   & 9.7   & 17.2  & 6.5   & 7.9   & 25.7  & 11.9  & 31.3  & 9.7 \\
			75    & 25    & 75    & 50    & 4.1   & 8.0   & 5.3   & 24.7  & 14.6  & 9.3   & 9.3   & 9.4   & 14.5  & 11.3  & 15.2  & 32.5  & 10.9  & 28.7  & 8.6 \\
			25    & 50    & 75    & 50    & 4.2   & 6.2   & 7.7   & 4.9   & 8.6   & 4.1   & 7.2   & 12.5  & 7.9   & 4.4   & 3.1   & 6.3   & 12.8  & 9.6   & 7.6 \\
			50    & 50    & 75    & 50    & 4.6   & 7.7   & 7.2   & 6.7   & 9.2   & 4.3   & 7.7   & 12.6  & 11.1  & 4.7   & 5.4   & 15.0  & 15.0  & 24.9  & 10.2 \\
			75    & 50    & 75    & 50    & 4.6   & 10.7  & 6.5   & 18.6  & 11.1  & 6.0   & 9.7   & 11.0  & 11.8  & 5.4   & 9.8   & 21.8  & 9.6   & 35.5  & 7.8 \\
			25    & 75    & 75    & 50    & 3.8   & 6.0   & 11.9  & 4.0   & 6.3   & 3.0   & 4.1   & 10.6  & 5.9   & 2.9   & 3.0   & 3.4   & 13.8  & 4.9   & 4.3 \\
			50    & 75    & 75    & 50    & 4.0   & 7.2   & 10.8  & 4.6   & 6.7   & 3.1   & 4.3   & 10.5  & 6.8   & 3.0   & 3.1   & 5.9   & 12.2  & 7.3   & 7.1 \\
			75    & 75    & 75    & 50    & 5.4   & 14.1  & 9.0   & 6.2   & 7.6   & 3.7   & 4.9   & 9.0   & 8.2   & 3.1   & 4.6   & 9.5   & 11.2  & 18.4  & 8.8 \\
			25    & 25    & 25    & 75    & 4.4   & 4.4   & 2.5   & 6.3   & 8.7   & 5.3   & 3.9   & 2.2   & 9.6   & 4.8   & 4.4   & 6.4   & 5.3   & 8.0   & 8.9 \\
			50    & 25    & 25    & 75    & 4.7   & 4.7   & 2.3   & 8.0   & 9.4   & 5.6   & 4.6   & 2.4   & 13.8  & 5.3   & 4.7   & 10.8  & 3.9   & 10.7  & 9.6 \\
			75    & 25    & 25    & 75    & 6.3   & 6.9   & 1.8   & 9.8   & 11.5  & 6.7   & 9.0   & 2.7   & 28.8  & 6.7   & 4.8   & 12.2  & 3.9   & 13.4  & 9.9 \\
			25    & 50    & 25    & 75    & 4.5   & 5.7   & 3.2   & 4.4   & 6.4   & 4.4   & 4.3   & 3.0   & 7.7   & 3.4   & 3.3   & 5.4   & 6.3   & 5.5   & 6.3 \\
			50    & 50    & 25    & 75    & 4.8   & 6.0   & 2.8   & 5.8   & 6.8   & 4.9   & 5.1   & 3.2   & 11.1  & 3.9   & 3.3   & 8.2   & 5.1   & 9.1   & 9.1 \\
			75    & 50    & 25    & 75    & 6.1   & 7.0   & 2.5   & 13.6  & 8.0   & 6.9   & 8.5   & 3.5   & 16.9  & 4.6   & 3.3   & 12.9  & 5.0   & 10.0  & 10.9 \\
			25    & 75    & 25    & 75    & 4.6   & 4.1   & 4.2   & 4.4   & 4.5   & 3.6   & 3.2   & 4.6   & 6.5   & 2.6   & 3.3   & 4.5   & 8.3   & 4.3   & 5.9 \\
			50    & 75    & 25    & 75    & 4.8   & 4.4   & 3.6   & 4.1   & 4.7   & 3.8   & 3.3   & 4.7   & 7.6   & 2.7   & 2.2   & 4.6   & 6.4   & 5.7   & 8.3 \\
			75    & 75    & 25    & 75    & 5.7   & 5.8   & 2.8   & 5.5   & 5.3   & 4.3   & 4.1   & 5.0   & 11.9  & 2.9   & 2.4   & 8.4   & 6.9   & 6.2   & 7.1 \\
			25    & 25    & 50    & 75    & 4.6   & 4.5   & 3.0   & 6.8   & 10.5  & 5.5   & 4.0   & 2.7   & 11.2  & 6.1   & 5.0   & 7.0   & 7.3   & 10.1  & 8.6 \\
			50    & 25    & 50    & 75    & 5.0   & 4.8   & 2.7   & 8.9   & 11.3  & 5.6   & 4.7   & 2.9   & 14.7  & 7.2   & 6.6   & 11.3  & 4.9   & 16.4  & 10.7 \\
			75    & 25    & 50    & 75    & 6.6   & 6.6   & 2.2   & 10.9  & 13.8  & 6.8   & 10.2  & 3.3   & 27.7  & 7.9   & 10.0  & 19.4  & 4.4   & 13.8  & 15.1 \\
			25    & 50    & 50    & 75    & 4.7   & 5.5   & 3.6   & 5.1   & 8.0   & 4.9   & 4.4   & 3.5   & 7.3   & 3.9   & 5.8   & 5.1   & 6.4   & 5.7   & 8.9 \\
			50    & 50    & 50    & 75    & 5.0   & 6.2   & 4.2   & 6.2   & 8.4   & 5.3   & 5.3   & 3.7   & 9.3   & 4.4   & 6.5   & 9.2   & 6.4   & 14.8  & 10.8 \\
			75    & 50    & 50    & 75    & 6.3   & 7.5   & 3.0   & 14.9  & 9.6   & 7.4   & 9.3   & 4.2   & 19.5  & 5.3   & 8.5   & 17.4  & 6.0   & 14.5  & 8.7 \\
			25    & 75    & 50    & 75    & 4.6   & 4.3   & 7.5   & 5.4   & 5.4   & 3.7   & 3.4   & 5.2   & 6.6   & 2.8   & 3.0   & 3.1   & 10.1  & 4.2   & 6.0 \\
			50    & 75    & 50    & 75    & 5.0   & 4.6   & 5.8   & 5.1   & 5.6   & 3.9   & 3.5   & 5.5   & 7.7   & 2.9   & 2.8   & 3.4   & 11.5  & 7.7   & 9.1 \\
			75    & 75    & 50    & 75    & 5.9   & 6.0   & 3.6   & 5.7   & 6.2   & 4.6   & 4.3   & 5.8   & 12.7  & 3.3   & 3.9   & 12.3  & 10.6  & 11.0  & 9.4 \\
			25    & 25    & 75    & 75    & 4.9   & 5.1   & 4.1   & 9.4   & 14.7  & 5.8   & 4.9   & 4.6   & 8.6   & 6.4   & 5.2   & 6.9   & 10.6  & 10.1  & 8.6 \\
			50    & 25    & 75    & 75    & 5.3   & 5.4   & 3.9   & 11.4  & 16.0  & 6.0   & 5.6   & 4.9   & 13.9  & 8.0   & 7.7   & 15.4  & 6.6   & 17.0  & 10.6 \\
			75    & 25    & 75    & 75    & 6.9   & 8.3   & 3.8   & 21.6  & 18.6  & 8.0   & 12.1  & 5.6   & 43.8  & 16.5  & 11.8  & 45.4  & 7.4   & 15.5  & 14.5 \\
			25    & 50    & 75    & 75    & 5.2   & 5.2   & 5.4   & 13.2  & 11.1  & 5.9   & 5.1   & 5.6   & 8.2   & 5.6   & 6.7   & 6.6   & 13.7  & 7.9   & 7.9 \\
			50    & 50    & 75    & 75    & 5.6   & 5.5   & 6.0   & 15.7  & 11.9  & 6.2   & 6.0   & 5.6   & 9.7   & 6.4   & 7.4   & 11.3  & 9.7   & 20.4  & 9.4 \\
			75    & 50    & 75    & 75    & 7.1   & 10.7  & 5.0   & 20.6  & 13.1  & 7.6   & 11.9  & 6.6   & 18.1  & 8.7   & 9.0   & 35.8  & 10.0  & 31.2  & 9.9 \\
			25    & 75    & 75    & 75    & 4.9   & 9.4   & 8.7   & 9.2   & 7.7   & 4.2   & 4.4   & 7.4   & 7.4   & 3.5   & 3.7   & 5.3   & 12.2  & 5.3   & 5.5 \\
			50    & 75    & 75    & 75    & 5.3   & 10.6  & 9.6   & 11.4  & 8.1   & 4.2   & 4.7   & 7.8   & 9.2   & 3.7   & 4.2   & 6.7   & 16.1  & 17.7  & 9.7 \\
			75    & 75    & 75    & 75    & 6.9   & 16.2  & 9.2   & 10.2  & 8.8   & 5.0   & 6.2   & 8.0   & 14.3  & 3.9   & 7.6   & 20.7  & 11.9  & 25.2  & 10.2 \\
			\bottomrule
		\end{tabular}%
		\label{tabA:MPSS}%
	\end{table}%
	
	\begin{table}[htbp]
		\tiny
		\centering
		\caption{Marginal Product of Capital}
		\begin{tabular}{rrrr|rrrrr|rrrrr|rrrrr}
			\toprule
			\multicolumn{4}{c|}{Percentile} & \multicolumn{5}{c|}{CAP}               & \multicolumn{5}{c|}{CAPNLS}            & \multicolumn{5}{c}{CNLS} \\
			\multicolumn{1}{l}{$K$} & \multicolumn{1}{l}{$L$} & \multicolumn{1}{l}{$E$} & \multicolumn{1}{l|}{$S$} & \multicolumn{1}{l}{2899} & \multicolumn{1}{l}{2010} & \multicolumn{1}{l}{2811} & \multicolumn{1}{l}{2520} & \multicolumn{1}{l|}{1541} & \multicolumn{1}{l}{2899} & \multicolumn{1}{l}{2010} & \multicolumn{1}{l}{2811} & \multicolumn{1}{l}{2520} & \multicolumn{1}{l|}{1541} & \multicolumn{1}{l}{2899} & \multicolumn{1}{l}{2010} & \multicolumn{1}{l}{2811} & \multicolumn{1}{l}{2520} & \multicolumn{1}{l}{1541} \\
			\midrule
			25    & 25    & 25    & 25    & 0.09  & 0.39  & 0.55  & 0.15  & 0.45  & 0.20  & 0.30  & 0.32  & 0.40  & 0.28  & 0.27  & 1.59  & 0.93  & 0.80  & 2.77 \\
			50    & 25    & 25    & 25    & 0.09  & 0.30  & 0.46  & -0.04 & 0.45  & 0.11  & 0.16  & 0.25  & 0.24  & 0.25  & 0.02  & 0.03  & 0.09  & 0.07  & 1.14 \\
			75    & 25    & 25    & 25    & 0.09  & 0.30  & 0.05  & -0.06 & 0.15  & 0.03  & 0.10  & 0.18  & 0.06  & 0.21  & 0.00  & 0.00  & 0.02  & 0.00  & 0.03 \\
			25    & 50    & 25    & 25    & 0.09  & 0.38  & 0.50  & 0.23  & 0.59  & 0.24  & 0.28  & 0.31  & 0.43  & 0.28  & 0.82  & 1.38  & 0.87  & 0.79  & 3.01 \\
			50    & 50    & 25    & 25    & 0.09  & 0.38  & 0.42  & 0.15  & 0.45  & 0.23  & 0.19  & 0.26  & 0.35  & 0.26  & 0.22  & 0.05  & 0.09  & 0.08  & 2.51 \\
			75    & 50    & 25    & 25    & 0.09  & 0.31  & 0.14  & 0.00  & 0.15  & 0.12  & 0.10  & 0.18  & 0.14  & 0.22  & 0.01  & 0.00  & 0.04  & 0.01  & 0.05 \\
			25    & 75    & 25    & 25    & 0.29  & 0.40  & 0.39  & 0.24  & 0.59  & 0.26  & 0.29  & 0.25  & 0.44  & 0.29  & 1.21  & 1.47  & 2.79  & 0.93  & 2.97 \\
			50    & 75    & 25    & 25    & 0.29  & 0.40  & 0.39  & 0.24  & 0.59  & 0.24  & 0.23  & 0.21  & 0.39  & 0.29  & 0.40  & 0.05  & 0.10  & 0.23  & 2.69 \\
			75    & 75    & 25    & 25    & 0.13  & 0.39  & 0.23  & 0.12  & 0.29  & 0.19  & 0.18  & 0.18  & 0.27  & 0.28  & 0.11  & 0.01  & 0.06  & 0.01  & 0.42 \\
			25    & 25    & 50    & 25    & 0.09  & 0.38  & 0.55  & 0.19  & 0.45  & 0.21  & 0.27  & 0.33  & 0.39  & 0.29  & 0.59  & 1.59  & 1.15  & 0.83  & 2.74 \\
			50    & 25    & 50    & 25    & 0.09  & 0.38  & 0.50  & -0.04 & 0.45  & 0.12  & 0.16  & 0.27  & 0.25  & 0.25  & 0.09  & 0.04  & 0.22  & 0.11  & 1.09 \\
			75    & 25    & 50    & 25    & 0.09  & 0.30  & 0.26  & -0.04 & 0.45  & 0.03  & 0.10  & 0.18  & 0.06  & 0.22  & 0.00  & 0.00  & 0.01  & 0.00  & 0.03 \\
			25    & 50    & 50    & 25    & 0.09  & 0.39  & 0.50  & 0.34  & 0.45  & 0.25  & 0.27  & 0.32  & 0.44  & 0.27  & 1.87  & 1.29  & 1.56  & 0.83  & 2.88 \\
			50    & 50    & 50    & 25    & 0.09  & 0.39  & 0.50  & 0.02  & 0.45  & 0.23  & 0.18  & 0.27  & 0.35  & 0.26  & 0.20  & 0.10  & 0.26  & 0.23  & 2.54 \\
			75    & 50    & 50    & 25    & 0.09  & 0.29  & 0.35  & 0.02  & 0.15  & 0.12  & 0.10  & 0.20  & 0.13  & 0.21  & 0.00  & 0.00  & 0.08  & 0.01  & 0.06 \\
			25    & 75    & 50    & 25    & 0.29  & 0.40  & 0.42  & 0.37  & 0.59  & 0.27  & 0.24  & 0.27  & 0.44  & 0.30  & 2.84  & 1.88  & 1.97  & 1.08  & 3.14 \\
			50    & 75    & 50    & 25    & 0.29  & 0.40  & 0.39  & 0.26  & 0.59  & 0.25  & 0.23  & 0.25  & 0.39  & 0.28  & 0.56  & 0.11  & 0.17  & 0.33  & 2.95 \\
			75    & 75    & 50    & 25    & 0.13  & 0.39  & 0.39  & 0.12  & 0.29  & 0.20  & 0.18  & 0.18  & 0.27  & 0.26  & 0.17  & 0.02  & 0.04  & 0.04  & 2.25 \\
			25    & 25    & 75    & 25    & 0.13  & 0.38  & 0.55  & 0.07  & 0.51  & 0.22  & 0.20  & 0.34  & 0.37  & 0.31  & 1.79  & 1.58  & 1.11  & 0.88  & 2.75 \\
			50    & 25    & 75    & 25    & 0.09  & 0.38  & 0.47  & 0.07  & 0.51  & 0.17  & 0.17  & 0.29  & 0.27  & 0.25  & 0.32  & 0.04  & 0.32  & 0.19  & 1.09 \\
			75    & 25    & 75    & 25    & 0.09  & 0.30  & 0.43  & -0.04 & 0.51  & 0.05  & 0.13  & 0.23  & 0.08  & 0.23  & 0.03  & 0.00  & 0.04  & 0.01  & 0.03 \\
			25    & 50    & 75    & 25    & 0.13  & 0.38  & 0.50  & 0.37  & 0.45  & 0.27  & 0.24  & 0.31  & 0.44  & 0.30  & 2.59  & 1.82  & 1.53  & 0.91  & 2.92 \\
			50    & 50    & 75    & 25    & 0.13  & 0.38  & 0.47  & 0.16  & 0.45  & 0.27  & 0.19  & 0.28  & 0.35  & 0.25  & 0.69  & 0.09  & 0.41  & 0.36  & 2.52 \\
			75    & 50    & 75    & 25    & 0.09  & 0.38  & 0.43  & 0.02  & 0.45  & 0.15  & 0.13  & 0.24  & 0.13  & 0.22  & 0.04  & 0.00  & 0.08  & 0.08  & 0.06 \\
			25    & 75    & 75    & 25    & 0.29  & 0.39  & 0.50  & 0.37  & 0.45  & 0.30  & 0.24  & 0.28  & 0.46  & 0.29  & 3.18  & 1.81  & 2.04  & 1.22  & 3.39 \\
			50    & 75    & 75    & 25    & 0.29  & 0.39  & 0.47  & 0.24  & 0.45  & 0.27  & 0.23  & 0.27  & 0.41  & 0.29  & 2.02  & 0.15  & 0.39  & 0.51  & 2.93 \\
			75    & 75    & 75    & 25    & 0.16  & 0.37  & 0.39  & 0.14  & 0.45  & 0.24  & 0.18  & 0.26  & 0.26  & 0.24  & 0.13  & 0.02  & 0.08  & 0.12  & 2.26 \\
			25    & 25    & 25    & 50    & 0.09  & 0.39  & 0.50  & 0.15  & 0.45  & 0.15  & 0.36  & 0.36  & 0.42  & 0.28  & 0.52  & 1.86  & 1.60  & 0.98  & 2.74 \\
			50    & 25    & 25    & 50    & 0.09  & 0.32  & 0.42  & 0.08  & 0.45  & 0.11  & 0.29  & 0.31  & 0.28  & 0.28  & 0.03  & 1.26  & 0.25  & 0.10  & 1.35 \\
			75    & 25    & 25    & 50    & 0.09  & 0.30  & 0.03  & -0.06 & 0.15  & 0.03  & 0.10  & 0.26  & 0.08  & 0.23  & 0.00  & 0.01  & 0.04  & 0.00  & 0.03 \\
			25    & 50    & 25    & 50    & 0.09  & 0.38  & 0.49  & 0.23  & 0.59  & 0.26  & 0.37  & 0.37  & 0.47  & 0.33  & 0.84  & 1.82  & 2.37  & 1.21  & 2.92 \\
			50    & 50    & 25    & 50    & 0.09  & 0.38  & 0.37  & 0.15  & 0.29  & 0.23  & 0.31  & 0.32  & 0.41  & 0.28  & 0.19  & 1.31  & 0.30  & 0.27  & 2.60 \\
			75    & 50    & 25    & 50    & 0.09  & 0.31  & 0.07  & 0.00  & 0.15  & 0.11  & 0.10  & 0.24  & 0.14  & 0.26  & 0.01  & 0.01  & 0.05  & 0.08  & 0.05 \\
			25    & 75    & 25    & 50    & 0.29  & 0.40  & 0.41  & 0.24  & 0.29  & 0.26  & 0.32  & 0.29  & 0.46  & 0.33  & 1.30  & 2.06  & 4.81  & 2.30  & 2.99 \\
			50    & 75    & 25    & 50    & 0.29  & 0.40  & 0.38  & 0.24  & 0.29  & 0.26  & 0.32  & 0.25  & 0.44  & 0.31  & 0.39  & 1.25  & 0.41  & 0.30  & 2.72 \\
			75    & 75    & 25    & 50    & 0.13  & 0.39  & 0.18  & 0.24  & 0.29  & 0.19  & 0.19  & 0.16  & 0.30  & 0.27  & 0.07  & 0.02  & 0.07  & 0.05  & 0.40 \\
			25    & 25    & 50    & 50    & 0.09  & 0.38  & 0.53  & 0.19  & 0.45  & 0.18  & 0.32  & 0.37  & 0.42  & 0.30  & 0.77  & 1.81  & 1.26  & 1.47  & 2.80 \\
			50    & 25    & 50    & 50    & 0.09  & 0.30  & 0.44  & 0.08  & 0.45  & 0.12  & 0.29  & 0.34  & 0.25  & 0.28  & 0.17  & 1.42  & 0.56  & 0.35  & 1.43 \\
			75    & 25    & 50    & 50    & 0.09  & 0.30  & 0.20  & -0.04 & 0.15  & 0.03  & 0.10  & 0.26  & 0.07  & 0.24  & 0.00  & 0.01  & 0.02  & 0.01  & 0.04 \\
			25    & 50    & 50    & 50    & 0.09  & 0.39  & 0.52  & 0.34  & 0.45  & 0.28  & 0.34  & 0.38  & 0.46  & 0.33  & 1.40  & 2.11  & 2.39  & 1.69  & 2.90 \\
			50    & 50    & 50    & 50    & 0.09  & 0.39  & 0.52  & 0.15  & 0.45  & 0.24  & 0.30  & 0.35  & 0.42  & 0.27  & 0.35  & 1.62  & 0.61  & 0.73  & 2.70 \\
			75    & 50    & 50    & 50    & 0.09  & 0.29  & 0.22  & 0.00  & 0.15  & 0.12  & 0.10  & 0.24  & 0.14  & 0.24  & 0.00  & 0.00  & 0.07  & 0.05  & 0.08 \\
			25    & 75    & 50    & 50    & 0.29  & 0.40  & 0.45  & 0.37  & 0.59  & 0.27  & 0.32  & 0.31  & 0.47  & 0.32  & 2.17  & 2.52  & 2.86  & 2.58  & 3.13 \\
			50    & 75    & 50    & 50    & 0.29  & 0.39  & 0.41  & 0.24  & 0.59  & 0.26  & 0.32  & 0.28  & 0.44  & 0.31  & 0.66  & 1.51  & 1.31  & 0.66  & 2.96 \\
			75    & 75    & 50    & 50    & 0.13  & 0.39  & 0.36  & 0.12  & 0.29  & 0.20  & 0.19  & 0.18  & 0.29  & 0.25  & 0.15  & 0.03  & 0.08  & 0.02  & 2.16 \\
			25    & 25    & 75    & 50    & 0.09  & 0.39  & 0.55  & 0.19  & 0.51  & 0.21  & 0.29  & 0.38  & 0.37  & 0.31  & 1.90  & 1.85  & 1.24  & 1.51  & 2.80 \\
			50    & 25    & 75    & 50    & 0.09  & 0.39  & 0.50  & 0.19  & 0.51  & 0.16  & 0.27  & 0.36  & 0.27  & 0.28  & 0.54  & 1.42  & 0.59  & 0.39  & 1.43 \\
			75    & 25    & 75    & 50    & 0.09  & 0.30  & 0.46  & -0.04 & 0.51  & 0.06  & 0.12  & 0.28  & 0.07  & 0.24  & 0.03  & 0.00  & 0.08  & 0.02  & 0.04 \\
			25    & 50    & 75    & 50    & 0.13  & 0.38  & 0.50  & 0.34  & 0.51  & 0.29  & 0.32  & 0.37  & 0.47  & 0.34  & 2.39  & 2.18  & 2.33  & 1.81  & 2.94 \\
			50    & 50    & 75    & 50    & 0.13  & 0.38  & 0.50  & 0.25  & 0.51  & 0.27  & 0.28  & 0.35  & 0.42  & 0.31  & 0.79  & 1.66  & 0.82  & 0.73  & 2.65 \\
			75    & 50    & 75    & 50    & 0.09  & 0.30  & 0.46  & 0.02  & 0.51  & 0.15  & 0.12  & 0.29  & 0.12  & 0.24  & 0.07  & 0.00  & 0.26  & 0.09  & 0.06 \\
			25    & 75    & 75    & 50    & 0.29  & 0.39  & 0.50  & 0.37  & 0.45  & 0.30  & 0.30  & 0.35  & 0.48  & 0.31  & 2.97  & 3.27  & 2.67  & 2.90  & 3.05 \\
			50    & 75    & 75    & 50    & 0.29  & 0.39  & 0.50  & 0.37  & 0.45  & 0.30  & 0.28  & 0.31  & 0.45  & 0.31  & 1.66  & 1.59  & 1.60  & 0.76  & 2.74 \\
			75    & 75    & 75    & 50    & 0.16  & 0.37  & 0.42  & 0.14  & 0.45  & 0.25  & 0.18  & 0.25  & 0.29  & 0.26  & 0.13  & 0.01  & 0.31  & 0.18  & 2.34 \\
			25    & 25    & 25    & 75    & 0.09  & 0.34  & 0.40  & 0.17  & 0.35  & 0.12  & 0.44  & 0.34  & 0.39  & 0.31  & 1.12  & 2.35  & 2.42  & 0.94  & 2.58 \\
			50    & 25    & 25    & 75    & 0.09  & 0.34  & 0.29  & 0.06  & 0.35  & 0.09  & 0.44  & 0.33  & 0.33  & 0.30  & 0.06  & 1.39  & 0.57  & 0.15  & 1.41 \\
			75    & 25    & 25    & 75    & 0.09  & 0.31  & 0.05  & 0.06  & 0.15  & 0.05  & 0.29  & 0.30  & 0.23  & 0.25  & 0.01  & 0.17  & 0.14  & 0.02  & 0.05 \\
			25    & 50    & 25    & 75    & 0.18  & 0.40  & 0.39  & 0.32  & 0.29  & 0.29  & 0.48  & 0.35  & 0.47  & 0.33  & 1.20  & 2.80  & 3.88  & 1.16  & 2.79 \\
			50    & 50    & 25    & 75    & 0.12  & 0.40  & 0.30  & 0.12  & 0.29  & 0.26  & 0.45  & 0.34  & 0.43  & 0.33  & 0.29  & 1.45  & 0.55  & 0.34  & 2.61 \\
			75    & 50    & 25    & 75    & 0.09  & 0.31  & 0.02  & 0.12  & 0.29  & 0.07  & 0.27  & 0.27  & 0.31  & 0.26  & 0.04  & 0.17  & 0.11  & 0.08  & 0.05 \\
			25    & 75    & 25    & 75    & 0.37  & 0.40  & 0.36  & 0.32  & 0.29  & 0.31  & 0.47  & 0.33  & 0.50  & 0.43  & 1.56  & 2.85  & 7.44  & 1.92  & 3.03 \\
			50    & 75    & 25    & 75    & 0.37  & 0.40  & 0.25  & 0.21  & 0.29  & 0.31  & 0.46  & 0.30  & 0.49  & 0.41  & 0.28  & 1.85  & 1.33  & 0.41  & 2.70 \\
			75    & 75    & 25    & 75    & 0.15  & 0.40  & 0.11  & 0.21  & 0.29  & 0.20  & 0.35  & 0.21  & 0.44  & 0.29  & 0.12  & 0.26  & 0.13  & 0.09  & 0.91 \\
			25    & 25    & 50    & 75    & 0.09  & 0.34  & 0.56  & 0.19  & 0.21  & 0.14  & 0.44  & 0.36  & 0.40  & 0.28  & 1.42  & 2.37  & 2.32  & 1.48  & 2.56 \\
			50    & 25    & 50    & 75    & 0.09  & 0.34  & 0.34  & 0.08  & 0.21  & 0.10  & 0.44  & 0.35  & 0.32  & 0.27  & 0.22  & 1.50  & 0.76  & 0.44  & 1.42 \\
			75    & 25    & 50    & 75    & 0.09  & 0.32  & 0.04  & 0.06  & 0.21  & 0.05  & 0.29  & 0.30  & 0.22  & 0.24  & 0.01  & 0.15  & 0.38  & 0.03  & 0.10 \\
			25    & 50    & 50    & 75    & 0.15  & 0.42  & 0.51  & 0.32  & 0.29  & 0.28  & 0.48  & 0.38  & 0.48  & 0.36  & 2.41  & 2.25  & 4.64  & 1.69  & 3.07 \\
			50    & 50    & 50    & 75    & 0.12  & 0.42  & 0.41  & 0.12  & 0.29  & 0.26  & 0.45  & 0.35  & 0.44  & 0.33  & 0.71  & 1.59  & 0.79  & 0.84  & 2.46 \\
			75    & 50    & 50    & 75    & 0.09  & 0.32  & 0.05  & 0.12  & 0.15  & 0.07  & 0.26  & 0.30  & 0.29  & 0.25  & 0.01  & 0.16  & 0.26  & 0.11  & 0.20 \\
			25    & 75    & 50    & 75    & 0.37  & 0.40  & 0.41  & 0.32  & 0.29  & 0.33  & 0.46  & 0.37  & 0.50  & 0.44  & 2.66  & 2.41  & 7.92  & 2.31  & 3.11 \\
			50    & 75    & 50    & 75    & 0.37  & 0.40  & 0.33  & 0.21  & 0.29  & 0.31  & 0.46  & 0.33  & 0.49  & 0.40  & 0.78  & 1.91  & 1.52  & 1.04  & 2.88 \\
			75    & 75    & 50    & 75    & 0.15  & 0.40  & 0.13  & 0.21  & 0.29  & 0.22  & 0.35  & 0.24  & 0.44  & 0.29  & 0.10  & 0.29  & 0.21  & 0.24  & 2.21 \\
			25    & 25    & 75    & 75    & 0.12  & 0.41  & 0.55  & 0.19  & 0.51  & 0.19  & 0.48  & 0.40  & 0.37  & 0.28  & 1.96  & 2.37  & 2.21  & 1.61  & 2.73 \\
			50    & 25    & 75    & 75    & 0.12  & 0.33  & 0.50  & 0.19  & 0.51  & 0.17  & 0.43  & 0.38  & 0.30  & 0.28  & 0.58  & 1.68  & 0.77  & 0.56  & 1.41 \\
			75    & 25    & 75    & 75    & 0.09  & 0.32  & 0.35  & 0.08  & 0.51  & 0.08  & 0.24  & 0.35  & 0.19  & 0.25  & 0.02  & 1.37  & 0.49  & 0.06  & 0.12 \\
			25    & 50    & 75    & 75    & 0.25  & 0.41  & 0.51  & 0.34  & 0.51  & 0.29  & 0.48  & 0.41  & 0.49  & 0.33  & 2.95  & 2.27  & 3.26  & 1.82  & 3.64 \\
			50    & 50    & 75    & 75    & 0.12  & 0.41  & 0.51  & 0.25  & 0.51  & 0.28  & 0.45  & 0.40  & 0.45  & 0.30  & 1.21  & 1.81  & 1.03  & 1.14  & 2.19 \\
			75    & 50    & 75    & 75    & 0.09  & 0.32  & 0.36  & 0.08  & 0.51  & 0.13  & 0.25  & 0.35  & 0.24  & 0.25  & 0.21  & 1.26  & 0.42  & 0.14  & 0.22 \\
			25    & 75    & 75    & 75    & 0.37  & 0.40  & 0.52  & 0.34  & 0.59  & 0.37  & 0.54  & 0.41  & 0.51  & 0.37  & 3.33  & 3.43  & 5.22  & 2.63  & 3.68 \\
			50    & 75    & 75    & 75    & 0.37  & 0.40  & 0.52  & 0.34  & 0.59  & 0.36  & 0.52  & 0.38  & 0.51  & 0.36  & 1.60  & 2.16  & 1.90  & 1.55  & 2.82 \\
			75    & 75    & 75    & 75    & 0.34  & 0.40  & 0.40  & 0.21  & 0.15  & 0.26  & 0.32  & 0.35  & 0.46  & 0.27  & 0.34  & 1.16  & 0.56  & 0.73  & 2.15 \\
			\bottomrule
		\end{tabular}%
		\label{tabA:MP_K}%
	\end{table}%

	\begin{table}[htbp]
		\tiny
		\centering
		\caption{Marginal Product of Labor}
		\begin{tabular}{rrrr|rrrrr|rrrrr|rrrrr}
			\toprule
			\multicolumn{4}{c|}{Percentile} & \multicolumn{5}{c|}{CAP}               & \multicolumn{5}{c|}{CAPNLS}            & \multicolumn{5}{c}{CNLS} \\
			\multicolumn{1}{l}{$K$} & \multicolumn{1}{l}{$L$} & \multicolumn{1}{l}{$E$} & \multicolumn{1}{l|}{$S$} & \multicolumn{1}{l}{2899} & \multicolumn{1}{l}{2010} & \multicolumn{1}{l}{2811} & \multicolumn{1}{l}{2520} & \multicolumn{1}{l|}{1541} & \multicolumn{1}{l}{2899} & \multicolumn{1}{l}{2010} & \multicolumn{1}{l}{2811} & \multicolumn{1}{l}{2520} & \multicolumn{1}{l|}{1541} & \multicolumn{1}{l}{2899} & \multicolumn{1}{l}{2010} & \multicolumn{1}{l}{2811} & \multicolumn{1}{l}{2520} & \multicolumn{1}{l}{1541} \\
			\midrule
			25    & 25    & 25    & 25    & 0.14  & 0.04  & 0.09  & 0.21  & 0.07  & 0.19  & 0.09  & 0.12  & 0.19  & 0.15  & 0.16  & 0.05  & 0.08  & 0.12  & 0.06 \\
			50    & 25    & 25    & 25    & 0.14  & 0.08  & 0.09  & 0.28  & 0.07  & 0.22  & 0.10  & 0.12  & 0.22  & 0.15  & 0.16  & 0.05  & 0.09  & 0.13  & 0.13 \\
			75    & 25    & 25    & 25    & 0.14  & 0.08  & 0.09  & 0.27  & 0.06  & 0.26  & 0.09  & 0.12  & 0.26  & 0.15  & 0.16  & 0.05  & 0.10  & 0.14  & 0.18 \\
			25    & 50    & 25    & 25    & 0.14  & 0.03  & 0.09  & 0.16  & 0.05  & 0.15  & 0.06  & 0.10  & 0.15  & 0.12  & 0.04  & 0.01  & 0.05  & 0.01  & 0.03 \\
			50    & 50    & 25    & 25    & 0.14  & 0.03  & 0.08  & 0.21  & 0.07  & 0.15  & 0.06  & 0.10  & 0.15  & 0.13  & 0.07  & 0.01  & 0.05  & 0.04  & 0.03 \\
			75    & 50    & 25    & 25    & 0.14  & 0.06  & 0.08  & 0.22  & 0.06  & 0.18  & 0.07  & 0.10  & 0.20  & 0.13  & 0.11  & 0.01  & 0.06  & 0.04  & 0.14 \\
			25    & 75    & 25    & 25    & 0.09  & 0.01  & 0.08  & 0.14  & 0.05  & 0.08  & 0.03  & 0.09  & 0.13  & 0.08  & 0.00  & 0.00  & 0.02  & 0.00  & 0.00 \\
			50    & 75    & 25    & 25    & 0.09  & 0.01  & 0.08  & 0.14  & 0.05  & 0.08  & 0.03  & 0.08  & 0.13  & 0.08  & 0.00  & 0.00  & 0.04  & 0.01  & 0.00 \\
			75    & 75    & 25    & 25    & 0.12  & 0.02  & 0.08  & 0.14  & 0.04  & 0.10  & 0.04  & 0.08  & 0.13  & 0.08  & 0.01  & 0.00  & 0.04  & 0.02  & 0.02 \\
			25    & 25    & 50    & 25    & 0.14  & 0.05  & 0.09  & 0.27  & 0.07  & 0.19  & 0.09  & 0.12  & 0.19  & 0.15  & 0.16  & 0.05  & 0.11  & 0.13  & 0.07 \\
			50    & 25    & 50    & 25    & 0.14  & 0.05  & 0.09  & 0.28  & 0.07  & 0.22  & 0.10  & 0.12  & 0.23  & 0.15  & 0.19  & 0.07  & 0.13  & 0.13  & 0.15 \\
			75    & 25    & 50    & 25    & 0.14  & 0.08  & 0.10  & 0.28  & 0.07  & 0.27  & 0.10  & 0.12  & 0.26  & 0.15  & 0.19  & 0.07  & 0.16  & 0.17  & 0.21 \\
			25    & 50    & 50    & 25    & 0.14  & 0.04  & 0.09  & 0.16  & 0.07  & 0.15  & 0.07  & 0.11  & 0.15  & 0.13  & 0.07  & 0.01  & 0.06  & 0.04  & 0.04 \\
			50    & 50    & 50    & 25    & 0.14  & 0.04  & 0.09  & 0.22  & 0.07  & 0.15  & 0.06  & 0.10  & 0.16  & 0.13  & 0.13  & 0.01  & 0.06  & 0.04  & 0.05 \\
			75    & 50    & 50    & 25    & 0.14  & 0.07  & 0.08  & 0.22  & 0.06  & 0.18  & 0.08  & 0.11  & 0.20  & 0.13  & 0.14  & 0.02  & 0.06  & 0.05  & 0.19 \\
			25    & 75    & 50    & 25    & 0.09  & 0.01  & 0.08  & 0.14  & 0.05  & 0.08  & 0.03  & 0.09  & 0.12  & 0.08  & 0.00  & 0.00  & 0.05  & 0.00  & 0.01 \\
			50    & 75    & 50    & 25    & 0.09  & 0.01  & 0.08  & 0.14  & 0.05  & 0.09  & 0.03  & 0.09  & 0.12  & 0.08  & 0.01  & 0.00  & 0.05  & 0.01  & 0.01 \\
			75    & 75    & 50    & 25    & 0.12  & 0.02  & 0.08  & 0.14  & 0.04  & 0.09  & 0.04  & 0.08  & 0.13  & 0.08  & 0.02  & 0.01  & 0.05  & 0.02  & 0.02 \\
			25    & 25    & 75    & 25    & 0.12  & 0.05  & 0.09  & 0.27  & 0.11  & 0.20  & 0.10  & 0.12  & 0.22  & 0.16  & 0.16  & 0.06  & 0.11  & 0.13  & 0.07 \\
			50    & 25    & 75    & 25    & 0.14  & 0.05  & 0.09  & 0.27  & 0.11  & 0.22  & 0.10  & 0.12  & 0.25  & 0.16  & 0.20  & 0.07  & 0.14  & 0.13  & 0.15 \\
			75    & 25    & 75    & 25    & 0.14  & 0.08  & 0.09  & 0.28  & 0.11  & 0.26  & 0.10  & 0.12  & 0.27  & 0.16  & 0.23  & 0.07  & 0.17  & 0.20  & 0.21 \\
			25    & 50    & 75    & 25    & 0.12  & 0.05  & 0.09  & 0.14  & 0.07  & 0.15  & 0.07  & 0.11  & 0.16  & 0.14  & 0.07  & 0.03  & 0.06  & 0.05  & 0.05 \\
			50    & 50    & 75    & 25    & 0.12  & 0.05  & 0.09  & 0.20  & 0.07  & 0.15  & 0.07  & 0.10  & 0.17  & 0.13  & 0.12  & 0.03  & 0.06  & 0.06  & 0.06 \\
			75    & 50    & 75    & 25    & 0.14  & 0.05  & 0.09  & 0.22  & 0.07  & 0.17  & 0.09  & 0.10  & 0.21  & 0.13  & 0.14  & 0.04  & 0.06  & 0.09  & 0.20 \\
			25    & 75    & 75    & 25    & 0.09  & 0.04  & 0.09  & 0.14  & 0.07  & 0.09  & 0.04  & 0.10  & 0.13  & 0.09  & 0.01  & 0.00  & 0.05  & 0.00  & 0.02 \\
			50    & 75    & 75    & 25    & 0.09  & 0.04  & 0.09  & 0.15  & 0.07  & 0.09  & 0.04  & 0.10  & 0.14  & 0.09  & 0.01  & 0.01  & 0.05  & 0.01  & 0.02 \\
			75    & 75    & 75    & 25    & 0.10  & 0.04  & 0.08  & 0.15  & 0.07  & 0.09  & 0.05  & 0.10  & 0.15  & 0.10  & 0.06  & 0.02  & 0.05  & 0.02  & 0.05 \\
			25    & 25    & 25    & 50    & 0.14  & 0.04  & 0.11  & 0.21  & 0.07  & 0.22  & 0.09  & 0.13  & 0.20  & 0.15  & 0.18  & 0.08  & 0.12  & 0.14  & 0.06 \\
			50    & 25    & 25    & 50    & 0.14  & 0.07  & 0.11  & 0.27  & 0.07  & 0.23  & 0.10  & 0.13  & 0.23  & 0.15  & 0.18  & 0.08  & 0.12  & 0.22  & 0.14 \\
			75    & 25    & 25    & 50    & 0.14  & 0.08  & 0.11  & 0.27  & 0.06  & 0.26  & 0.10  & 0.13  & 0.27  & 0.16  & 0.18  & 0.08  & 0.12  & 0.25  & 0.18 \\
			25    & 50    & 25    & 50    & 0.14  & 0.03  & 0.09  & 0.16  & 0.05  & 0.15  & 0.06  & 0.11  & 0.16  & 0.13  & 0.05  & 0.01  & 0.07  & 0.10  & 0.04 \\
			50    & 50    & 25    & 50    & 0.14  & 0.03  & 0.08  & 0.21  & 0.04  & 0.16  & 0.05  & 0.11  & 0.16  & 0.13  & 0.09  & 0.01  & 0.09  & 0.13  & 0.04 \\
			75    & 50    & 25    & 50    & 0.14  & 0.06  & 0.10  & 0.22  & 0.06  & 0.18  & 0.07  & 0.11  & 0.21  & 0.13  & 0.12  & 0.02  & 0.09  & 0.12  & 0.15 \\
			25    & 75    & 25    & 50    & 0.09  & 0.01  & 0.08  & 0.14  & 0.04  & 0.08  & 0.03  & 0.09  & 0.13  & 0.08  & 0.01  & 0.00  & 0.01  & 0.00  & 0.00 \\
			50    & 75    & 25    & 50    & 0.09  & 0.01  & 0.08  & 0.14  & 0.04  & 0.08  & 0.03  & 0.09  & 0.13  & 0.08  & 0.01  & 0.00  & 0.04  & 0.01  & 0.00 \\
			75    & 75    & 25    & 50    & 0.12  & 0.02  & 0.08  & 0.14  & 0.04  & 0.10  & 0.03  & 0.08  & 0.14  & 0.08  & 0.02  & 0.00  & 0.05  & 0.02  & 0.02 \\
			25    & 25    & 50    & 50    & 0.14  & 0.05  & 0.10  & 0.27  & 0.07  & 0.22  & 0.10  & 0.13  & 0.20  & 0.16  & 0.19  & 0.09  & 0.17  & 0.16  & 0.07 \\
			50    & 25    & 50    & 50    & 0.14  & 0.08  & 0.10  & 0.27  & 0.07  & 0.23  & 0.11  & 0.13  & 0.25  & 0.16  & 0.20  & 0.10  & 0.17  & 0.29  & 0.15 \\
			75    & 25    & 50    & 50    & 0.14  & 0.08  & 0.10  & 0.28  & 0.06  & 0.27  & 0.10  & 0.13  & 0.27  & 0.16  & 0.21  & 0.10  & 0.19  & 0.41  & 0.24 \\
			25    & 50    & 50    & 50    & 0.14  & 0.04  & 0.09  & 0.16  & 0.07  & 0.16  & 0.06  & 0.12  & 0.16  & 0.13  & 0.12  & 0.03  & 0.07  & 0.10  & 0.05 \\
			50    & 50    & 50    & 50    & 0.14  & 0.04  & 0.09  & 0.21  & 0.07  & 0.15  & 0.06  & 0.11  & 0.16  & 0.13  & 0.15  & 0.03  & 0.11  & 0.12  & 0.06 \\
			75    & 50    & 50    & 50    & 0.14  & 0.07  & 0.09  & 0.22  & 0.06  & 0.18  & 0.08  & 0.12  & 0.21  & 0.13  & 0.17  & 0.04  & 0.12  & 0.12  & 0.21 \\
			25    & 75    & 50    & 50    & 0.09  & 0.01  & 0.08  & 0.14  & 0.05  & 0.08  & 0.03  & 0.10  & 0.13  & 0.08  & 0.01  & 0.00  & 0.04  & 0.00  & 0.01 \\
			50    & 75    & 50    & 50    & 0.09  & 0.02  & 0.08  & 0.14  & 0.05  & 0.08  & 0.03  & 0.09  & 0.13  & 0.08  & 0.01  & 0.00  & 0.05  & 0.01  & 0.01 \\
			75    & 75    & 50    & 50    & 0.12  & 0.02  & 0.08  & 0.14  & 0.04  & 0.09  & 0.03  & 0.08  & 0.13  & 0.09  & 0.03  & 0.00  & 0.06  & 0.03  & 0.03 \\
			25    & 25    & 75    & 50    & 0.14  & 0.06  & 0.09  & 0.27  & 0.11  & 0.22  & 0.11  & 0.13  & 0.23  & 0.16  & 0.19  & 0.10  & 0.17  & 0.16  & 0.08 \\
			50    & 25    & 75    & 50    & 0.14  & 0.06  & 0.09  & 0.27  & 0.11  & 0.24  & 0.11  & 0.13  & 0.26  & 0.16  & 0.23  & 0.11  & 0.17  & 0.29  & 0.16 \\
			75    & 25    & 75    & 50    & 0.14  & 0.08  & 0.09  & 0.28  & 0.11  & 0.26  & 0.10  & 0.13  & 0.27  & 0.16  & 0.32  & 0.11  & 0.21  & 0.41  & 0.24 \\
			25    & 50    & 75    & 50    & 0.12  & 0.05  & 0.09  & 0.16  & 0.11  & 0.15  & 0.07  & 0.12  & 0.17  & 0.15  & 0.11  & 0.05  & 0.08  & 0.12  & 0.07 \\
			50    & 50    & 75    & 50    & 0.12  & 0.05  & 0.09  & 0.21  & 0.11  & 0.15  & 0.07  & 0.12  & 0.17  & 0.15  & 0.15  & 0.07  & 0.13  & 0.12  & 0.07 \\
			75    & 50    & 75    & 50    & 0.14  & 0.08  & 0.09  & 0.22  & 0.11  & 0.18  & 0.08  & 0.11  & 0.24  & 0.15  & 0.19  & 0.11  & 0.17  & 0.14  & 0.23 \\
			25    & 75    & 75    & 50    & 0.09  & 0.04  & 0.09  & 0.14  & 0.07  & 0.09  & 0.04  & 0.10  & 0.14  & 0.10  & 0.04  & 0.00  & 0.05  & 0.00  & 0.05 \\
			50    & 75    & 75    & 50    & 0.09  & 0.04  & 0.09  & 0.14  & 0.07  & 0.09  & 0.04  & 0.10  & 0.14  & 0.10  & 0.06  & 0.03  & 0.06  & 0.03  & 0.05 \\
			75    & 75    & 75    & 50    & 0.10  & 0.04  & 0.08  & 0.15  & 0.07  & 0.09  & 0.05  & 0.10  & 0.14  & 0.10  & 0.08  & 0.03  & 0.07  & 0.04  & 0.05 \\
			25    & 25    & 25    & 75    & 0.14  & 0.09  & 0.13  & 0.26  & 0.08  & 0.25  & 0.09  & 0.13  & 0.21  & 0.15  & 0.28  & 0.08  & 0.18  & 0.14  & 0.07 \\
			50    & 25    & 25    & 75    & 0.14  & 0.09  & 0.13  & 0.26  & 0.08  & 0.26  & 0.09  & 0.13  & 0.23  & 0.15  & 0.29  & 0.11  & 0.20  & 0.22  & 0.14 \\
			75    & 25    & 25    & 75    & 0.14  & 0.08  & 0.12  & 0.26  & 0.06  & 0.26  & 0.10  & 0.13  & 0.27  & 0.15  & 0.31  & 0.11  & 0.20  & 0.27  & 0.19 \\
			25    & 50    & 25    & 75    & 0.11  & 0.05  & 0.12  & 0.16  & 0.04  & 0.17  & 0.08  & 0.12  & 0.17  & 0.14  & 0.05  & 0.02  & 0.07  & 0.11  & 0.04 \\
			50    & 50    & 25    & 75    & 0.13  & 0.05  & 0.11  & 0.21  & 0.04  & 0.18  & 0.09  & 0.12  & 0.19  & 0.14  & 0.03  & 0.06  & 0.11  & 0.14  & 0.05 \\
			75    & 50    & 25    & 75    & 0.14  & 0.06  & 0.12  & 0.21  & 0.04  & 0.24  & 0.09  & 0.12  & 0.22  & 0.15  & 0.08  & 0.08  & 0.11  & 0.17  & 0.17 \\
			25    & 75    & 25    & 75    & 0.06  & 0.02  & 0.09  & 0.16  & 0.04  & 0.07  & 0.03  & 0.10  & 0.15  & 0.09  & 0.02  & 0.00  & 0.01  & 0.06  & 0.01 \\
			50    & 75    & 25    & 75    & 0.06  & 0.02  & 0.09  & 0.16  & 0.04  & 0.07  & 0.04  & 0.10  & 0.15  & 0.09  & 0.02  & 0.00  & 0.05  & 0.07  & 0.01 \\
			75    & 75    & 25    & 75    & 0.11  & 0.02  & 0.10  & 0.16  & 0.04  & 0.07  & 0.04  & 0.10  & 0.16  & 0.09  & 0.02  & 0.01  & 0.07  & 0.06  & 0.02 \\
			25    & 25    & 50    & 75    & 0.14  & 0.09  & 0.12  & 0.27  & 0.10  & 0.25  & 0.09  & 0.13  & 0.21  & 0.16  & 0.32  & 0.15  & 0.21  & 0.17  & 0.11 \\
			50    & 25    & 50    & 75    & 0.14  & 0.09  & 0.13  & 0.27  & 0.10  & 0.26  & 0.09  & 0.13  & 0.24  & 0.16  & 0.37  & 0.14  & 0.24  & 0.35  & 0.17 \\
			75    & 25    & 50    & 75    & 0.14  & 0.09  & 0.11  & 0.26  & 0.10  & 0.26  & 0.10  & 0.13  & 0.27  & 0.16  & 0.40  & 0.11  & 0.24  & 0.42  & 0.24 \\
			25    & 50    & 50    & 75    & 0.12  & 0.06  & 0.11  & 0.16  & 0.04  & 0.18  & 0.08  & 0.12  & 0.17  & 0.14  & 0.16  & 0.05  & 0.13  & 0.12  & 0.06 \\
			50    & 50    & 50    & 75    & 0.13  & 0.06  & 0.11  & 0.21  & 0.04  & 0.19  & 0.09  & 0.12  & 0.18  & 0.14  & 0.18  & 0.07  & 0.14  & 0.14  & 0.08 \\
			75    & 50    & 50    & 75    & 0.14  & 0.07  & 0.11  & 0.21  & 0.06  & 0.25  & 0.09  & 0.13  & 0.23  & 0.15  & 0.22  & 0.09  & 0.14  & 0.21  & 0.21 \\
			25    & 75    & 50    & 75    & 0.06  & 0.02  & 0.09  & 0.16  & 0.04  & 0.07  & 0.04  & 0.11  & 0.15  & 0.09  & 0.02  & 0.00  & 0.04  & 0.07  & 0.02 \\
			50    & 75    & 50    & 75    & 0.06  & 0.02  & 0.09  & 0.16  & 0.04  & 0.07  & 0.04  & 0.10  & 0.15  & 0.10  & 0.02  & 0.00  & 0.08  & 0.10  & 0.03 \\
			75    & 75    & 50    & 75    & 0.11  & 0.02  & 0.10  & 0.16  & 0.04  & 0.07  & 0.04  & 0.10  & 0.16  & 0.10  & 0.03  & 0.01  & 0.11  & 0.14  & 0.04 \\
			25    & 25    & 75    & 75    & 0.13  & 0.06  & 0.12  & 0.27  & 0.11  & 0.24  & 0.09  & 0.13  & 0.24  & 0.16  & 0.30  & 0.14  & 0.22  & 0.16  & 0.13 \\
			50    & 25    & 75    & 75    & 0.13  & 0.09  & 0.11  & 0.27  & 0.11  & 0.24  & 0.09  & 0.14  & 0.26  & 0.16  & 0.39  & 0.14  & 0.24  & 0.42  & 0.18 \\
			75    & 25    & 75    & 75    & 0.14  & 0.09  & 0.12  & 0.27  & 0.11  & 0.26  & 0.10  & 0.13  & 0.28  & 0.17  & 0.49  & 0.13  & 0.25  & 0.45  & 0.25 \\
			25    & 50    & 75    & 75    & 0.11  & 0.06  & 0.10  & 0.16  & 0.11  & 0.19  & 0.09  & 0.13  & 0.17  & 0.16  & 0.17  & 0.08  & 0.16  & 0.13  & 0.08 \\
			50    & 50    & 75    & 75    & 0.13  & 0.06  & 0.10  & 0.21  & 0.11  & 0.19  & 0.09  & 0.13  & 0.19  & 0.16  & 0.21  & 0.09  & 0.18  & 0.16  & 0.11 \\
			75    & 50    & 75    & 75    & 0.14  & 0.09  & 0.11  & 0.27  & 0.11  & 0.24  & 0.10  & 0.13  & 0.26  & 0.16  & 0.27  & 0.10  & 0.19  & 0.38  & 0.23 \\
			25    & 75    & 75    & 75    & 0.06  & 0.04  & 0.09  & 0.16  & 0.05  & 0.08  & 0.04  & 0.11  & 0.15  & 0.11  & 0.09  & 0.01  & 0.06  & 0.08  & 0.06 \\
			50    & 75    & 75    & 75    & 0.06  & 0.04  & 0.09  & 0.16  & 0.05  & 0.08  & 0.05  & 0.11  & 0.15  & 0.11  & 0.09  & 0.02  & 0.07  & 0.12  & 0.06 \\
			75    & 75    & 75    & 75    & 0.06  & 0.04  & 0.09  & 0.16  & 0.06  & 0.06  & 0.06  & 0.11  & 0.16  & 0.11  & 0.13  & 0.07  & 0.13  & 0.13  & 0.10 \\
			\bottomrule
		\end{tabular}%
		\label{tabA:MP_L}%
	\end{table}%

	\begin{table}[htbp]
		\tiny
		\centering
		\caption{Marginal Rate of Technical Substitution ($K/L$)}
		\begin{tabular}{rrrr|rrrrr|rrrrr|rrrrr}
			\toprule
			\multicolumn{4}{c|}{Percentile} & \multicolumn{5}{c|}{CAP}               & \multicolumn{5}{c|}{CAPNLS}            & \multicolumn{5}{c}{CNLS} \\
			\multicolumn{1}{l}{$K$} & \multicolumn{1}{l}{$L$} & \multicolumn{1}{l}{$E$} & \multicolumn{1}{l|}{$S$} & \multicolumn{1}{l}{2899} & \multicolumn{1}{l}{2010} & \multicolumn{1}{l}{2811} & \multicolumn{1}{l}{2520} & \multicolumn{1}{l|}{1541} & \multicolumn{1}{l}{2899} & \multicolumn{1}{l}{2010} & \multicolumn{1}{l}{2811} & \multicolumn{1}{l}{2520} & \multicolumn{1}{l|}{1541} & \multicolumn{1}{l}{2899} & \multicolumn{1}{l}{2010} & \multicolumn{1}{l}{2811} & \multicolumn{1}{l}{2520} & \multicolumn{1}{l}{1541} \\
			\midrule
			25    & 25    & 25    & 25    & 0.67  & 8.85  & 6.12  & 0.68  & 6.50  & 1.02  & 3.42  & 2.58  & 2.16  & 1.91  & 1.8   & 33.7  & 11.0  & 6.7   & 46.4 \\
			50    & 25    & 25    & 25    & 0.67  & 3.75  & 5.16  & -0.15 & 6.50  & 0.49  & 1.56  & 2.07  & 1.08  & 1.70  & 0.1   & 0.7   & 1.0   & 0.5   & 8.5 \\
			75    & 25    & 25    & 25    & 0.67  & 3.75  & 0.55  & -0.23 & 2.41  & 0.11  & 1.07  & 1.54  & 0.24  & 1.40  & 0.0   & 0.0   & 0.2   & 0.0   & 0.2 \\
			25    & 50    & 25    & 25    & 0.67  & 11.58 & 5.82  & 1.43  & 12.51 & 1.66  & 4.61  & 3.00  & 2.98  & 2.27  & 22.2  & 166.6 & 16.3  & 61.7  & 102.1 \\
			50    & 50    & 25    & 25    & 0.67  & 11.58 & 5.21  & 0.68  & 6.50  & 1.49  & 3.35  & 2.47  & 2.44  & 1.98  & 3.2   & 8.2   & 1.6   & 2.2   & 73.8 \\
			75    & 50    & 25    & 25    & 0.67  & 5.14  & 1.67  & 0.01  & 2.41  & 0.66  & 1.48  & 1.73  & 0.72  & 1.69  & 0.0   & 0.1   & 0.7   & 0.3   & 0.4 \\
			25    & 75    & 25    & 25    & 3.24  & 34.64 & 4.96  & 1.75  & 12.51 & 3.18  & 9.35  & 2.80  & 3.48  & 3.90  & 685.0 & 3233.1 & 168.4 & 1072.2 & 1407.9 \\
			50    & 75    & 25    & 25    & 3.24  & 34.64 & 4.96  & 1.75  & 12.51 & 2.81  & 7.14  & 2.59  & 3.10  & 3.89  & 104.9 & 54.2  & 2.6   & 32.0  & 1222.8 \\
			75    & 75    & 25    & 25    & 1.09  & 21.11 & 3.02  & 0.81  & 7.10  & 1.96  & 5.05  & 2.19  & 2.04  & 3.69  & 10.1  & 2.4   & 1.5   & 0.8   & 23.6 \\
			25    & 25    & 50    & 25    & 0.67  & 7.01  & 6.12  & 0.70  & 6.50  & 1.09  & 3.01  & 2.77  & 2.04  & 1.90  & 3.6   & 29.9  & 10.8  & 6.5   & 38.1 \\
			50    & 25    & 50    & 25    & 0.67  & 7.01  & 5.82  & -0.15 & 6.50  & 0.53  & 1.57  & 2.27  & 1.06  & 1.66  & 0.5   & 0.6   & 1.7   & 0.9   & 7.3 \\
			75    & 25    & 50    & 25    & 0.67  & 3.75  & 2.72  & -0.15 & 6.50  & 0.12  & 1.05  & 1.55  & 0.24  & 1.44  & 0.0   & 0.0   & 0.1   & 0.0   & 0.1 \\
			25    & 50    & 50    & 25    & 0.67  & 8.85  & 5.82  & 2.10  & 6.50  & 1.63  & 3.90  & 3.00  & 2.86  & 2.10  & 27.3  & 123.3 & 27.2  & 21.9  & 65.8 \\
			50    & 50    & 50    & 25    & 0.67  & 8.85  & 5.82  & 0.10  & 6.50  & 1.49  & 2.84  & 2.63  & 2.26  & 1.97  & 1.5   & 7.3   & 4.0   & 6.3   & 46.5 \\
			75    & 50    & 50    & 25    & 0.67  & 4.22  & 4.26  & 0.10  & 2.41  & 0.67  & 1.34  & 1.84  & 0.64  & 1.58  & 0.0   & 0.0   & 1.2   & 0.1   & 0.3 \\
			25    & 75    & 50    & 25    & 3.24  & 34.64 & 5.21  & 2.60  & 12.51 & 3.23  & 7.37  & 2.87  & 3.55  & 3.88  & 1426.7 & 1366.6 & 41.3  & 2805.8 & 496.3 \\
			50    & 75    & 50    & 25    & 3.24  & 34.64 & 4.96  & 1.83  & 12.51 & 2.90  & 7.28  & 2.71  & 3.16  & 3.71  & 95.5  & 27.5  & 3.7   & 38.8  & 345.2 \\
			75    & 75    & 50    & 25    & 1.09  & 21.11 & 4.96  & 0.81  & 7.10  & 2.09  & 4.86  & 2.18  & 2.07  & 3.15  & 8.9   & 3.1   & 1.0   & 1.5   & 92.8 \\
			25    & 25    & 75    & 25    & 1.09  & 7.01  & 6.12  & 0.24  & 4.74  & 1.06  & 2.08  & 2.85  & 1.70  & 1.92  & 10.9  & 27.8  & 10.1  & 6.9   & 36.7 \\
			50    & 25    & 75    & 25    & 0.67  & 7.01  & 5.50  & 0.24  & 4.74  & 0.77  & 1.77  & 2.46  & 1.07  & 1.58  & 1.6   & 0.6   & 2.3   & 1.5   & 7.1 \\
			75    & 25    & 75    & 25    & 0.67  & 3.75  & 4.84  & -0.15 & 4.74  & 0.20  & 1.28  & 1.99  & 0.28  & 1.43  & 0.1   & 0.0   & 0.2   & 0.1   & 0.1 \\
			25    & 50    & 75    & 25    & 1.09  & 7.01  & 5.82  & 2.60  & 6.50  & 1.85  & 3.35  & 2.89  & 2.74  & 2.16  & 39.6  & 65.4  & 25.9  & 18.7  & 64.4 \\
			50    & 50    & 75    & 25    & 1.09  & 7.01  & 5.50  & 0.79  & 6.50  & 1.84  & 2.77  & 2.71  & 2.09  & 1.88  & 5.7   & 3.4   & 6.4   & 5.7   & 42.7 \\
			75    & 50    & 75    & 25    & 0.67  & 7.01  & 4.84  & 0.10  & 6.50  & 0.88  & 1.55  & 2.30  & 0.59  & 1.66  & 0.3   & 0.0   & 1.4   & 0.9   & 0.3 \\
			25    & 75    & 75    & 25    & 3.24  & 8.85  & 5.82  & 2.60  & 6.50  & 3.36  & 5.65  & 2.85  & 3.41  & 3.19  & 335.3 & 529.7 & 42.5  & 2894.3 & 213.4 \\
			50    & 75    & 75    & 25    & 3.24  & 8.85  & 5.50  & 1.65  & 6.50  & 2.93  & 5.25  & 2.75  & 2.98  & 3.15  & 236.3 & 16.8  & 7.7   & 80.7  & 143.1 \\
			75    & 75    & 75    & 25    & 1.58  & 8.65  & 4.86  & 0.92  & 6.50  & 2.54  & 3.69  & 2.63  & 1.75  & 2.48  & 2.1   & 1.3   & 1.7   & 5.6   & 43.9 \\
			25    & 25    & 25    & 50    & 0.67  & 8.85  & 4.73  & 0.68  & 6.50  & 0.67  & 3.83  & 2.87  & 2.11  & 1.87  & 2.9   & 23.1  & 13.5  & 7.2   & 47.8 \\
			50    & 25    & 25    & 50    & 0.67  & 4.46  & 3.92  & 0.30  & 6.50  & 0.46  & 2.81  & 2.42  & 1.21  & 1.84  & 0.2   & 16.6  & 2.0   & 0.4   & 9.9 \\
			75    & 25    & 25    & 50    & 0.67  & 3.75  & 0.26  & -0.23 & 2.41  & 0.11  & 1.01  & 2.02  & 0.30  & 1.48  & 0.0   & 0.1   & 0.3   & 0.0   & 0.1 \\
			25    & 50    & 25    & 50    & 0.67  & 11.58 & 5.26  & 1.43  & 12.51 & 1.70  & 6.34  & 3.24  & 2.89  & 2.51  & 17.3  & 142.4 & 32.6  & 12.1  & 79.1 \\
			50    & 50    & 25    & 50    & 0.67  & 11.58 & 4.40  & 0.68  & 7.10  & 1.51  & 5.79  & 2.81  & 2.52  & 2.18  & 2.2   & 87.2  & 3.4   & 2.0   & 60.3 \\
			75    & 50    & 25    & 50    & 0.67  & 5.14  & 0.75  & 0.01  & 2.41  & 0.60  & 1.39  & 2.07  & 0.68  & 2.02  & 0.0   & 0.5   & 0.6   & 0.7   & 0.3 \\
			25    & 75    & 25    & 50    & 3.24  & 34.64 & 5.18  & 1.75  & 7.10  & 3.19  & 10.16 & 3.23  & 3.57  & 4.17  & 175.7 & 3873.3 & 352.3 & 1008.8 & 1028.9 \\
			50    & 75    & 25    & 50    & 3.24  & 34.64 & 4.82  & 1.75  & 7.10  & 3.17  & 10.15 & 2.95  & 3.44  & 4.07  & 41.7  & 959.2 & 9.5   & 36.7  & 854.2 \\
			75    & 75    & 25    & 50    & 1.09  & 21.11 & 2.26  & 1.75  & 7.10  & 1.99  & 5.64  & 1.91  & 2.17  & 3.33  & 4.4   & 9.3   & 1.5   & 3.0   & 20.7 \\
			25    & 25    & 50    & 50    & 0.67  & 7.01  & 5.26  & 0.70  & 6.50  & 0.83  & 3.03  & 2.86  & 2.15  & 1.91  & 4.1   & 19.6  & 7.4   & 9.3   & 38.0 \\
			50    & 25    & 50    & 50    & 0.67  & 3.75  & 4.39  & 0.30  & 6.50  & 0.51  & 2.69  & 2.58  & 1.00  & 1.80  & 0.8   & 13.8  & 3.3   & 1.2   & 9.5 \\
			75    & 25    & 50    & 50    & 0.67  & 3.75  & 1.91  & -0.15 & 2.41  & 0.11  & 1.00  & 2.00  & 0.27  & 1.48  & 0.0   & 0.1   & 0.1   & 0.0   & 0.2 \\
			25    & 50    & 50    & 50    & 0.67  & 8.85  & 6.01  & 2.10  & 6.50  & 1.77  & 5.70  & 3.21  & 2.87  & 2.47  & 12.2  & 70.9  & 32.7  & 16.1  & 54.2 \\
			50    & 50    & 50    & 50    & 0.67  & 8.85  & 6.01  & 0.68  & 6.50  & 1.60  & 4.87  & 3.10  & 2.56  & 2.04  & 2.3   & 47.5  & 5.3   & 6.2   & 42.4 \\
			75    & 50    & 50    & 50    & 0.67  & 4.22  & 2.45  & 0.01  & 2.41  & 0.64  & 1.30  & 2.04  & 0.67  & 1.83  & 0.0   & 0.1   & 0.6   & 0.4   & 0.4 \\
			25    & 75    & 50    & 50    & 3.24  & 34.64 & 5.55  & 2.60  & 12.51 & 3.25  & 10.29 & 3.26  & 3.56  & 4.05  & 251.3 & 1989.3 & 63.9  & 775.8 & 241.4 \\
			50    & 75    & 50    & 50    & 3.24  & 21.11 & 5.18  & 1.75  & 12.51 & 3.23  & 10.28 & 3.02  & 3.43  & 3.97  & 59.8  & 1168.0 & 24.7  & 68.8  & 199.5 \\
			75    & 75    & 50    & 50    & 1.09  & 21.11 & 4.60  & 0.81  & 7.10  & 2.08  & 5.54  & 2.15  & 2.14  & 2.85  & 5.6   & 9.4   & 1.5   & 0.8   & 76.5 \\
			25    & 25    & 75    & 50    & 0.67  & 6.08  & 5.77  & 0.70  & 4.74  & 0.97  & 2.64  & 3.00  & 1.60  & 1.93  & 10.2  & 18.3  & 7.3   & 9.6   & 35.3 \\
			50    & 25    & 75    & 50    & 0.67  & 6.08  & 5.48  & 0.70  & 4.74  & 0.67  & 2.43  & 2.83  & 1.05  & 1.69  & 2.3   & 13.4  & 3.5   & 1.3   & 9.2 \\
			75    & 25    & 75    & 50    & 0.67  & 3.75  & 4.86  & -0.15 & 4.74  & 0.24  & 1.19  & 2.16  & 0.27  & 1.46  & 0.1   & 0.0   & 0.4   & 0.1   & 0.2 \\
			25    & 50    & 75    & 50    & 1.09  & 7.01  & 5.82  & 2.10  & 4.74  & 1.89  & 4.56  & 3.06  & 2.81  & 2.30  & 22.3  & 46.7  & 29.3  & 15.6  & 42.5 \\
			50    & 50    & 75    & 50    & 1.09  & 7.01  & 5.82  & 1.18  & 4.74  & 1.81  & 3.86  & 2.94  & 2.45  & 2.02  & 5.4   & 25.3  & 6.2   & 6.2   & 35.9 \\
			75    & 50    & 75    & 50    & 0.67  & 3.75  & 5.16  & 0.10  & 4.74  & 0.84  & 1.42  & 2.53  & 0.51  & 1.58  & 0.4   & 0.0   & 1.6   & 0.7   & 0.3 \\
			25    & 75    & 75    & 50    & 3.24  & 8.85  & 5.82  & 2.60  & 6.50  & 3.49  & 6.98  & 3.38  & 3.40  & 3.02  & 77.2  & 1159.7 & 54.1  & 649.5 & 60.0 \\
			50    & 75    & 75    & 50    & 3.24  & 8.85  & 5.82  & 2.60  & 6.50  & 3.45  & 6.43  & 3.10  & 3.30  & 3.01  & 29.4  & 46.5  & 28.8  & 25.3  & 58.4 \\
			75    & 75    & 75    & 50    & 1.58  & 8.65  & 5.21  & 0.92  & 6.50  & 2.73  & 3.68  & 2.55  & 2.00  & 2.52  & 1.6   & 0.3   & 4.6   & 4.6   & 43.4 \\
			25    & 25    & 25    & 75    & 0.67  & 4.00  & 3.14  & 0.64  & 4.43  & 0.47  & 4.77  & 2.62  & 1.86  & 2.06  & 4.0   & 28.4  & 13.2  & 6.8   & 35.7 \\
			50    & 25    & 25    & 75    & 0.67  & 4.00  & 2.33  & 0.23  & 4.43  & 0.36  & 4.65  & 2.53  & 1.43  & 2.04  & 0.2   & 12.5  & 2.8   & 0.7   & 10.0 \\
			75    & 25    & 25    & 75    & 0.67  & 3.69  & 0.43  & 0.23  & 2.41  & 0.18  & 2.94  & 2.29  & 0.88  & 1.62  & 0.0   & 1.5   & 0.7   & 0.1   & 0.3 \\
			25    & 50    & 25    & 75    & 1.62  & 8.46  & 3.33  & 2.04  & 7.10  & 1.71  & 5.96  & 2.81  & 2.75  & 2.35  & 26.6  & 130.3 & 57.2  & 10.7  & 62.4 \\
			50    & 50    & 25    & 75    & 0.98  & 8.46  & 2.70  & 0.60  & 7.10  & 1.40  & 5.10  & 2.75  & 2.28  & 2.28  & 8.8   & 23.1  & 4.8   & 2.4   & 52.5 \\
			75    & 50    & 25    & 75    & 0.67  & 5.06  & 0.20  & 0.60  & 7.10  & 0.28  & 3.02  & 2.18  & 1.38  & 1.73  & 0.5   & 2.1   & 1.0   & 0.5   & 0.3 \\
			25    & 75    & 25    & 75    & 6.16  & 24.60 & 3.86  & 2.04  & 7.10  & 4.33  & 13.53 & 3.33  & 3.25  & 4.69  & 84.8  & 1012.0 & 512.6 & 33.0  & 360.8 \\
			50    & 75    & 25    & 75    & 6.16  & 24.60 & 2.67  & 1.34  & 7.10  & 4.29  & 12.75 & 3.01  & 3.16  & 4.70  & 14.0  & 3579.5 & 25.5  & 6.2   & 221.3 \\
			75    & 75    & 25    & 75    & 1.30  & 24.60 & 1.18  & 1.34  & 7.10  & 2.90  & 9.89  & 2.20  & 2.76  & 3.13  & 5.5   & 22.2  & 1.9   & 1.5   & 41.4 \\
			25    & 25    & 50    & 75    & 0.67  & 4.00  & 4.70  & 0.70  & 2.08  & 0.58  & 4.77  & 2.73  & 1.87  & 1.78  & 4.5   & 15.8  & 10.9  & 8.9   & 22.3 \\
			50    & 25    & 50    & 75    & 0.67  & 4.00  & 2.68  & 0.30  & 2.08  & 0.37  & 4.65  & 2.64  & 1.31  & 1.69  & 0.6   & 10.8  & 3.1   & 1.3   & 8.5 \\
			75    & 25    & 50    & 75    & 0.67  & 3.44  & 0.34  & 0.23  & 2.08  & 0.18  & 2.94  & 2.28  & 0.82  & 1.47  & 0.0   & 1.4   & 1.6   & 0.1   & 0.4 \\
			25    & 50    & 50    & 75    & 1.25  & 7.64  & 4.43  & 2.04  & 7.10  & 1.54  & 5.96  & 3.03  & 2.79  & 2.57  & 14.9  & 48.0  & 36.6  & 13.9  & 52.8 \\
			50    & 50    & 50    & 75    & 0.98  & 7.15  & 3.62  & 0.60  & 7.10  & 1.43  & 4.94  & 2.79  & 2.38  & 2.28  & 4.0   & 24.2  & 5.6   & 6.0   & 32.2 \\
			75    & 50    & 50    & 75    & 0.67  & 4.39  & 0.46  & 0.60  & 2.41  & 0.27  & 2.86  & 2.37  & 1.26  & 1.67  & 0.0   & 1.8   & 1.9   & 0.5   & 1.0 \\
			25    & 75    & 50    & 75    & 6.16  & 24.60 & 4.48  & 2.04  & 7.10  & 4.50  & 12.75 & 3.47  & 3.27  & 4.67  & 130.9 & 1516.1 & 188.1 & 34.6  & 127.4 \\
			50    & 75    & 50    & 75    & 6.16  & 24.60 & 3.80  & 1.34  & 7.10  & 4.26  & 12.75 & 3.26  & 3.17  & 4.00  & 37.4  & 489.4 & 20.2  & 10.2  & 108.8 \\
			75    & 75    & 50    & 75    & 1.30  & 24.60 & 1.39  & 1.34  & 7.10  & 3.31  & 9.74  & 2.51  & 2.80  & 2.88  & 3.5   & 32.4  & 2.0   & 1.8   & 55.7 \\
			25    & 25    & 75    & 75    & 0.91  & 6.28  & 4.69  & 0.70  & 4.74  & 0.77  & 5.27  & 3.02  & 1.54  & 1.69  & 6.5   & 16.6  & 10.3  & 9.8   & 21.7 \\
			50    & 25    & 75    & 75    & 0.91  & 3.61  & 4.42  & 0.70  & 4.74  & 0.68  & 4.58  & 2.81  & 1.17  & 1.69  & 1.5   & 12.1  & 3.2   & 1.3   & 7.9 \\
			75    & 25    & 75    & 75    & 0.67  & 3.60  & 2.85  & 0.30  & 4.74  & 0.31  & 2.39  & 2.59  & 0.67  & 1.51  & 0.0   & 10.8  & 1.9   & 0.1   & 0.5 \\
			25    & 50    & 75    & 75    & 2.37  & 6.28  & 4.84  & 2.10  & 4.74  & 1.52  & 5.43  & 3.25  & 2.81  & 2.09  & 17.4  & 28.4  & 20.6  & 14.2  & 45.8 \\
			50    & 50    & 75    & 75    & 0.91  & 6.28  & 4.84  & 1.18  & 4.74  & 1.51  & 4.90  & 3.17  & 2.37  & 1.88  & 5.9   & 19.4  & 5.7   & 7.0   & 19.4 \\
			75    & 50    & 75    & 75    & 0.67  & 3.60  & 3.26  & 0.30  & 4.74  & 0.53  & 2.59  & 2.72  & 0.92  & 1.55  & 0.8   & 12.6  & 2.2   & 0.4   & 0.9 \\
			25    & 75    & 75    & 75    & 6.16  & 9.44  & 5.67  & 2.10  & 12.51 & 4.37  & 12.17 & 3.58  & 3.36  & 3.43  & 36.5  & 554.3 & 88.3  & 34.9  & 63.3 \\
			50    & 75    & 75    & 75    & 6.16  & 9.44  & 5.67  & 2.10  & 12.51 & 4.29  & 11.01 & 3.40  & 3.32  & 3.39  & 17.1  & 90.9  & 28.8  & 12.8  & 45.3 \\
			75    & 75    & 75    & 75    & 6.14  & 9.44  & 4.37  & 1.34  & 2.41  & 3.98  & 5.69  & 3.13  & 2.88  & 2.40  & 2.6   & 16.8  & 4.4   & 5.7   & 22.6 \\
			\bottomrule
		\end{tabular}%
		\label{tabA:MRTS}%
	\end{table}%

	\clearpage
	\bibliographystyle{chicago}
	
	\bibliography{reference}

\end{document}